\documentclass[%
 reprint,
nofootinbib,
 amsmath,amssymb,
 aps, onecolumn
]{revtex4-2}

\newenvironment{changemargin}[2]{%
\begin{list}{}{%
\setlength{\topsep}{0pt}%
\setlength{\leftmargin}{#1}%
\setlength{\rightmargin}{#2}%
\setlength{\listparindent}{\parindent}%
\setlength{\itemindent}{\parindent}%
\setlength{\parsep}{\parskip}%
}%
\item[]}{\end{list}}

\usepackage{graphicx}% Include figure files
\usepackage{dcolumn}% Align table columns on decimal point
\usepackage{bm}% bold math
\begin{document}

\preprint{APS/123-QED}

\title{Stochastic inflation at all order in slow-roll parameters:\\ foundations}% Force line breaks with \\

\author{Diego Cruces}
 \email{dcruces@ub.edu}%Lines break automatically or can be forced with \\
\author{Cristiano Germani}%
 \email{germani@icc.ub.edu}
\affiliation{%
 Institut de Ciencies del Cosmos (ICCUB), Universitat de Barcelona, \\Martí i Franquès 1,
E08028 Barcelona, Spain\\
Departement de F\'isica Qu\`antica i Astrofisica, Universitat de Barcelona, Mart\'i i Franqu\`es 1, 08028 Barcelona, Spain
}%

\begin{abstract}
In this paper we develop the formalism for the stochastic approach to inflation at all order in slow-roll parameters. This is done by including the momentum and Hamiltonian constraints into the stochastic equations. We then specialise to the widely used Starobinski approximation where interactions between IR and UV modes are neglected. We show that, whenever this approximation holds, no significant deviations are observed when comparing the two-point correlation functions (power spectrum) calculated with stochastic methods, to the ones calculated with the QFT approach to linear theory. As a byproduct, we argue that: a) the approaches based on the Starobinski approximation, generically, do not capture any loop effects of the quantum scalar-gravity system; b) correlations functions can only be calculated in the linear theory regimes, thus, no non-perturbative statistics can be extracted within this approximation, as commonly claimed.
\end{abstract}

%\keywords{Suggested keywords}%Use showkeys class option if keyword
                              %display desired
\maketitle

%\tableofcontents

\section{Introduction}
\label{Sec_intro}

The possibility that Primordial Black Holes (PBHs) can be a significant fraction (if not all) of the Dark Matter (DM) has been a source of interest for almost 50 years \cite{Chapline:1975ojl}. 

A possible PBHs formation mechanism is through the gravitational collapse of large (non-perturbative)
over-densities generated during an inflationary epoch of the universe. Those perturbations are produced by the quantum fluctuations of the inflaton and are exponentially rare \cite{ilia}. Thus, in order to predict the abundance of PBHs, a precise statistical knowledge of inflationary perturbations is highly desirable.

The hope of the stochastic approach to inflation is that it incorporates quantum corrections to the inflationary dynamics in a non-perturbative way \cite{Starobinsky:1994bd}. However, as we shall see in this paper, the current realisation of it generically fails to achieve this goal. 

In this approach, wavelengths that are well outside the cosmological horizon (the horizon from now on) are approximated in powers of spatial gradients rather than on amplitudes (as in linear theory). At the same time though, those modes are influenced by the quantum sector by receiving quantum-kicks from stochastic forces generated by the perturbative sub-horizon modes. The success of the stochastic formalism resides on the fact that it allows to reduce a quantum problem into a statistical one and it has been widely used in the literature \cite{Casini:1998wr,Pattison:2017mbe,Pattison:2021oen,Firouzjahi:2018vet,Prokopec:2019srf,Ballesteros:2020sre,Assadullahi:2016gkk,Clesse:2010iz}. 

By exploring the foundation of the stochastic formalism, we will identify two important limitation that the current realization of it has:
\begin{enumerate}
\item The use of the separate universe approach, which makes the formalism generically valid only at leading order in slow-roll parameters.
\item The white noises construction, which is only possible in the linear regime.
\end{enumerate}

While we will leave the issue related with the white noise for future work, in this paper we will solve the first limitation, improvin the original formulation of the stochastic approach to a novel one that leads to unprecedentedly precise predictions for the perturbative statistics of inflation.

\section{Gradient expansion at all order in slow-roll}
\label{Sec_gradexp}

In the ADM $(3+1)$-decomposition, the metric related to scalar sources is expressed as \cite{Arnowitt:1962hi}:
\begin{equation}
ds^2=g_{\mu\nu}dx^{\mu}dx^{\nu}=-\alpha^2dt^2+\gamma_{ij}(dx^i+\beta^idt)(dx^j+\beta^jdt),
\label{ADM_metric}
\end{equation}
where the spatial metric can be redefined as $\gamma_{ij}=a^2e^{2\zeta}\tilde{\gamma}_{ij}$ with $\det(\tilde{\gamma}_{ij})=1$. Here $a=a(t)$ is the scale factor.

Using the metric \eqref{ADM_metric}, the Einstein-Hilbert action with a minimally coupled scalar field, takes the following form \footnote{We are using units $c=1$.} 
\begin{equation}
S=\frac{1}{2}\int \sqrt{h}\left[\frac{M_{PL}^2}{2}\left(R^{(3)}+\alpha^{-1}(K_{ij}K_{ij}-K^2)\right)-2\alpha V+\alpha^{-1}(\dot{\phi}-\beta^i\partial_i\phi)^2-\alpha\gamma^{ij}\partial_{i}\phi\partial_{j}\phi\right],
\label{ADM_action}
\end{equation} 
where $R_{ij}^{(3)}$ is the Ricci tensor of the spatial metric, $K_{ij}$ is its extrinsic curvature and $M_{PL}$ is the Planck mass. Explicitly 
\begin{equation}
K_{ij}=\frac{1}{2\alpha}\left(\partial_t\gamma_{ij}-D_i \beta_j-D_j\beta_i\right); \hspace{1 cm} K=\gamma^{ij}K_{ij},
\label{extcurv}
\end{equation}
where $D_i$ represents the covariant derivative with respect to the spatial metric $\gamma_{ij}$. 

It is convenient to decompose the extrinsic curvature into its trace and traceless part as 
\begin{equation}
K_{ij}\equiv \frac{\gamma_{ij}}{3}K+a^2 {A}_{ij}\ ,
\label{extcurv_decomp}
\end{equation}
where $\gamma^{ij}{A}_{ij}=0$. We will also define $\tilde A_{ij}\equiv e^{-2\zeta}A_{ij}$ for later purposes.

In the ADM formalism, $\gamma_{ij}$ and $\phi$ are the dynamical variables. Whereas, $\alpha$ (the lapse) and $\beta^i$ (the shift vector) are Lagrange multipliers for the action \eqref{ADM_action} which generate the following Hamiltonian and momentum constraints of the scalar-gravity system
\begin{equation}
R^{(3)}-\tilde{A}_{ij}\tilde{A}^{ij}+\frac{2}{3}K^2=16\pi G E,
\label{ADM_HAM_main}
\end{equation}
\begin{equation}
D^j\tilde{A}_{i j}-\frac{2}{3}D_iK=8\pi G J_i,
\label{ADM_MOM_main}
\end{equation}
where $E\equiv T_{\mu\nu}n^{\mu}n^{\nu}$ and $J_i\equiv -T_{\mu j}n^{\mu}\gamma^{j}_i$ and $n_\mu=(-\alpha,0,0,0)$ is the form orthogonal to the time slice. 

For completeness the basic equations of this formalism are written in appendix \ref{AppADM}.

\subsection{Inflationary scenarios and slow-roll parameters}
\label{Sec_inflation}
As we have already mentioned, PBHs represent natural candidates for dark matter (DM) (latest constraints on this idea can be found in \cite{constraints}). However, to statistically generate enough PBHs for this to hold one needs, at least, a power spectrum of primordial curvature perturbations several order of magnitudes larger than the one observed in the cosmic microwave background (CMB). 

It is known that a period of Slow-Roll (SR), of which the predictions of the CMB are based upon, cannot lead to the appropriate power spectrum necessary to generate enough PBHs to match the DM abundance \cite{hu}\footnote{For the non-linear relation between the inflationary power spectrum and PBHS abundance, under the assumption of gaussianity, the interested reader can see  \cite{germani-sheth}.}. Thus, one necessarily needs an inflationary epoch evolving beyond SR. A possibility is the introduction of an inflection point in the inflationary potential \cite{Germani:2017bcs}. This, leads the inflaton to undergo a so-called Ultra-Slow-Roll (USR) phase of inflation \cite{Kinney:2005vj, Martin:2012pe}. Taking into account that the statistics of PBHs from non-gaussian fluctuations has yet to be fully developed, the single field USR option with standard kinetic term seems then to be the best \cite{atal}.  

The evolution of a scalar field $\left(\phi\right)$ in an exactly homogeneous and isotropic universe described by a Friedman-Lemaitre-Robertson-Walker (FLRW) metric 
\begin{equation}
ds^2=-dt^2+a(t)^2 d\vec{x}\cdot d\vec{x}\ ,
\end{equation}
has the following equation of motion:
\begin{equation}
\ddot{\phi}+3 H\dot{\phi} + V_{\phi}\left(\phi\right)=0\,,
\label{KG_test}
\end{equation}
where $ V_{\phi}\equiv \frac{\partial V}{\partial \phi}$, $H\equiv \frac{\dot{a}}{a}$ is the Hubble parameter, and a dot denotes a derivative with respect to the cosmic time $t$.

The Friedmann equation is
\begin{equation}
H^2=\frac{1}{3M_{PL}^2}\left(\frac{\dot{\phi}^2}{2}+V\left(\phi\right)\right)\,.
\label{HAM_test}
\end{equation}

The Slow-Roll (SR) parameters $\epsilon_i$ define the rate of change of the Hubble parameter:
\begin{equation}
\epsilon_1 \equiv - \frac{\dot{H}}{H^2} = \frac{\dot{\phi}^2}{2H^2M_{PL}^2} \ll 1 \,; \qquad \epsilon_{i+1} \equiv \frac{\dot{\epsilon}_i}{H \epsilon_i} \quad \text{with} \quad i \geq 1,
\label{epsilon_def}
\end{equation}
where, to write the final expressions, we have used the Friedmann equation and the equation of motion of the field.

We can now define different inflationary regimes depending on the values of the $\epsilon_i$s:

\begin{itemize}
\item Slow-Roll (SR): The field is slowly rolling down a potential with an almost constant velocity, which makes the acceleration negligible. In this case the equation of motion \eqref{KG_test} are approximately
\begin{equation}
3 H\dot{\phi} + V_{\phi}\left(\phi\right) \simeq 0\,.
\label{KG_test_SR}
\end{equation}

The SR parameters are much smaller than one ($\epsilon_i \ll 1$) and can be written in terms of the potential as

\begin{equation}
\epsilon_1^{SR} \simeq \frac{1}{2M_{PL}^2}\left(\frac{V_{\phi}}{V}\right)^2\,; \qquad \epsilon_2^{SR} \simeq \frac{2}{M_{PL}^2}\left(\frac{V_{\phi\phi}}{V}\right)-4\epsilon_1^{SR}\,.
\label{epsilon_SR}
\end{equation}

\item Ultra-Slow-Roll (USR): The field is moving along an exactly flat potential $\left(V_{\phi}=0\right)$, which makes the acceleration relevant. In this case the equation of motion \eqref{KG_test} is
\begin{equation}
\ddot{\phi}+3H\dot{\phi}=0\,.
\label{KG_test_USR}
\end{equation}

From \eqref{KG_test_USR} one can infer that the velocity of the field (and hence $\epsilon_1$) exponentially decreases, which makes some $\epsilon_i \sim \mathcal{O}(1)$. More precisely:

\begin{align} \nonumber
\epsilon_i^{USR}=-6+2\epsilon_1^{USR} & \qquad \text{when i even}. \\
\epsilon_i^{USR}=2\epsilon_1^{USR} & \qquad\text{when i $>$ 1 and odd}.
\label{epsilon_USR}
\end{align}

An exponential decrease of $\epsilon_1$ makes the power spectrum of curvature perturbation increase.

\item Both SR and USR are, at least approximately, sub-cases of Constant-Roll (CR). Here $\frac{\ddot{\phi}}{H\dot{\phi}} = \kappa$ where $\kappa$ is a constant. SR is realized when $\kappa=0$ while USR when $\kappa=-3$. We will not analyse further this generic case. 
\end{itemize}

It is important to remark that, given a potential related to PBH formation, the SR and USR phases alternate. Thus, the equations of motion \eqref{KG_test_SR} and \eqref{KG_test_USR} will always be an approximation of the system.

\subsection{Gradient expansion}
\label{Sec_gradexp_sepuni}

The gradient expansion approximation  \cite{Salopek:1990jq,Lyth:2004gb}, consists in considering small patches of the Universe which can be approximately described by a local Friedman geometry. By choosing some local coordinates $(t_l,\vec{x}_l)$, this geometry may be described by a FLRW metric 
\begin{equation}
ds_l^2=-dt_l^2+a_l(t_l)^2\delta_{ij}dx_l^idx_l^j\ .
\label{patch_metric}
\end{equation}
We define the local Hubble expansion as $H_l\equiv \frac{\partial_{t_l}a_l}{a_l}$.

The patch is chosen in a way that the characteristic scale of inhomogeneities, which we call it $L$, is much larger than $H_l^{-1}$. One can then define an expansion parameter $\sigma\equiv (H_lL)^{-1}\ll 1$.

Reversing the argument, at leading order in $\sigma$, each patch of the universe of size $\left(\sigma H_l\right)^{-1}$ (the coarse grained scale) is approximately described by an homogeneous Friedman universe. Higher order terms in $\sigma$ expansion will instead capture local inhomogeneities. 

Contrary to the linear theory approach to cosmological perturbations, the gradient expansion is valid for any amplitude of local over-densities, as long as the patch is taken small enough for the gradients to be negligible. This aspect, lead many authors to claim that the stochastic approach to cosmological perturbations may give non-perturbative information on scalar correlations functions \cite{Vennin:2015hra,Grain:2017dqa,Pattison:2017mbe,Clesse:2015wea,Kunze:2006tu}. However, as we shall discuss, the way those correlations are calculated via the stochastic methods, can only give information about the linear approximation regimes.   

We conclude this section by stressing that the assumption of which the gradient expansion is based upon, is that a patch can be found such that any spatial gradient would only introduces an order $\sigma$. In other words, for any generic function $X$, $\partial_i X \sim H_l\,X \times \mathcal{O}(\sigma)$ in the patch chosen. 

\subsection{Background versus local metrics}
\label{bg_vs_loc}

In absence of quantum fluctuations of the scalar-metric system (we have in mind inflation), one can define a global background metric with coordinates $t$ and $x^i$:
\begin{equation}
ds^2_b=-dt^{2} + a(t)^2 \delta_{ij}dx^idx^j\ ,
\label{BG_metric}
\end{equation}
where generically $a$ does not coincide with $a_l$.

It is straightforward to show that in any local patch, by considering only scalar perturbations and in isotropic spatial coordinates, we can always rewrite the metric \eqref{patch_metric} as 
\begin{equation}
	ds^2_{l}=-_{(0)}\alpha^2 dt_p^{2} + _{(0)}\gamma_{ij}\left(dx_p^i + _{(0)}\beta^i dt_p\right)\left(dx_p^j + _{(0)}\beta^j dt_p\right)\ ,
	\label{mp}
\end{equation}
with the conditions 
\begin{enumerate}
	\item $_{(0)}\alpha = _{(0)}\alpha (t_p)$,
	\item $_{(0)}\beta^i = b(t_p)\ x^i_p$, and finally
	\item $_{(0)}\gamma_{ij}=\gamma(t_p)\delta_{ij}$\ .
	\label{coord_transf_cond2}
\end{enumerate}
In the metric \eqref{mp} we have used the sub-script $_{(0)}$ to remind the reader that we are at zeroth order in gradient expansion and $(t_p,\vec{x}_p)$ to define generic coordinates for the patch chosen.

The functions $_{(0)}\alpha(t_p),\ \gamma(t_p)$ and $b(t_p)$ depend on the gauge chosen and the solution of the Einstein equations.

There is no loss of generality in defining $\gamma(t_p)=a(t_p)^2 e^{2_{(0)}\zeta(t_p)}$ where $a(t_p)$ has the same functional form of the background scale factor and $_{(0)}\zeta(t_p)$ (called curvature perturbation in linear theory) is a generically non-vanishing function introduced because, generically, $\gamma(t_p)\neq a(t_p)$. 

Note that the local and background metrics live on two different spaces. Thus, there is no any coordinate transformation relating them. However, to simplify notation, from now on we will set $x^\mu_p\rightarrow x^\mu$ being careful to treat $a(t)$ as the solution of the Einstein equations in absence of any quantum over-density. With this, we also define the ``background'' Hubble parameter $H^b(t)\equiv\frac{\partial_t a(t)}{a(t)}$.

With those definitions, the local patch metric reads
\begin{equation}
	ds^2_{l}=-_{(0)}\alpha^2 dt^{2} + a(t)^2 e^{2_{(0)}\zeta}\delta_{ij}\left(dx^i + _{(0)}\beta^i dt\right)\left(dx^j + _{(0)}\beta^j dt\right)\ .
	\label{patch_metric_BG}
\end{equation}
Because of the over-densities, the metric \eqref{patch_metric_BG} and \eqref{BG_metric} must differ:
Suppose that \eqref{patch_metric_BG} was obtained by employing the so-called spatially flat gauge, where $_{(0)}\zeta=0$. Then, one immediately see that either $_{(0)}\alpha$ or $b(t)$ (or both) must be different from their background value. This is a well known result from perturbation theory: the perturbed lapse and shift functions have, generically, an homogeneous time dependent part (see e.g. \cite{Maldacena:2002vr}). Setting $_{(0)}\alpha = 1$, $_{(0)}\zeta = 0$ and $_{(0)}\beta^i = 0$, as they are in the local patch coordinates, would then introduce errors which we will quantify later on.

At the next to leading order in gradient expansion, the metric in the local patch can still be written in the ADM form. With the identification $\gamma_{ij}\equiv a(t)^2 e^{2\zeta}\tilde\gamma_{ij}$, we have
\begin{equation}
	ds^2_{l}=-\alpha^2 dt^{2} + a(t)^2 e^{2\zeta}\tilde\gamma_{ij}\left(dx^i + \beta^i dt\right)\left(dx^j + \beta^j dt\right)\ ,
	\label{patch_metric_full}
\end{equation}
where, as we have already discussed,
\begin{align} \nonumber
_{(0)}\alpha (t) \sim \mathcal{O}(\sigma^0) \qquad _{(0)} \zeta (t) \sim \mathcal{O}(\sigma^0) \qquad & _{(0)}\partial_i\beta^i (x^i,t)\sim \mathcal{O}(\sigma^{0}) \\
 \qquad \tilde{\gamma}_{ij} - \delta_{ij} \sim \mathcal{O}(\sigma) \qquad _{(0)}\phi \sim \mathcal{O}(\sigma^0)\ . &
\label{order_estimation}
\end{align}

The last term has been added to take into account the expansion of the scalar field, which is generically non-zero at the background level.

\subsection{The importance of constraints}
\label{Sec_const}

In this section we will briefly justify why the momentum constraint plays a very important role to capture slow-roll suppressed terms.

Let us do a step back and considering the case of linear perturbation theory. Suppose again we consider the spatially flat gauge ($\gamma_{ij}=a^2\delta_{ij}$ for any order in gradient expansion): the only remaining scalar degree of freedom is $\delta\phi$. The Hamiltonian and momentum constraints are simultaneously satisfied if and only if \cite{Maldacena:2002vr}:
 \begin{equation}
 \alpha \simeq 1 + \sqrt{\frac{\epsilon_1}{2M_{PL}^2}}\,\delta\phi
 \label{alphalin}
 \end{equation} 
  and 
 \begin{equation}
 \partial_i\beta^i \simeq - \left(a H^b\right)\sqrt{\frac{\epsilon_1}{2M_{PL}^2}}\left[\delta\dot{\phi}-H^b\frac{\epsilon_2}{2}\delta\phi\right]\,,
 \label{betalin}
 \end{equation}  
  which obviously contain long wavelength terms which are the would-be equivalent to $_{(0)}\alpha$ and $_{(0)}\partial_i {\beta}^i$.

Let us now go back to gradient expansion. At next to leading order in $\sigma$ the momentum constraint, as we shall show later on, has the following functional form  

\begin{equation}
\partial_i \left(F\left(\alpha, \partial_i\beta^i, \phi\right)\right) = \partial_i \left(G\left(\alpha, \partial_i\beta^i, \phi\right)\right) + \mathcal{O}(\sigma^2)\ ,
\label{ADM_MOM_funct}
\end{equation}
where $F$ and $G$ are generic functions that we do not specify here. It would seem reasonable that the momentum constraint, at zeroth order in $\sigma$, is automatically satisfied leading to $_{(0)}\alpha = 1$, $_{(0)}\zeta = 0$ and $_{(0)}\beta^i = 0$ from the Hamiltonian constraint. However, this would be in clear contradiction to the results of perturbation theory and with the fact that \eqref{patch_metric_BG} and \eqref{BG_metric} must differ.

Thus, necessarily, the momentum constraint must contain terms at order zero in gradient expansion. This is what actually happens in perturbation theory. Generically, the shift, entering in the momentum constraint, is non-local \cite{Maldacena:2002vr}. 

\subsubsection{SR as an exception (separate universe approach)}
\label{Sec_const_SR}
The way the difference between \eqref{patch_metric_BG} and \eqref{BG_metric} is usually introduced in the literature \cite{Sugiyama:2012tj} is by setting $\partial_i\,_{(0)}\beta^i=0$ while allowing $\,_{(0)}\alpha$ to be a homogeneous function of time. As we have already said this is in odds with \eqref{betalin}. Nevertheless, during a SR regime, one can check that (see equations \eqref{QdotQ_exp}-\eqref{nu_SR} of appendix \ref{AppLin_uniH} for details):
$$\frac{\delta\dot{\phi}}{H^b\delta\phi} \simeq \frac{\epsilon_2}{2}+\mathcal{O}(\epsilon_i^2)\,,$$
 which makes $\partial_i\,_{(0)}\beta^i$ to be of higher order in gradient expansion up to an accuracy of next-to leading order in $\epsilon_1$. Thus, in this case, the momentum constraint give information only at next-to-next-to leading order in $\epsilon_1$ and can therefore be discarded within the SR approximation. The same is not true in regimes beyond SR where $\partial_i\,_{(0)}\beta^i$ is of the same order as $\,_{(0)}\alpha - 1$. 
 
Three very important aspects are worthy to remark here:

\begin{enumerate}
\item Whenever we are setting  $\partial_i\,_{(0)}\beta^i=0$, we are also discarding the momentum constraint. This is because otherwise, momentum and Hamiltonian constraints on superhorizon scales would be incompatible.
\item Discarding the momentum constraint introduces an error in the system that can be quantified in terms of the slow-roll parameters and it depends on the regime of inflation:
\begin{itemize}
\item During a SR regime, the error appears at next-to-next-to leading order in $\epsilon_1$.
\item During any other regime, the error appears at leading order in $\epsilon_1$, which is equivalent to not considering gravity backreactions.
\end{itemize}
\item Discarding the momentum constraint also means that we are not considering information about the interaction between the different Hubble patches. Neglecting the momentum constraint at leading order in gradient expansion is then equivalent to study an ensemble of Hubble patches that evolve as separate FLRW universes. This approximation is the so-called \textit{separate universe approach}.

Although it could seem that the separate universe approach has nothing to do with the expansion in slow-roll parameters, we have seen in this subsection that they are closely related. This is because the separate universe approach resides in two main approximations: a) leading order in gradient expansion and b) discarding the momentum constraint. While the former is valid at all orders in slow roll parameters, the second is not. This statement is not in disagreement with some previous works remarking the wide applicability of the separate universe approach \cite{Tanaka:2021dww,Garriga:2015tea}: the conclusions of \cite{Tanaka:2021dww} are valid under the assumption of locality while the conclusions of \cite{Garriga:2015tea} can only be applied to an attractor infationary regime, both conditions are not generically satisfied. Specifically, for example in the spatially flat gauge, the shift vector is non-local \cite{Maldacena:2002vr} and an USR (or constant-roll) regime does not have an attractor behaviour. In slow-roll, however, the non-locality appears only at next-to-next to leading order in $\epsilon_1$.

\end{enumerate}

\section{Stochastic formalism: foundations}
\label{Sec_sto}
The idea of the stochastic approach to inflation is to reduce the evolution of the full quantum scalar-gravity system to an equivalent stochastic problem \cite{Starobinsky:1986fx}. Considering the Fourier decomposition of the metric and scalar field, this is done by splitting the variables of interest (let us say $X$) into two parts: a long-wavelength part (also said Infra-Red (IR)) in which $\frac{k}{\sigma aH}< 1$ (where $k$ is the Fourier mode of the function $X$), and a short-wavelength part (also said Ultra-Violet (UV)). The UV part evolves well inside the Hubble radius and, in agreement with the on-set of inflation, is perturbatively small. Thus, one can use linear perturbation theory for the UV. 

The IR part instead can be large. It can be thought that the IR part is a result of a ``condensate'' of UV modes. However, because the IR part only contains long-wavelengths, the gradient expansion can be there used. In principle, the gradient expansion can give information at all order in perturbation theory or even about non-perturbative regimes (which are relevant for PBHs). This is the reason why the local universe approach is so appealing.

\subsection{An explanatory example: stochastic formalism in spatially flat gauge}
\label{Sec_sto_flat}

The Stochastic formalism is based on three main approximations. To illustrate this we will consider the Hamiltonian constraint \eqref{ADM_HAM_main} in spatially flat gauge where $\gamma_{ij}=a^2\delta_{ij}$.

We have 
\begin{equation}
-\left(\tilde{A}_{\text{f}}\right)_{ij}\left(\tilde{A}_{\text{f}}\right)^{ij}+\frac{2}{3}K_{\text{f}}^2 - \frac{2}{M_{PL}^2}\left(T_{\text{f}}\right)_{\mu\nu}n^{\mu}n^{\nu}=0\ ,
\label{ADM_HAM_flat}
\end{equation}
where $n^{\mu}=g^{\mu\nu}n_{\nu}=\left(\frac{1}{\alpha_{\text{f}}},-\frac{\left(\beta^i\right)_{\text{f}}}{\alpha_{\text{f}}}\right)$. Note that we have introduced a sub-index ``f'' to specify that all quantities are calculated in the spatially flat gauge.

Eq. \eqref{ADM_HAM_flat} can be written in terms of the metric variables $\alpha_{\text{f}}$ and $\left( \beta_{\text{f}}\right)_i$ and the scalar field $\phi_{\text{f}}$, using the definitions of $\tilde{A}_{ij}$ and $K$ given in the introduction. The result is the following:

\begin{changemargin}{-1.5cm}{-1cm} 
\begin{align} \nonumber
- &\frac{1}{4 \alpha^2_{\text{f}}}\left[\delta^{ik}\partial^j \left(\beta_{\text{f}}\right)_k + \delta^{jk}\partial^i \left(\beta_{\text{f}}\right)_k - \frac{2}{3}\delta^{ij}\partial^k \left(\beta_{\text{f}}\right)_k\right] \left[\delta_{ik}\partial_j \left(\beta_{\text{f}}\right)^k + \delta_{jk}\partial_i \left(\beta_{\text{f}}\right)^k - \frac{2}{3}\delta_{ij}\partial_k \left(\beta_{\text{f}}\right)^k\right] \\ \nonumber
+ & \frac{2}{3}\left(-3 \frac{H^b}{\alpha_{\text{f}}} + \frac{1}{\alpha_{\text{f}}}\partial_k \left(\beta_{\text{f}}\right)^k \right)^2 \\ 
- & \frac{2}{M_{PL}^2} \left[\frac{\dot{\phi}^2_{\text{f}}}{2\alpha^2_{\text{f}}} - \frac{\dot{\phi}_{\text{f}}\left(\beta_{\text{f}}\right)^i\partial_i \phi_{\text{f}}}{\alpha^2_{\text{f}}} + \frac{\left(\beta_{\text{f}}\right)^i\left(\beta_{\text{f}}\right)^j \partial_i \phi_{\text{f}}\partial_j\phi_{\text{f}}}{2\alpha^2_{\text{f}}} + V\left(\phi_{\text{f}}\right)+ \frac{\partial^i \phi_{\text{f}}\partial_i\phi_{\text{f}}}{2}\right]=0 \ .
\label{ADM_HAM_flat_expanded}
\end{align}
\end{changemargin}

Equation \eqref{ADM_HAM_flat_expanded} is a bit cumbersome but it is very helpful to understand the way stochastic approach to inflation is constructed. As anticipated at the beginning of this section, the first thing to do is to split the variables of interest into their $IR$ and $UV$ parts. 

In spatially flat gauge we have only three variables to split:
\begin{align}  \nonumber
\alpha_{\text{f}}=\alpha^{IR}_{\text{f}}+\alpha^{UV}_{\text{f}} \\ \nonumber
\left(\beta_{\text{f}}\right)^i=\left(\beta^{IR}_{\text{f}}\right)^i+\left(\beta^{UV}_{\text{f}}\right)^i \\ 
\phi_{\text{f}}=\phi^{IR}_{\text{f}}+\phi^{UV}_{\text{f}}
\label{split}
\end{align}
We are now ready to construct the stochastic system:

\begin{itemize}
\item Due to the perturbative nature of the $X^{UV}$ variables, we will expand the Hamiltonian constraint keeping only linear terms in UV and isolate them in the right hand side of the equation, getting \footnote{We thank Aichen Li for pointing out a typo in the equation \eqref{ADM_HAM_flat_lin}.}

\begin{changemargin}{-3cm}{-1cm} 
\begin{align}
 \nonumber
- &\frac{1}{4 \left(\alpha^{IR}_{\text{f}}\right)^2}\left[\delta^{ik}\partial^j \left(\beta^{IR}_{\text{f}}\right)_k + \delta^{jk}\partial^i \left(\beta^{IR}_{\text{f}}\right)_k - \frac{2}{3}\delta^{ij}\partial^k \left(\beta^{IR}_{\text{f}}\right)_k\right] \left[\delta_{ik}\partial_j \left(\beta^{IR}_{\text{f}}\right)^k + \delta_{jk}\partial_i \left(\beta^{IR}_{\text{f}}\right)^k - \frac{2}{3}\delta_{ij}\partial_k \left(\beta^{IR}_{\text{f}}\right)^k\right] \\ \nonumber
+ & \frac{2}{3}\left(-3 \frac{H^b}{\alpha^{IR}_{\text{f}}} + \frac{1}{\alpha^{IR}_{\text{f}}}\partial_k \left(\beta_{\text{f}}\right)^k \right)^2 \\ \nonumber
- & \frac{2}{M_{PL}^2} \left[\frac{\left(\dot{\phi}^{IR}_{\text{f}}\right)^2}{2\left(\alpha^{IR}_{\text{f}}\right)^2} - \frac{\dot{\phi}^{IR}_{\text{f}}\left(\beta^{IR}_{\text{f}}\right)^i\partial_i \phi^{IR}_{\text{f}}}{\left(\alpha^{IR}_{\text{f}}\right)^2} + \frac{\left(\beta^{IR}_{\text{f}}\right)^i\left(\beta^{IR}_{\text{f}}\right)^j \partial_i \phi^{IR}_{\text{f}}\partial_j\phi^{IR}_{\text{f}}}{2\left(\alpha^{IR}_{\text{f}}\right)^2} + \frac{\partial^i \phi^{IR}_{\text{f}}\partial_i\phi^{IR}_{\text{f}}}{2} + V\left(\phi^{IR}_{\text{f}}\right)\right] \\
\nonumber
=&\frac{\alpha_{\text{f}}^{UV}}{2(\alpha_{\text{f}}^{IR})^{3}}\big[\delta^{jk}\partial^{i}(\beta_{\text{f}}^{IR})_{k}+\delta^{ik}\partial^{j}(\beta_{\text{f}}^{IR})_{k}\big]\big[\delta_{jl}\partial_{i}(\beta_{\text{f}}^{IR})^{l}+\delta_{il}\partial_{j}(\beta_{\text{f}}^{IR})^{l}\big]+\frac{2}{3}\frac{\alpha^{UV}}{(\alpha_{\text{f}}^{IR})^{3}}(\partial_{i}(\beta_{\text{f}}^{IR})^{i})^{2}-\frac{2}{3}\frac{\partial_{i}(\beta_{\text{f}}^{IR})^{i}}{(\alpha_{\text{f}}^{IR})^{2}}\partial_{j}(\beta_{\text{f}}^{UV})^{j}\\
\nonumber
&-\frac{1}{2(\alpha_{\text{f}}^{IR})^{2}}\big[\delta^{jk}\partial^{i}(\beta_{\text{f}}^{UV})_{k}+\delta^{ik}\partial^{j}(\beta_{\text{f}}^{UV})_{k}\big]\big[\delta_{jl}\partial_{i}(\beta_{\text{f}}^{IR})^{l}+\delta_{il}\partial_{j}(\beta_{\text{f}}^{IR})^{l}\big]+\left( \frac{12(H^{b})^{2}}{(\alpha_{\text{f}}^{IR})^{3}}-\frac{8H^{b}\partial_{i}(\beta_{\text{f}}^{IR})^{i}}{(\alpha_{\text{f}}^{IR})^{3}}\right)\alpha_{\text{f}}^{UV}+\frac{4H^{b}\partial_{i}(\beta_{\text{f}}^{UV})^{i}}{(\alpha_{\text{f}}^{IR})^{2}}\\
\nonumber
&+  \frac{2}{M_{PL}^{2}}\bigg[\partial^{j}\phi_{\text{f}}^{IR}\partial_{j}\phi_{\text{f}}^{UV}+\frac{\dot{\phi}_{\text{f}}^{IR}}{(\alpha_{\text{f}}^{IR})^{2}}\dot{\phi}_{\text{f}}^{UV}-\frac{(\dot{\phi}_{\text{f}}^{IR})^{2}}{(\alpha_{\text{f}}^{IR})^{3}}\alpha_{\text{f}}^{UV}+V^{\prime}(\phi_{\text{f}}^{IR})\phi_{\text{f}}^{UV}+\frac{2\dot{\phi}_{\text{f}}^{IR}(\beta_{\text{f}}^{IR})^{i}\partial_{i}\phi_{\text{f}}^{IR}}{(\alpha_{\text{f}}^{IR})^{3}}\alpha_{\text{f}}^{UV}-\frac{(\beta_{\text{f}}^{IR})^{i}\partial_{i}\phi_{\text{f}}^{IR}}{(\alpha_{\text{f}}^{IR})^{2}}\dot{\phi}_{\text{f}}^{UV}\\
\nonumber
&-\frac{\dot{\phi}_{\text{f}}^{IR}(\beta_{\text{f}}^{IR})^{i}}{(\alpha_{\text{f}}^{IR})^{2}}\partial_{i}\phi_{\text{f}}^{UV}-\frac{\dot{\phi}_{\text{f}}^{IR}\partial_{i}\phi_{\text{f}}^{IR}}{(\alpha_{\text{f}}^{IR})^{2}}(\beta_{\text{f}}^{UV})^{i}+\frac{\partial_{j}\phi_{\text{f}}^{IR}(\beta_{\text{f}}^{IR})^{i}(\beta_{\text{f}}^{IR})^{j}}{(\alpha_{\text{f}}^{IR})^{2}}\partial_{i}\phi_{\text{f}}^{UV}+\frac{\partial_{i}\phi_{\text{f}}^{IR}\partial_{j}\phi_{\text{f}}^{IR}(\beta_{\text{f}}^{IR})^{j}}{(\alpha_{\text{f}}^{IR})^{2}}(\beta_{\text{f}}^{UV})^{i}\\
&-\frac{\partial_{i}\phi_{\text{f}}^{IR}\partial_{j}\phi_{\text{f}}^{IR}(\beta_{\text{f}}^{IR})^{i}(\beta_{\text{f}}^{IR})^{j}}{(\alpha_{\text{f}}^{IR})^{3}}\alpha_{\text{f}}^{UV}\bigg]
\label{ADM_HAM_flat_lin}
\end{align}
\end{changemargin}

\item Secondly, since the IR variables are well outside the Hubble horizon, a gradient expansion can be performed over them. Keeping ourselves at leading order in gradient expansion (see section \ref{Sec_gradexp_sepuni} for details), equation \eqref{ADM_HAM_flat_lin} becomes:

\begin{changemargin}{-3cm}{-1cm} 
\begin{align}
 \nonumber
- &\frac{1}{4 \left(_{(0)}\alpha^{IR}_{\text{f}}\right)^2}\left[\delta^{ik}\partial^j \left(_{(0)}\beta^{IR}_{\text{f}}\right)_k + \delta^{jk}\partial^i \left(_{(0)}\beta^{IR}_{\text{f}}\right)_k - \frac{2}{3}\delta^{ij}\partial^k \left(_{(0)}\beta^{IR}_{\text{f}}\right)_k\right] \\ \nonumber
\times & \left[\delta_{ik}\partial_j \left(_{(0)}\beta^{IR}_{\text{f}}\right)^k + \delta_{jk}\partial_i \left(_{(0)}\beta^{IR}_{\text{f}}\right)^k - \frac{2}{3}\delta_{ij}\partial_k \left(_{(0)}\beta^{IR}_{\text{f}}\right)^k\right] \\ \nonumber
+ & \frac{2}{3}\left(-3 \frac{H^b}{_{(0)}\alpha^{IR}_{\text{f}}} + \frac{1}{_{(0)}\alpha^{IR}_{\text{f}}}\partial_k \left(_{(0)}\beta^{IR}_{\text{f}}\right)^k \right)^2 \\
\nonumber
-  & \frac{2}{M_{PL}^2} \left[\frac{\left(_{(0)}\dot{\phi}^{IR}_{\text{f}}\right)^2}{2\left(_{(0)}\alpha^{IR}_{\text{f}}\right)^2} - \frac{_{(0)}\dot{\phi}^{IR}_{\text{f}}\left(_{(0)}\beta^{IR}_{\text{f}}\right)^i\partial_i \left(_{(0)}\phi^{IR}_{\text{f}}\right)}{\left(_{(0)}\alpha^{IR}_{\text{f}}\right)^2} + \frac{\left(_{(0)}\beta^{IR}_{\text{f}}\right)^i\left(_{(0)}\beta^{IR}_{\text{f}}\right)^j \partial_i \left(_{(0)}\phi^{IR}_{\text{f}}\right) \partial_j \left(_{(0)}\phi^{IR}_{\text{f}}\right)}{2\left(_{(0)}\alpha^{IR}_{\text{f}}\right)^2} + V\left(_{(0)}\phi^{IR}_{\text{f}}\right)\right] \\
\nonumber
=&\frac{\alpha_{\text{f}}^{UV}}{2(_{(0)}\alpha_{\text{f}}^{IR})^{3}}\big[\delta^{jk}\partial^{i}({}_{(0)}\beta_{\text{f}}^{IR})_{k}+\delta^{ik}\partial^{j}({}_{(0)}\beta_{\text{f}}^{IR})_{k}\big]\big[\delta_{jl}\partial_{i}({}_{(0)}\beta_{\text{f}}^{IR})^{l}+\delta_{il}\partial_{j}({}_{(0)}\beta_{\text{f}}^{IR})^{l}\big]\\	
\nonumber
&-\frac{1}{2(_{(0)}\alpha_{\text{f}}^{IR})^{2}}\big[\delta_{jl}\partial_{i}({}_{(0)}\beta_{\text{f}}^{IR})^{l}+\delta_{il}\partial_{j}({}_{(0)}\beta_{\text{f}}^{IR})^{l}\big]\big[\delta^{jk}\partial^{i}(\beta_{\text{f}}^{UV})_{k}+\delta^{ik}\partial^{j}(\beta_{\text{f}}^{UV})_{k}\big]\\	
\nonumber
&+\frac{2}{3}\frac{(\partial_{i}({}_{(0)}\beta_{\text{f}}^{IR})^{i})^{2}}{(_{(0)}\alpha_{\text{f}}^{IR})^{3}}\alpha^{UV}-\frac{2}{3}\frac{\partial_{i}({}_{(0)}\beta_{\text{f}}^{IR})^{i}}{(_{(0)}\alpha_{\text{f}}^{IR})^{2}}\partial_{j}(\beta_{\text{f}}^{UV})^{j}
+\left( \frac{12(H^{b})^{2}}{(_{(0)}\alpha_{\text{f}}^{IR})^{3}}-\frac{8H^{b}\partial_{i}({}_{(0)}\beta_{\text{f}}^{IR})^{i}}{(_{(0)}\alpha_{\text{f}}^{IR})^{3}} \right)\alpha_{\text{f}}^{UV}+\frac{4H^{b}}{(_{(0)}\alpha_{\text{f}}^{IR})^{2}}\partial_{i}(\beta_{\text{f}}^{UV})^{i}\\
\nonumber
&+\frac{2}{M_{PL}^{2}}\bigg[\frac{_{(0)}\dot{\phi}_{\text{f}}^{IR}}{(_{(0)}\alpha_{\text{f}}^{IR})^{2}}\dot{\phi}_{\text{f}}^{UV}-\frac{(_{(0)}\dot{\phi}_{\text{f}}^{IR})^{2}}{(_{(0)}\alpha_{\text{f}}^{IR})^{3}}\alpha_{\text{f}}^{UV}+V_{\phi}({}_{(0)}\phi_{\text{f}}^{IR})\phi_{\text{f}}^{UV} + \frac{2_{(0)}\dot{\phi}_{\text{f}}^{IR}\left(_{(0)}\beta_{\text{f}}^{IR}\right)^{i}\partial_{i}\left(_{(0)}\phi_{\text{f}}^{IR}\right)}{\left(_{(0)}\alpha_{\text{f}}^{IR}\right)^{3}}\alpha_{\text{f}}^{UV}-\frac{\left(_{(0)}\beta_{\text{f}}^{IR}\right)^{i}\partial_{i}\left(_{(0)}\phi_{\text{f}}^{IR}\right)}{\left(_{(0)}\alpha_{\text{f}}^{IR}\right)^{2}}\dot{\phi}_{\text{f}}^{UV}\\
&-\frac{_{(0)}\dot{\phi}_{\text{f}}^{IR}({}_{(0)}\beta_{\text{f}}^{IR})^{i}}{(_{(0)}\alpha_{\text{f}}^{IR})^{2}}\partial_{i}\phi_{\text{f}}^{UV}+\frac{\partial_{j}({}_{(0)}\phi_{\text{f}}^{IR})({}_{(0)}\beta_{\text{f}}^{IR})^{i}({}_{(0)}\beta_{\text{f}}^{IR})^{j})}{(_{(0)}\alpha_{\text{f}}^{IR})^{2}}\partial_{i}\phi_{\text{f}}^{UV}-\frac{\partial_{i}\left(_{(0)}\phi_{\text{f}}^{IR}\right)\partial_{j}\left(_{(0)}\phi_{\text{f}}^{IR}\right)\left(_{(0)}\beta_{\text{f}}^{IR}\right)^{i}\left(_{(0)}\beta_{\text{f}}^{IR}\right)^{j}}{\left(_{(0)}\alpha_{\text{f}}^{IR}\right)^{3}}\alpha_{\text{f}}^{UV}\bigg] 
\label{ADM_HAM_flat_grad}
\end{align}
\end{changemargin}

where we have inserted and extra subindex $\,_{(0)}$ to indicate that we are at leading order in gradient expansion.

\end{itemize}

Using Fourier analysis, we can now define more rigorously the IR and UV modes. If we choose the Heaviside theta as the window function to split these two modes (as done in the stochastic approaches to inflation) we have

\begin{align} \nonumber
X^{IR}(t,\mathbf{x})\equiv\int\frac{d \mathbf{k}}{(2\pi)^{3/2}}\Theta ( \sigma a_l(t)H_l(t)-k )\mathcal{X}_{\mathbf{k}}(\mathbf{x},t),\\
X^{UV}(t,\mathbf{x})\equiv\int\frac{d \mathbf{k}}{(2\pi)^{3/2}}\Theta ( k - \sigma a_l(t)H_l(t) )\mathcal{X}_{\mathbf{k}}(\mathbf{x},t)\ .
\label{Splitting}
\end{align}
Note that, in the spirit of gradient expansion, the splitting is done on the local cosmological horizon $a_l H_l$ which generically differs from the one of the background.

 Note also that in spatially flat gauge and at leading order in gradient expansion we have the following:

\begin{equation}
a_l= a(t)\, e^{-\int \frac{1}{3}\partial_i\left(_{(0)}\beta^{IR}_{\text{f}}\right)^i dt } ;  \qquad H_l=\frac{H^b}{_{(0)}\alpha_{\text{f}}} - \frac{1}{3 \,_{(0)}\alpha_{\text{f}}}\partial_i\left(_{(0)}\beta^{IR}_{\text{f}}\right)^i\,.
\end{equation}
This has been obtained by noticing that the three-dimensional scalar $K\equiv -3 H_l$.

Inserting the definition of $X^{UV}$ of equation \eqref{Splitting} into \eqref{ADM_HAM_flat_grad} we get the following expression:
 
 \begin{changemargin}{-2.5cm}{-1cm} 
\begin{align}
 \nonumber
 \qquad \quad & -  \frac{1}{4 \left(_{(0)}\alpha^{IR}_{\text{f}}\right)^2} \left[\delta^{ik}\partial^j \left(_{(0)}\beta^{IR}_{\text{f}}\right)_k + \delta^{jk}\partial^i \left(_{(0)}\beta^{IR}_{\text{f}}\right)_k - \frac{2}{3}\delta^{ij}\partial^k \left(_{(0)}\beta^{IR}_{\text{f}}\right)_k\right] \\ \nonumber
& \times  \left[\delta_{ik}\partial_j \left(_{(0)}\beta^{IR}_{\text{f}}\right)^k + \delta_{jk}\partial_i \left(_{(0)}\beta^{IR}_{\text{f}}\right)^k - \frac{2}{3}\delta_{ij}\partial_k \left(_{(0)}\beta^{IR}_{\text{f}}\right)^k\right] \\ \nonumber
 & +  \frac{2}{3}\left(-3 \frac{H^b}{_{(0)}\alpha^{IR}_{\text{f}}} + \frac{1}{_{(0)}\alpha^{IR}_{\text{f}}}\partial_k \left(_{(0)}\beta^{IR}_{\text{f}}\right)^k \right)^2 -  \frac{2}{M_{PL}^2} \left[\frac{\left(_{(0)}\dot{\phi}^{IR}_{\text{f}}\right)}{2\left(_{(0)}\alpha^{IR}_{\text{f}}\right)^2} + V\left(_{(0)}\phi^{IR}_{\text{f}}\right)\right] \\ \nonumber
& =  -\partial_{t}\big(\sigma a_{l}H_{l}\big)\int\frac{d\mathbf{k}}{(2\pi)^{3/2}}\delta(k-\sigma a_{l}H_{l})(\varphi_{\mathbf{k}}^{UV})_{\text{f}}\bigg\{\frac{2}{M_{PL}^{2}}\frac{_{(0)}\dot{\phi}_{\text{f}}^{IR}}{(_{(0)}\alpha_{\text{f}}^{IR})^{2}} - \frac{\left(_{(0)}\beta_{\text{f}}^{IR}\right)^{i}\partial_{i}\left(_{(0)}\phi_{\text{f}}^{IR}\right)}{\left(_{(0)}\alpha_{\text{f}}^{IR}\right)^{2}}\bigg\}
\\ \nonumber 
 & +\int\frac{d\mathbf{k}}{(2\pi)^{3/2}}\Theta(k-\sigma a_{l}H_{l})\bigg\{\frac{\big[\delta^{jk}\partial^{i}({}_{(0)}\beta_{\text{f}}^{IR})_{k}+\delta^{ik}\partial^{j}({}_{(0)}\beta_{\text{f}}^{IR})_{k}\big]\big[\delta_{jl}\partial_{i}({}_{(0)}\beta_{\text{f}}^{IR})^{l}+\delta_{il}\partial_{j}({}_{(0)}\beta_{\text{f}}^{IR})^{l}\big]}{2(_{(0)}\alpha_{\text{f}}^{IR})^{3}}\left(\boldsymbol{\alpha}_{\mathbf{k}}^{UV}\right)_{\text{f}} \\ 
 \nonumber
 &-\frac{\big[\delta_{jl}\partial_{i}({}_{(0)}\beta_{\text{f}}^{IR})^{l}+\delta_{il}\partial_{j}({}_{(0)}\beta_{\text{f}}^{IR})^{l}\big]}{2(_{(0)}\alpha_{\text{f}}^{IR})^{2}}\big[\delta^{jk}\partial^{i}\left(\left(\boldsymbol{\beta}_{\mathbf{k}}^{UV}\right)_{\text{f}}\right)_{k}+\delta^{ik}\partial^{j}\left(\left(\boldsymbol{\beta}_{\mathbf{k}}^{UV}\right)_{\text{f}}\right)_{k}\big]+\frac{2}{3}\frac{(\partial_{i}({}_{(0)}\beta_{\text{f}}^{IR})^{i})^{2}}{(_{(0)}\alpha_{\text{f}}^{IR})^{3}}\left(\boldsymbol{\alpha}_{\mathbf{k}}^{UV}\right)_{\text{f}} \\
 \nonumber
 &-\frac{2}{3}\frac{\partial_{i}({}_{(0)}\beta_{\text{f}}^{IR})^{i}}{(_{(0)}\alpha_{\text{f}}^{IR})^{2}}\partial_{j}\left(\left(\boldsymbol{\beta}_{\mathbf{k}}^{UV}\right)_{\text{f}}\right)^{j}+\left( \frac{12(H^{b})^{2}-8H^{b}\partial_{i}({}_{(0)}\beta_{\text{f}}^{IR})^{i}}{(_{(0)}\alpha_{\text{f}}^{IR})^{3}} \right)\left(\boldsymbol{\alpha}_{\mathbf{k}}^{UV}\right)_{\text{f}}+\frac{4H^{b}}{(_{(0)}\alpha_{\text{f}}^{IR})^{2}}\partial_{i}\left(\left(\boldsymbol{\beta}_{\mathbf{k}}^{UV}\right)_{\text{f}}\right)^{i}\\
 \nonumber
  &+\frac{2}{M_{PL}^{2}}\bigg[\frac{_{(0)}\dot{\phi}_{\text{f}}^{IR}}{(_{(0)}\alpha_{\text{f}}^{IR})^{2}}(\dot{\varphi}_{\mathbf{k}}^{UV})_{\text{f}}-\frac{(_{(0)}\dot{\phi}_{\text{f}}^{IR})^{2}}{(_{(0)}\alpha_{\text{f}}^{IR})^{3}}\left(\boldsymbol{\alpha}_{\mathbf{k}}^{UV}\right)_{\text{f}}+V_{\phi}({}_{(0)}\phi_{\text{f}}^{IR})(\varphi_{\mathbf{k}}^{UV})_{\text{f}} + \frac{2_{(0)}\dot{\phi}_{\text{f}}^{IR}\left(_{(0)}\beta_{\text{f}}^{IR}\right)^{i}\partial_{i}\left(_{(0)}\phi_{\text{f}}^{IR}\right)}{\left(_{(0)}\alpha_{\text{f}}^{IR}\right)^{3}}\left(\boldsymbol{\alpha}_{\mathbf{k}}^{UV}\right)_{\text{f}}-\frac{\left(_{(0)}\beta_{\text{f}}^{IR}\right)^{i}\partial_{i}\left(_{(0)}\phi_{\text{f}}^{IR}\right)}{\left(_{(0)}\alpha_{\text{f}}^{IR}\right)^{2}}(\dot{\varphi}_{\mathbf{k}}^{UV})_{\text{f}}\\ \nonumber
  &-\frac{_{(0)}\dot{\phi}_{\text{f}}^{IR}({}_{(0)}\beta_{\text{f}}^{IR})^{i}}{(_{(0)}\alpha_{\text{f}}^{IR})^{2}}\partial_{i}(\varphi_{\mathbf{k}}^{UV})_{\text{f}}+\frac{\partial_{j}({}_{(0)}\phi_{\text{f}}^{IR})({}_{(0)}\beta_{\text{f}}^{IR})^{i}({}_{(0)}\beta_{\text{f}}^{IR})^{j})}{(_{(0)}\alpha_{\text{f}}^{IR})^{2}}\partial_{i}(\varphi_{\mathbf{k}}^{UV})_{\text{f}}-\frac{\partial_{i}\left(_{(0)}\phi_{\text{f}}^{IR}\right)\partial_{j}\left(_{(0)}\phi_{\text{f}}^{IR}\right)\left(_{(0)}\beta_{\text{f}}^{IR}\right)^{i}\left(_{(0)}\beta_{\text{f}}^{IR}\right)^{j}}{\left(_{(0)}\alpha_{\text{f}}^{IR}\right)^{3}}\left(\boldsymbol{\alpha}_{\mathbf{k}}^{UV}\right)_{\text{f}}\bigg]\bigg\}\\
\label{ADM_HAM_flat_fou}
\end{align}
\end{changemargin}
where $\left(\varphi_{\mathbf{k}}\right)_{\zeta}$, $\left(\boldsymbol{\alpha}^{UV}_{\mathbf{k}}\right)_{\text{f}}$ and $\partial_k \left(\left(\boldsymbol{\beta}^{UV}_{\mathbf{k}}\right)_{\text{f}}\right)^k$ are operators defined as in \eqref{Quant_operator}.

The right hand side of \eqref{ADM_HAM_flat_fou} has two different terms:
\begin{itemize}
\item The second integral (terms multiplying $\Theta(k- \sigma a_l H_l)$) is the Hamiltonian constraint at sub-horizon scales. Assuming it is satisfied once the Bunch-Davies vacuum is chosen, it can be consistently set to zero. 

\item In this respect, the first integral, proportional to a Dirac delta, can be seen as a boundary condition for the IR Hamiltonian.
\end{itemize}

 We then get  
 \begin{changemargin}{-2.5cm}{-1cm} 
	\begin{align}
		\nonumber
		- &\frac{1}{4 \left(_{(0)}\alpha^{IR}_{\text{f}}\right)^2}\left[\delta^{ik}\partial^j \left(_{(0)}\beta^{IR}_{\text{f}}\right)_k + \delta^{jk}\partial^i \left(_{(0)}\beta^{IR}_{\text{f}}\right)_k - \frac{2}{3}\delta^{ij}\partial^k \left(_{(0)}\beta^{IR}_{\text{f}}\right)_k\right] \\ \nonumber
		\times & \left[\delta_{ik}\partial_j \left(_{(0)}\beta^{IR}_{\text{f}}\right)^k + \delta_{jk}\partial_i \left(_{(0)}\beta^{IR}_{\text{f}}\right)^k - \frac{2}{3}\delta_{ij}\partial_k \left(_{(0)}\beta^{IR}_{\text{f}}\right)^k\right] \\ \nonumber
		+ & \frac{2}{3}\left(-3 \frac{H^b}{_{(0)}\alpha^{IR}_{\text{f}}} + \frac{1}{_{(0)}\alpha^{IR}_{\text{f}}}\partial_k \left(_{(0)}\beta^{IR}_{\text{f}}\right)^k \right)^2 -  \frac{2}{M_{PL}^2} \left[\frac{\left(_{(0)}\dot{\phi}^{IR}_{\text{f}}\right)}{2\left(_{(0)}\alpha^{IR}_{\text{f}}\right)^2} + V\left(_{(0)}\phi^{IR}_{\text{f}}\right)\right] \\ =&-\partial_{t}\big(\sigma a_{l}H_{l}\big)\int\frac{d\mathbf{k}}{(2\pi)^{3/2}}\delta(k-\sigma a_{l}H_{l})(\varphi_{\mathbf{k}}^{UV})_{\text{f}}\bigg\{\frac{2}{M_{PL}^{2}}\frac{_{(0)}\dot{\phi}_{\text{f}}^{IR}}{(_{(0)}\alpha_{\text{f}}^{IR})^{2}} - \frac{\left(_{(0)}\beta_{\text{f}}^{IR}\right)^{i}\partial_{i}\left(_{(0)}\phi_{\text{f}}^{IR}\right)}{\left(_{(0)}\alpha_{\text{f}}^{IR}\right)^{2}}\bigg\}
		\label{above}
	\end{align}
\end{changemargin}
In order to interpret \eqref{above} as a stochastic equation, the right hand side should be, at least approximately, a white noise. 

The IR and UV modes are coupled in the Hamiltonian constraint of $\left(\varphi^{UV}_{\mathbf{k}}\right)_{\text{f}}$ . Thus, at least at the exact level, one cannot interpret $\left(\varphi^{UV}_{\mathbf{k}}\right)_{\text{f}}$ as a quantum Gaussian variable, as it would be in linear theory.

More technically, the dynamics of the system becomes non-Markovian, namely, the noises themselves, modify the local background in which they are computed. This is very difficult to treat and, as far as we know, the only attempt to do that has been done in \cite{Figueroa:2020jkf}. 

To circumvent this problem, it is commonly assumed (we will call this the Starobinski approximation \cite{Starobinsky:1986fx}) that  
$Y^{IR}X^{UV}=Y^{b}X^{UV}+{\cal O}{\left((X^{UV})^2\right)}$. Here, $X^{UV}$ and $Y^{IR}$ are any UV and IR functions. We then define $Y^b$ as the equivalent background function of $Y^{IR}$. As an example, under this assumption, the Hamiltonian term $$V_{\phi}\left(\,_{(0)}\phi_{\text{f}}^{IR}\right)\phi_{\text{f}}^{UV}= V_{\phi}\left(\phi^b\right)\phi_{\text{f}}^{UV}+{\cal O}{\left((\phi_{\text{f}}^{UV})^2\right)}\ ,$$ where $V_{\phi}\left(\phi^b\right)$ is calculated in the global background \eqref{BG_metric}.

The Starobinski approximation is equivalent to state that any $Y^{IR}-Y^b={\cal O}(X^{UV})$. Thus, we immediately see that if this approximation holds, stochastic inflation can only exactly reproduce the results of linear theory in the linear regime. We will discuss specific examples later on. 

\subsubsection{White noise}
\label{Sec_white_noise}

Having adopted the Starobinski approximation we are now ready to interpret \eqref{above} as a stochastic equation. First of all we now define ``the noise'' as
\begin{equation}
	\left.\xi_1(t)\right|_{\text{f}}=-\sigma a \left(H^b\right)^2 \left(1-\epsilon_1\right)\int \frac{d^3 k}{(2\pi)^{3/2}}\delta(k-\sigma a H^b)\left.\delta\varphi_{\mathbf{k}}\right|_{\text{f}},
	\label{Field_noi_old}\ ,
\end{equation}
    where $\delta\varphi |_{\text{f}}$ is the fluctuation of the scalar field on the background calculated in the spatially flat gauge. With this we have that \eqref{above} can be approximately written as
\begin{changemargin}{-2.5cm}{-1cm} 
	\begin{align}
		\nonumber
		- &\frac{1}{4 \left(_{(0)}\alpha^{IR}_{\text{f}}\right)^2}\left[\delta^{ik}\partial^j \left(_{(0)}\beta^{IR}_{\text{t}}\right)_k + \delta^{jk}\partial^i \left(_{(0)}\beta^{IR}_{\text{t}}\right)_k - \frac{2}{3}\delta^{ij}\partial^k \left(_{(0)}\beta^{IR}_{\text{f}}\right)_k\right] \\ \nonumber
		\times & \left[\delta^{ik}\partial^j \left(_{(0)}\beta^{IR}_{\text{f}}\right)_k + \delta^{jk}\partial^i \left(_{(0)}\beta^{IR}_{\text{f}}\right)_k - \frac{2}{3}\delta^{ij}\partial^k \left(_{(0)}\beta^{IR}_{\text{f}}\right)_k\right] \\ 
		+ & \frac{2}{3}\left(-3 \frac{H^b}{_{(0)}\alpha^{IR}_{\text{f}}} + \frac{1}{_{(0)}\alpha^{IR}_{\text{f}}}\partial_k \left(_{(0)}\beta^{IR}_{\text{f}}\right)^k \right)^2 -  \frac{2}{M_{PL}^2} \left[\frac{\left(_{(0)}\dot{\phi}^{IR}_{\text{f}}\right)^2}{2\left(_{(0)}\alpha^{IR}_{\text{f}}\right)^2} + V\left(_{(0)}\phi^{IR}_{\text{f}}\right)\right] = \frac{2}{M_{PL}^{2}}\frac{_{(0)}\dot{\phi}_{\text{f}}^{IR}}{(_{(0)}\alpha_{\text{f}}^{IR})^{2}}\left.\xi_1(t)\right|_{\text{f}} \,.
		\label{ADM_HAM_flat_final}
	\end{align}
\end{changemargin}
Because we are in inflation, the UV modes evolve fully quantum-mechanically and the IR ones do it stochastically. The reason is the well known fact that at super-horizon scales the quantum system is in a squeezed state \cite{Kiefer:2008ku, Grishchuk:1990bj}.

The noise can then be calculated by considering that the quantum evolution of the UV modes are generically defined by the hermitian operator $\mathcal{X}^q_{\mathbf{k}}(\mathbf{x},t)$:
\begin{equation}
	\mathcal{X}^q_{\mathbf{k}}(\mathbf{x},t)=e^{-i\mathbf{k}\cdot \mathbf{x}}X_{\mathbf{k}}(t)a_{\mathbf{k}}+e^{i\mathbf{k}\cdot \mathbf{x}}X^{\star}_{\mathbf{k}}(t)a^{\dagger}_{\mathbf{k}} \, ,
	\label{Quant_operator}
\end{equation}
where $a_{\mathbf{k}}$ and $a^{\dagger}_{\mathbf{k}}$ are the usual creation and annihilation operators related to $\mathcal{X}^q_{\mathbf{k}}(\mathbf{x},t)$. Finally, $X_{\mathbf{k}}(t)$ is the solution of the evolution equations in the global background and at deep sub-horizon scales.

The integral of \eqref{Field_noi_old} evaluates the field fluctuations at the coarse-grained scale. Here, the $UV$ perturbations, that started from a coherent vacuum sate, have evolved into a highly squeezed state in which the variable $\left(\varphi^{UV}_{\mathbf{k}}\right)_{\text{f}}$ can take any value with corresponding probability $\left|\left(\phi^{UV}_{\mathbf{k}}\right)_{\text{f}}\right|^2$ \cite{Kiefer:2008ku, Grishchuk:1990bj}. 

We would like to end this section by warning the reader that the stochastic equation \eqref{ADM_HAM_flat_final} is still slightly inconsistent. The point is that, by the same Starobinski approximation adopted on the right hand side, the left hand side should also be linearized. We will nevertheless bare this inconsistency as long as the correlations functions calculated with stochastic means will coincide, up to second order in perturbation theory, to the once calculated in linear perturbation theory with QFT methods. On the contrary, while the result of the stochastic method will be un-physical, inconsistencies between the two approaches will signal the break-down of perturbation theory.

From now on, the rest of the paper is devoted to check whether, in all cases of interest, the stochastic formalism is equivalent to linear perturbation theory at all order in slow roll parameters.

By the help of the momentum constraint, we will then construct a stochastic formalism which is valid at all orders in $\epsilon_i$ and will call this the ``new'' stochastic formalism. Our ``new'' stochastic formalism can be thought as a nontrivial check of the linear QFT results and, at the same time, it is easier to implement numerically, especially for the calculation of higher correlations functions.

Before doing so, we will illustrate the stochastic method to inflation used so far and call it the ``old'' stochastic formalism. As we shall see, this method make use of the separate universe approach and hence it is generically only valid at leading order in $\epsilon$ as explained in section \ref{Sec_const_SR}

\subsection{``Old'' stochastic formalism: spatially flat gauge with $\partial_i\left(_{(0)}\beta_{\text{f}}\right)^i=0$.}
\label{Sec_flat_beta0}

The fact that under the Starobinski approximation, stochastic inflation can only reproduce linear perturbation theory, has not been made explicit until now, as far as we know. This lack of awareness, has made of the stochastic formalism one of the most used framework to study non-linear effects during inflation. It has been used in the cases of Slow-Roll \cite{Casini:1998wr,Pattison:2017mbe} and beyond Slow-Roll \cite{Pattison:2021oen,Firouzjahi:2018vet,Prokopec:2019srf,Ballesteros:2020sre} for single field inflationary models, or in multifield inflation \cite{Assadullahi:2016gkk,Clesse:2010iz}.

The gauge typically used for the stochastic approach to inflation is the spatially flat gauge \cite{Ramos:2013nsa,Grain:2017dqa} with the further approximation $\partial_i\left(_{(0)}\beta_{\text{f}}\right)^i=0$. As we have already mention in section \ref{Sec_const_SR}, this further approximation is only consistent in a SR regime up to next-to leading order in $\epsilon_1$, although it has been used, inconsistently, also in other contexts. Another approach used in the literature has been to compute the noises within spatially flat gauge while using the uniform $N$ (number of e-foldings for the perturbed system) gauge in the IR part \cite{Pattison:2019hef}. This is consistent only at zeroth order in $\epsilon_1$, where the uniform $N$ and spatially flat gauges coincide. Thus, we can simply consider the spatially flat gauge with $\partial_i\left(_{(0)}\beta_{\text{f}}\right)^i=0$.

Using the background number of e-folds $dN=H^b dt$ as a time variable and neglecting all terms proportional to $\epsilon_1$ (in order to be consistent with $\partial_i\left(_{(0)}\beta_{\text{f}}\right)^i=0$), we will arrive to the following SR stochastic equation (see Appendix \ref{AppOld} for the derivation):
	\begin{equation}
		\frac{\partial \phi^{IR}_{\text{f}}}{\partial N}+3M_{PL}^2\frac{V_{\phi}\left(\phi^{IR}_{\text{f}}\right)}{V\left(\phi^{IR}_{\text{f}}\right)}=\frac{H^b}{2\pi}\xi(N)|_{\text{f}}\,.
		\label{STO_old_SR}
	\end{equation}
where $\langle\xi(N_1)|_{\text{f}} \xi(N_2)|_{\text{f}}\rangle=\delta \left(N_1-N_2\right)$. Note that, in order to be consistent with the Starobinsky approximation the noises has been calculated at leading order in $\epsilon_1$.
	
We would like to stress once more that this equation is only valid in the linear regime in which $\phi_\text{f}^{IR}-\phi_b=\cal{O}(\delta\phi|_\text{f})$ and at leading order in slow roll. Thus it is less precise than perturbation theory, contrary to what commonly stated.

At zeroth order in $\epsilon_1$ for the lapse, i.e. taking $_{(0)}\alpha_{\text{f}}= 1$, the approximation $\partial_i\left(_{(0)}\beta_{\text{f}}\right)^i=0$ is also always consistent. In this respect, one can write the stochastic USR system as
	\begin{align}
		\nonumber
		\,_{(0)}\pi^{IR}_{\text{f}}&=\frac{\partial \,_{(0)}\phi^{IR}_{\text{f}}}{\partial N} + \frac{H^b}{2\pi}\xi(N)|_{\text{f}}\,, \\ 
		\frac{\partial \,_{(0)} \pi^{IR}_{\text{f}}}{\partial N} & = - 3\,_{(0)}\pi^{IR}_{\text{f}}\,.
		\label{STO_old_USR}
	\end{align}
 
In the next section we will construct a ``new'' stochastic formalism which is valid at all order in $\epsilon_1$. This, as we have already mentioned, will be achieved by making use of the momentum constraint. It turns out that the simplest gauge to study the momentum constraint is the uniform Hubble gauge, thus, our stochastic equations will be written in this gauge rather than the spatially flat one, as used until now in the literature.

\section{``New'' stochastic formalism}
\label{Sec_uniH}
In this section we will use the so called uniform Hubble gauge \cite{Tanaka:2007gh}, where $K=-3\frac{\dot{a}}{a}=-3H^b$. This gauge does not fix the coordinates uniquely and one can further impose $\beta^i=0$. Note that this gauge bypasses immediately all the issues related to the correct estimation of the gradient order of the shift vector. 

The procedure to follow is exactly the same as the one explained in the section \ref{Sec_sto_flat} and the details can be found in the appendix \ref{AppNew}. Here we will only write the main results. Moreover, to make reading easier, we are also going to suppress the sub-indices and super-indices indicating the gauge and the gradient expansion orders. This means that an $\alpha$ here will mean $\,_{(0)}\alpha^{IR}_{\delta K}$, where $\delta K\equiv K+3 H^b=0$ specifies the gauge, and so on. Moreover, as a further simplification of notation we will now use $H\equiv H^b$ 

The full stochastic system is the following set of coupled equations: 
\begin{itemize}
\item The evolution equation for the spatial metric \eqref{ADM_EVSPA}:
\begin{equation}
\frac{\partial\zeta}{\partial N} - \left(\alpha-1\right) =  - \left. \xi_{4}(N)\right|_{\delta K=0}\,,
\label{STO_EVSPA_H}
\end{equation}
where $\left.\xi_4(N)\right|_{\delta K=0}$ is given in appendix \ref{AppNew}.

\item Scalar field equation for the field \eqref{ADM_KG}
\begin{changemargin}{-1.5cm}{-1cm} 
\begin{equation}
\frac{\partial \pi}{\partial N} + (3\alpha - \epsilon_1)\frac{\partial \tilde{\phi}}{\partial N} + \alpha \frac{V_{\phi}(\phi)}{H^2}= - \left(3-\epsilon_1\right)\left.\xi_1(N)\right|_{\delta K=0} - \left.\xi_2(N)\right|_{\delta K=0} + \frac{\partial \phi^b}{\partial N}\left(\left.\xi_3(N)\right|_{\delta K=0}+3\left.\xi_4(N)\right|_{\delta K=0}\right),
\label{STO_KG_H}
\end{equation}
\end{changemargin}
where $\left.\xi_1(N)\right|_{\delta K=0}$, $\left.\xi_2(N)\right|_{\delta K=0}$ and $\left.\xi_3(N)\right|_{\delta K=0}$ are again given in the appendix \ref{AppNew}. In \eqref{STO_KG_H} we have also used the following redefinition:
\begin{equation}
\pi\equiv\frac{1}{\alpha}\frac{\partial \phi}{\partial N} + \left.\xi_1(N)\right|_{\delta K=0}=\frac{\partial \tilde{\phi}}{\partial N} + \left.\xi_1(N)\right|_{\delta K=0}.
\label{STO_redef_H}
\end{equation}

\item Hamiltonian constraint \eqref{ADM_HAM}
\begin{equation}
H^2=\frac{V(\phi)}{3M_{PL}^2-\frac{1}{2}\left(\frac{\partial \tilde{\phi}}{\partial N}\right)^2 - \frac{\partial \phi^b}{\partial N}\left.\xi_1(N)\right|_{\delta K=0}}=\frac{V(\phi^b)}{3M_{PL}^2-\frac{1}{2}\left(\frac{\partial \phi^b}{\partial N}\right)^2} \,.
\label{STO_HAM_H}
\end{equation}

\item Evolution equation for the trace of the extrinsic curvature \eqref{ADM_EVEXT}
\begin{equation}
\left(\frac{\partial \phi^b}{\partial N}\right)^2=\alpha\left(\frac{\partial \tilde{\phi}}{\partial N}\right)^2 + \frac{2}{3}\left(2+\alpha\right)\frac{\partial \phi^b}{\partial N}\left.\xi_1(N)\right|_{\delta K=0} \,.
\label{STO_EVEXT_H}
\end{equation}

This equation, together with the Hamiltonian constraint \eqref{STO_HAM_H}, gives an exact solution for $\alpha$:

\begin{equation}
\alpha = \frac{3 \left(\frac{\partial \phi^b}{\partial N}\right)^2 - 4\frac{\partial \phi^b}{\partial N}\left.\xi_1(N)\right|_{\delta K=0}}{18 M_{PL}^2\left(1-\frac{V\left(\phi\right)}{V\left(\phi^b\right)}\right)+ \frac{V\left(\phi\right)}{V\left(\phi^b\right)}\left(\frac{\partial \phi^b}{\partial N}\right)^2 - 4\frac{\partial \phi^b}{\partial N}\left.\xi_1(N)\right|_{\delta K=0}} \,.
\label{alpha}
\end{equation}

Making use of the Starobinski approximation we get 
\begin{equation}
\alpha=1+\left(1-\frac{V\left(\phi\right)}{V\left(\phi^b\right)}\right)-\frac{6 M_{PL}^2}{\left(\frac{\partial \phi^b}{\partial N}\right)^2}\left(1-\frac{V\left(\phi\right)}{V\left(\phi^b\right)}\right)
\label{alpha_exp}
\end{equation}

\item Because of the novelty of the momentum constraint in the stochastic framework we will here make explicit all steps. We start from \eqref{ADM_MOM}. As explained in section \ref{bg_vs_loc}, the way of extracting some information from the momentum constraint at leading order in gradient expansion is by going to next to leading order in gradient expansion and then applying the limit $\sigma \rightarrow 0$ (see \cite{Hamazaki:2008mh}).  This is the reason we consider the exact momentum constraint valid at all orders in $\sigma$:
\begin{equation}
D^j\,_{(n)}\tilde{A}_{i j}= - H^b\frac{\frac{\partial \,_{(n)}\phi}{\partial N}}{\,_{(n)}\alpha M_{PL}^2}\partial_i\,_{(n)}\phi\ .
\label{STO_MOM_H}
\end{equation}
The sub-index $\,_{(n)}$ means that we are at all orders in gradient expansion.

The momentum constraint can be written in a more convenient way as
\begin{equation}
e^{-3\,_{(n)}\zeta}\partial^j\left(e^{3\,_{(n)}\zeta}\,_{(n)}\tilde{A}_{ij}\right)= - H^b\frac{\frac{\partial \,_{(n)}\phi}{\partial N}}{\,_{(n)}\alpha M_{PL}^2}\partial_i\,_{(n)}\phi.
\label{STO_MOM_H_1}
\end{equation}
Moreover, $\,_{(n)}\tilde{A}_{ij}$ can be written in terms of the time derivative of $\,_{(n)}\tilde{\gamma}_{ij}$ using \eqref{ADM_EVSPA2} getting:
\begin{equation}
\frac{1}{2}\left[3\partial^j\,_{(n)}\zeta\partial_N\left(\,_{(n)}\tilde{\gamma}_{ij}-\delta_{ij}\right)-\frac{\partial^j\,_{(n)}\alpha}{\,_{(n)}\alpha}\partial_N\left(\,_{(n)}\tilde{\gamma}_{ij}-\delta_{ij}\right) + \partial_N\partial^j\left(\,_{(n)}\tilde{\gamma}_{ij}-\delta_{ij}\right)\right] =  \frac{\frac{\partial \,_{(n)}\phi}{\partial N}}{M_{PL}^2}\partial_i\,_{(n)}\phi .
\label{STO_MOM_H_3}
\end{equation}

We now split between $IR$ and $UV$ by at the same time keeping only terms up to $\mathcal{O}(\sigma)$. The result is:

\begin{equation}
\frac{1}{2}\left[ \partial_N\partial^j\left(\,_{(1)}\tilde{\gamma}_{ij}-\delta_{ij}\right)\right] - \frac{\frac{\partial\phi}{\partial N}}{M_{PL}^2}\partial_i(\phi) = \frac{2}{3}\partial_i \left.\xi_5(N)\right|_{\delta K=0} + \mathcal{O}(\sigma^2),
\label{STO_MOM_H_noi}
\end{equation}
where $\left.\xi_5(N)\right|_{\delta K=0}$ is given in the appendix \ref{AppNew}.

Note that in the previous computation  $\partial^j\left(\,_{(1)}\tilde{\gamma}_{ij}-\delta_{ij}\right)\sim \mathcal{O}(\sigma)$. At a first look this would seem incorrect, however it is correct. By using the definition of the logarithm of a matrix together with fact that $\det{\tilde{\gamma}_{ij}}=1$, we can write the following identity : 
$$\left(\,_{(1)}\tilde{\gamma}_{ij}-\delta_{ij}\right) \simeq \log \left(\,_{(1)}\tilde{\gamma}_{ij}\right)= - M_{ij}$$
where $M_{ij}$ is a traceless matrix that can be written as $2\left(\partial_i\partial_j C - \frac{1}{3}\delta_{ij}\nabla^2 C\right)$, where $C$ is the scalar mode. 

The order estimation of Eq. \eqref{order_estimation} implies $\partial_i\partial_j C - \frac{1}{3}\delta_{ij}\nabla^2 C\sim \mathcal{O}(\sigma)$. However, this is not in contradiction (once isotropic coordinates are chosen) with $\nabla^2 C\sim\mathcal{O}(\sigma^0)$\footnote{Take for example $C= \textbf{x}\cdot \textbf{x}  \, g(t, \sigma \textbf{x})$, where $g$ is an arbitrary function. In this case we have:

$$\partial_i \partial_j C - \frac{1}{3}\delta_{ij}\nabla^2 C = \mathcal{O}(\sigma)$$
$$\frac{1}{3}\nabla^2 C=2 g(t,0) + \mathcal{O}(\sigma)$$

 and hence both $\partial_i \partial_j C - \frac{1}{3}\delta_{ij}\nabla^2 C$ and $\partial ^j\left(\partial_i \partial_j C - \frac{1}{3}\delta_{ij}\nabla^2 C\right)=\partial^j\left(\frac{2}{3}\nabla^2 C\right)$ are of order $\sigma$.}. This last term is precisely the contribution of the momentum constraint at leading order in gradient expansion and in the $\delta K=0$ gauge. With this knowledge, \eqref{STO_MOM_H_noi} can be written as \footnote{Note that, when talking about non-linear variables, we cannot write $\frac{\frac{\partial \phi}{\partial N}}{M_{PL}^2}\partial_i(\phi)$ as a total derivative, this is only true under the Starobinsky approximation, where $\frac{\partial \phi}{\partial N} \phi \simeq \frac{\partial \phi^b}{\partial N} \phi$.}:

\begin{equation}
\partial_i \left(\frac{\partial}{\partial N}\nabla^2 C + \frac{3}{2}\frac{\frac{\partial \phi^b}{\partial N}}{M_{PL}^2}\phi \right) = - \partial_i \left.\xi_5(N)\right|_{\delta K=0}\,.\label{k0}
\end{equation}
The key point now is that the leading order in gradient expansion is related to the longest wavelength of the perturbations. In Fourier modes, this means that we are considering the limit $k\rightarrow 0$ while keeping $k\neq 0$. Thus, \eqref{k0} is not identically satisfied. Because the $k=0$ mode represents the background, the solution of \eqref{k0} is 
\begin{equation}
\frac{\partial}{\partial N}\nabla^2 C  + \frac{3}{2}\frac{\frac{\partial \phi^b}{\partial N}}{M_{PL}^2}(\phi-\phi^b) = - \left.\xi_5(N)\right|_{\delta K=0}\,.
\label{C_def}
\end{equation}

We will see that, although this information does not enter in the evolution equations at zeroth order in gradient expansion, it does in the calculation of the perturbations correlators via the variable \cite{Takamizu:2013gy,Takamizu:2010xy}:

\begin{equation}
\zeta^{NL}\equiv\zeta+\frac{\nabla^2 C}{3}\ .
\label{NL_curv_def}
\end{equation}
The reason is that the correlators we want to calculate, like the power spectrum, carry information about the long wavelength limit of the curvature perturbations, which is precisely the information stored in the momentum constraint. 
\end{itemize}

Equations \eqref{STO_EVSPA_H},\eqref{STO_KG_H},\eqref{STO_HAM_H},\eqref{STO_EVEXT_H} and \eqref{C_def} represent a closed and solvable system of stochastic equations. It is interesting to note that in this gauge, contrary to the spatially flat one, the momentum constraint is decoupled from the rest of the stochastic system. 

The ``new'' stochastic framework we have worked out is now valid at all order in $\epsilon_i$, which represents one of the main results of this paper. In the spatially flat gauge, the same would have been only achieved by considering  $\partial_i\beta^i ={\cal O}(\sigma^0)\neq 0$, whereas by fixing $\partial_i\beta^i = 0$, one easily get the old stochastic formalism at leading order in $\epsilon_i$ as shown in appendix \ref{AppOld}. 

\subsection{Non-linear curvature perturbation}
\label{Sec_QNL}

Now that we managed to write the evolution equations for the stochastic system we need to define an observable.

In linear perturbation theory where the scalarly perturbed metric can be written in the form
\begin{equation}
	ds^2=-(1+2 A)dt^2+2a\partial_iBdx^idt+a^2\left[(1+2 D)\delta_{ij} - 2 E_{ij}^s\right]dx^idx^j,
\end{equation}
it exists a gauge invariant variable that encompasses all the scalar perturbations, this variable is called the Mukhanov-Sasaki variable and it is defined as:
\begin{equation}
Q^{lin}\equiv\delta\phi + \frac{\partial \phi^b}{\partial N}\left(D+\frac{1}{3}\nabla^2 E\right)\ .
\label{Lin_Q_def_p}
\end{equation}
A non-linear gauge invariant variable at leading order in gradient expansion was defined in  \cite{Langlois:2005ii, Rigopoulos:2004gr} as:
\begin{equation}
Q^{NL}_i=\partial_i \phi + \frac{1}{\alpha}\frac{\partial \phi}{\partial N}\partial_i \zeta \,.
\label{Q_gerasimos}
\end{equation}
However, the \eqref{Q_gerasimos}, in its linearisation, does not include the term $\propto\nabla^2 E$. 

One can however straightforwardly generalise \eqref{Q_gerasimos} by replacing $\zeta\rightarrow \zeta^{NL}$ as suggested in \cite{Takamizu:2013gy,Takamizu:2010xy,Wang:2013ic} and define:
\begin{equation}
Q^{IR}=\phi-\phi^b + \frac{1}{\alpha}\frac{\partial \phi}{\partial N}\zeta^{NL}\ .
\label{Q_nonlinear}
\end{equation}
It is then straightforward to check that, once the background is substracted, this new variable precisely matches \eqref{Lin_Q_def_p}. Whether or not $Q^{IR}$ is the corresponding non-linear generalisation of \eqref{Lin_Q_def_p} is not our concern here. As stressed many times, we can indeed only trust our stochastic equations in the linear regime (or in the Starobinski approximation) where $\langle Q^{lin}\ldots Q^{lin}\rangle\simeq \langle Q^{IR}\ldots Q^{IR}\rangle$. Thus, whenever this holds, the $Q^{IR}$ constructed from our non-linear stochastic variables, will be approximately the same observable as \eqref{Lin_Q_def_p}. 

In section \ref{Sec_checks}, we will apply this ``new'' stochastic formalism to different regimes of inflation comparing it with the ``old'' stochastic formalism and linear perturbation theory. 

\section{Numerical implementation}
\label{Sec_num}
It is important to realize that we will be comparing two different results from different theories: (a) linear perturbation theory and (b) stochastic approaches. The results coming from the ``old'' stochastic formalism must be compared with linear perturbation theory at leading order in $\epsilon_1$ whereas results coming from the ``new'' stochastic formalism must be compared with linear perturbation theory at all orders in $\epsilon_i$.

The quantity we want to compute in both theories is the real space correlator of the Long-Wavelength scalar variable $Q^{IR}$ as a function of the number of e-folds $N$. In particular in this paper we will focus to the two -point correlation function (related to the power spectrum in Fourier space).
 
\subsection{Linear perturbation theory}
\label{Sec_num_lin}

In linear perturbation theory we have:
\begin{equation}
\langle Q^{lin}(N, \mathbf{x}) Q^{lin}(N, \mathbf{x})\rangle=\int_{\sigma a(N=0) H(N=0)}^{\sigma a(N)H(N)}\frac{dk}{k}\mathcal{P}_{Q}(k,N)=\int_{\log(\sigma H(0))}^{\log(\sigma a(N)H(N))}\mathcal{P}_{Q}(k,N)d\log k,
\label{CORR_lin}
\end{equation}
where we are introducing the power spectrum evaluated at the same spatial point $\mathbf{x}$.
\begin{equation} 
\mathcal{P}_{Q}(k,N)=\frac{k^3}{2\pi^2}\left|Q_\mathbf{k}(N)\right|^2,
\label{Power_spect}
\end{equation}
where $Q_{\mathbf{k}}$ is the solution of the Mukhanov-Sasaki (MS) equation for the scalar perturbations (see \eqref{MS_Q_tau} in appendix \ref{AppLin_uniH}). The limits in \eqref{CORR_lin} correspond to the selection of modes inside the coarse grained scale (defined by $k=\sigma a(N)H(N)$) from the beginning of inflation ($N=0$). This anti-Fourier transformation from the power spectrum is needed in order to compare \eqref{CORR_lin} with the real space correlator coming from the stochastic formalism. 

\begin{center}
\begin{figure}[h]
    \includegraphics[scale=0.5]{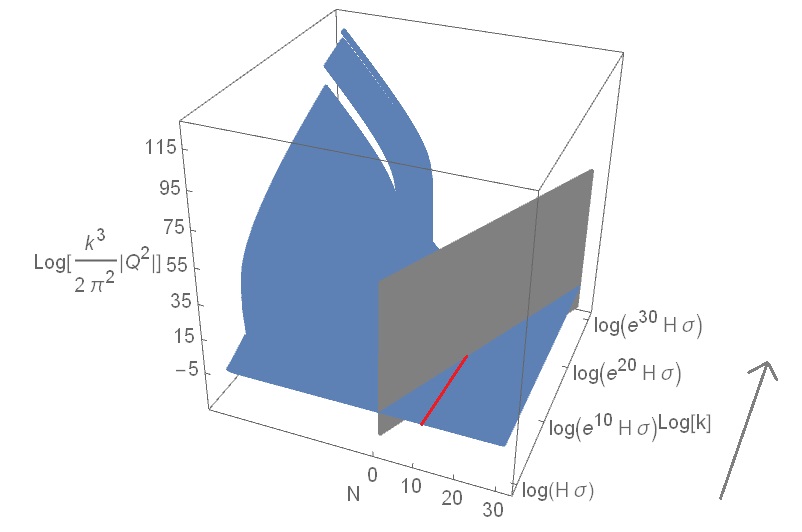}
    \caption{Numerical procedure followed in order to compute \eqref{CORR_lin}: each blue line corresponds to the solution of the MS equation $Q_{\mathbf{k}}(N)$ with fixed wave number $k$  in a generic Slow-Rolling background. The grey plane represents the plane in which each $k$-mode exits the coarse-grained scale. The idea is to integrate from $k=\sigma H(N=0)$ to $k=\sigma a(N) H(N)=\sigma e^{N} H(N)$ i.e. in the direction followed by the grey arrow. This means that the value of $\langle Q^{lin}(N_{\star})Q^{lin}(N_{\star})\rangle$ at time $N_{\star}$ will be the integral of the exponential of the blue surface (it is the exponential because we have plotted the log of the power spectrum for better visualization) from the $\log (\sigma H(0))$ plane up to the grey plane along the line where $N=N_{\star}$. For example, for $N_{\star}=10$ we will be integrating the red line.\label{Fig4.1}}
\end{figure}
\end{center}

In order to find \eqref{Power_spect} we numerically solve the MS equation for many values of $k$ between the two integration limits in \eqref{CORR_lin}. After that, we perform a numerical integration in the $k$ direction. In Fig. \ref{Fig4.1} this procedure is explained.

In the stochastic formalism, the $IR$ part of the field receives stochastic kicks from $N=0$ onward. Thus the first $k$-mode from which the $IR$ field receives a kick is the one with $k=\sigma a(N=0) H(N=0)$. 
Whenever $\mathcal{P}_{Q}(k,N)$ does not depend on $N$, one can do a very useful approximation, which consists in evaluating the power spectrum at coarse-grained scale crossing, i.e. at $k=\sigma a H$, and assume that this value does not change with time. This would allow us to write \eqref{CORR_lin} as
\begin{equation}
\langle Q^{lin}(N)Q^{lin}(N)\rangle=\int_0^N \mathcal{P}(k=\sigma a(N')H(N'))d N'.
\label{CORR_lin_app}
\end{equation}

In this case one could write the power spectrum as the derivative with respect of the number of e-folds $N$ of the correlator in real space.
\begin{equation}
\mathcal{P}(k)=\frac{d}{d N} \langle Q^{lin}(N)Q^{lin}(N)\rangle,
\label{Power_spect_app}
\end{equation}

Graphically, this would correspond to perform the integral \eqref{CORR_lin_app} in the direction $N$ (x axis in Fig \ref{Fig4.1}) by calculating the value of power spectrum only in the point in which blue and grey surfaces of Fig. \ref{Fig4.1} cross.  

However, this technique cannot be used if the power spectrum evolves with time. Thus, unfortunately, the approximation \eqref{Power_spect_app} cannot be used with the full numerical result. However, it can be used at zeroth order in $\epsilon_1$ in SR and USR but not in any transition between them.

\subsection{Stochastic evolutions}
\label{Sec_num_sto} 
In the stochastic approach, where the variables are statistical and non-linear, we can define a ``non-linear'' perturbation as $\Delta Q^{IR}=Q^{IR}-\overline{Q^{IR}}$, where $\overline{Q^{IR}}$ is the mean value of the variable $Q^{IR}$. With this definition it is clear that the correlator in real space at the same time $N$ is the statistical variance of the stochastic variable $Q^{IR}$.
  
We will compute $\text{Var}(Q^{IR}(N))$ by simulating the system of stochastic equations many times where the noises will take values distributed gaussianly with variances defined in \eqref{Noises_old} (if we are using the ``old'' stochastic formalism) or the corresponding ones in uniform Hubble gauge (if we are using the ``new'' stochastic formalism)\footnote{Within the ``new'' stochastic formalism, the variances of the noises are calculated numerically. For example, the variance of the noise $\left.\xi_1\right|_{\delta K=0}$ at time $N$ is the (numerically obtained) power spectrum of $\left.\delta\phi_{\mathbf{k}}\right|_{\delta K=0}$ evaluated at the wave number $k=\sigma a (N) H(N)$ and at coarse-grained crossing time, i.e. when  blue and grey surfaces of Fig. \ref{Fig4.1} cross.}. 

We will then run the system of stochastic equations many times until we have enough statistics  to give a trustworthy value for $\text{Var}(Q^{IR}(N))$. 

Since with the ``new'' stochastic formalism we are able to compute variables with precision $\epsilon_1 \ll 1$, we will use a Runge-Kutta method of third order adapted for stochastic equations, which was first developed in \cite{Robler:2010}. Note that the adaptation of Runge-Kutta methods to stochastic equations is not trivial \cite{Kloeden:1992}. We write down in \ref{Sec4.2.1} the algorithm used in our simulation, where the noises are always additive (meaning that their variance only depend on the time variable and not on the stochastic variables themselves) and completely correlated (which means that there is effectively only one noise).

\subsubsection{Numerical algorithm for the stochastic simulation}
\label{Sec4.2.1}

We denote by $X=\left( X_t \right)_{t \in \mathcal{I}}$ (where $\mathcal{I}=\left[ t_0 , T \right]$ for some $0\leq t_0 < T < \infty$) the solution of the d-dimensional system of stochastic differential equations (SDE) \eqref{SDE_example}.

\begin{equation}
\label{SDE_example}
X_t=X_{t_0}+\int_{t_0}^{t}a(s, X_s)ds + \sum_{j=1}^{m}\int_{t_0}^{t} b^j(s,X_s)dW_s^j,
\end{equation}
with an m-dimensional driving Weiner process $\left(W_t\right)_{t \geq 0}=\left(\left(W_t^1,... , W_t^m\right)^{T}\right)_{t \geq 0}$. 

In our case we have completely correlated noises and hence $m=1$. A further simplification can be done to \eqref{SDE_example} by imposing the additivity of the noises, which translates into $b(s, X_s)=b(s)$. Under these simplifications, the algorithm used in order to numerically solve \eqref{SDE_example} is an order 1.5 strong Stochastic Runge- Kutta (SRK) method defined by $Y_0=X_{t_0}$ and:

\begin{equation}
\label{Algorithm}
Y_{n+1}=Y_n+ \sum_{i=1}^{s} \alpha_i a\left(t_n+c_i^{(0)}h_n, H_i^{(0)}\right) + \sum_{i=1}^s \left(\beta_i^{(1)}I_{(1)}+\beta_i^{(2)}\frac{I_{(1,0)}}{h_n}\right)b\left(t_n+c_i^{(1)}h_n\right),
\end{equation}
for $n=0,1,...,N-1$ with stages
\begin{equation}
\label{Algorithm2}
H_i^{(0)}=Y_n+\sum_{j=1}^{s}A_{ij}^{(0)} a\left(t_n+c_j^{(0)}h_n, H_j^{(0)}\right)h_{n}+\sum_{j=1}^{s}B_{ij}^{(0)} b\left(t_n+c_j^{(1)}h_n\right)\frac{I_{(1,0)}}{h_n},
\end{equation}
for $i=1,...,s$. In the algorithm described above $h_n$ is the time step, $I_{(1)}$ and $I_{(1,0)}$ are some Îto stochastic integrals that will be specified in \eqref{Ito_integral}, and $\alpha_i$, $c_{i}^{(0)}$, $c_{i}^{(1)}$, $\beta_i^{(1)}$, $\beta_i^{(2)}$, $A_{ij}^{(0)}$ and $B_{ij}^{(0)}$ are some constants that characterize the method, they are usually written in a compact way using the so-called Butcher tableau:

\begin{table}[h!]
  \begin{center}
    \begin{tabular}{c|c|c|c}
      $c^{(0)}$ & $A^{(0)}$ & $B^{(0)}$ & $c^{(1)}$ \\
      \hline
       & $\alpha^{T}$ & $\beta^{(1)^T}$  & $\beta^{(2)^T}$  \\
    \end{tabular}
    \caption{}
   \label{table1}
  \end{center}
\end{table}

The specific entries of the Butcher tableau of TABLE \ref{table1} used in the SRK method of order 3 (in the deterministic part) are written down in TABLE \ref{table2}:

\begin{table}[h!]
  \begin{center}
    \begin{tabular}{ c | c c c | c c c | c c c}
      $0$ & & & & & & & $1$ & & \\
      $1$ & $1$ & & & $0$ & & & $0$ & & \\
      $\frac{1}{2}$ & $\frac{1}{4}$ & $\frac{1}{4}$ & & $1$ & $\frac{1}{2}$ & & $0$ & & \\
      \hline
       & $\frac{1}{6}$ & $\frac{1}{6}$ & $\frac{2}{3}$ & $1$ & $0$ & $0$ & $1$ & $-1$ & $0$  \\
    \end{tabular}
    \caption{}
   \label{table2}
  \end{center}
\end{table}

Once the Butcher tableau is specified, the only thing left is to define the stochastic Îto integrals $I_{(1)}$ and $I_{(1,0)}$

\begin{equation}
I_{(1)}=\int_{t_n}^{t_{n+1}}d W_s ; \qquad I_{(1,0)}=\int_{t_n}^{t_{n+1}}\int_{t_n}^s dW_{u} d s.
\label{Ito_integral}
\end{equation}

One can easily compute the expected value, the variance and the correlation of the integrals defined in \eqref{Ito_integral} getting:

\begin{align}
\nonumber
E\left(I_{(1)}\right)=0 \qquad E\left(I^2_{(1)}\right)=h_n^2 & \\
E\left(I_{(1,0)}\right)=0 \qquad E\left(I^2_{(1,0)}\right)=\frac{1}{3}h_n^3 & \qquad E\left(I_{(1,0)}I_{(1)}\right)=\frac{1}{2}h_n^2.
\label{Ito_moments}
\end{align}

The statistical behavior of \eqref{Ito_moments} can be implemented numerically by defining two independent $N(0;1)$\footnote{$N(0;1)$ refers to a random variable that follows a normal distribution with mean $0$ and variance $1$} random variables $U_1$ and $U_2$. In this case we have:

\begin{equation}
I_{(1)}=U_1\sqrt{h_n} \qquad I_{(1,0)}=\frac{1}{2}h_n^{3/2}\left(U_1+\frac{1}{\sqrt{3}}U_2\right)
\label{Ito_simul}
\end{equation}

\begin{center}
\begin{figure}[h]
    \includegraphics[scale=0.3]{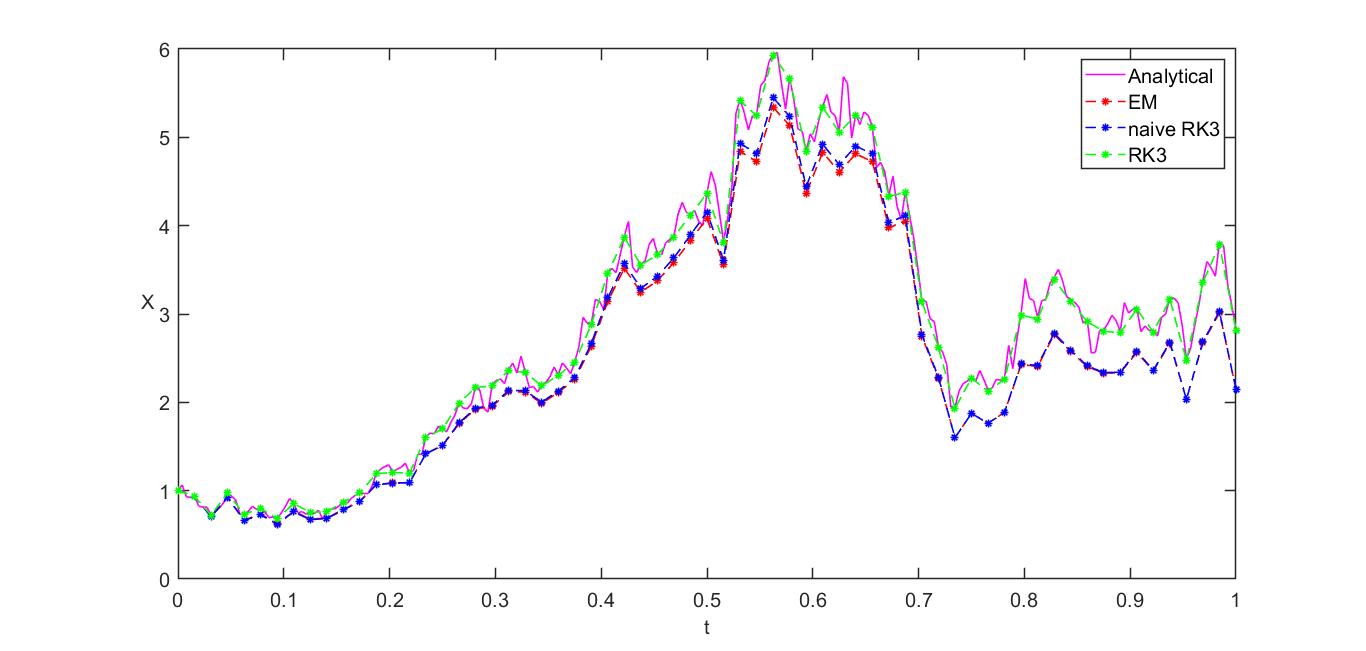}
    \caption{Analytical solution (in magenta) of the stochastic equation $d X(t)=\lambda X(t)dt + \nu X(t) dW_t$ where $\lambda=2$, $\nu=1$ and $X(0)=1$, $W_t$ represents a brownian motion. The dashed lines represent sumerical simulations of the same equation. One can clearly see that both the Euler-Maruyama method (red line) and a naive stochastic extension of the Runge-Kutta of third order for deterministic equations (blue line) give a similar precision. The precision is highly improved if we use the stochastic Runge-Kutta method proposed in \cite{Robler:2010}, which is the method used in this paper. \label{Fig4.2}}
\end{figure}
\end{center}

It is important to remark that if one do a naive extension of the Runge-Kutta method from deterministic equations to stochastic equations one would get a precision similar to the Euler-Maruyama method, which is of weak order 1. This was firstly noticed in \cite{Burrage:2006} and it can be numerically seen in Fig. \ref{Fig4.2}.

\section{Comparisons}
\label{Sec_checks}

In this section we will compare the real space power spectrum of the ``old'' and the ``new'' stochastic frameworks with linear perturbation theory. 

For each model of inflation we consider, that are characterized by the inflationary potential, we will show at least the following two figures: a) the comparison between the four different real space correlators; b) the relative difference between the real space correlator calculated using the ``new'' stochastic formalism and the one using numerical linear perturbation theory for two different $M$, where $M$ is the number of stochastic realizations to be averaged. This plot has the goal of showing that the more precise is the stochastic correlator (higher $M$), the closer we are to the linear perturbation theory correlator.

\subsection{Quadratic Slow Roll}
\label{Sec_checks_QSR}

The first model is the prototype of SR inflation in which the potential is 
\begin{equation}
V(\phi)=\frac{1}{2}m^2 \phi^2\ .
\label{SR_potential}
\end{equation}
In the numerical implementation we have chosen $m^2=1 \times 10^{-9}$, and, from now on, we use units $M_{PL}=1$. 

\begin{center}
\begin{figure}[h]
    \includegraphics[scale=0.5]{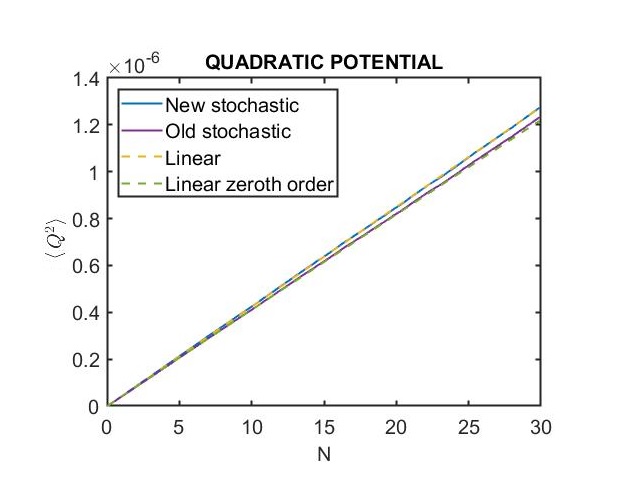}
    \caption{During quadratic slow roll, no important differences are seen between linear perturbation theory at all orders in $\epsilon_1$(yellow dashed line) and ``new'' stochastic formalism (blue solid line). Stochastic correlator at zeroth order (purple line) and linear correlator at zeroth order (dashed green line) both slightly differ from the two correlators at all orders in $\epsilon_1$, this difference is due to the dependence of numerical noises on the coarse grained scale $\sigma$ and it will be studied in section \ref{pert_diff}. Finally, one can observe a tiny difference between the purple and the dashed green lines, which will be also studied later on, in section \ref{lin_diff}. \label{CORR_QSR}}
\end{figure}
\end{center}

In this case the real space correlator $\langle \delta \phi^2\rangle$ calculated using the ``old'' stochastic formalism (purple line of Fig. \ref{CORR_QSR}) is simulated using equation \eqref{STO_old_SR}. The real space correlator $\langle \delta \phi^2_{lin}\rangle$ calculated using linear perturbation theory at zeroth order in $\epsilon_1$ (green dashed line of Fig. \ref{CORR_QSR}) is:

\begin{equation}
\langle \delta \phi^2_{lin}(N)\rangle=\int_{\sigma a(N=0) H}^{\sigma a(N)H}\frac{dk}{k}\mathcal{P}_{Q}(k,N) = \int_{\sigma a(N=0) H}^{\sigma a(N)H}\frac{dk}{k}\frac{k^3}{2\pi^2}\left|Q_\mathbf{k}(N)\right|^2 =\left(\frac{H}{2\pi}\right)^2 N 
\label{CORR_lin_zeroth}
\end{equation}
where $H$ is taken as a constant, consistently to the zeroth order in $\epsilon_1$. In passing we note that the solution \eqref{CORR_lin_zeroth} is valid for any slow-roll potential (like the Hilltop potential studied in subsection \ref{Sec_checks_HT}), but also for a USR regime.

\begin{center}
\begin{figure}[h]
    \includegraphics[scale=0.5]{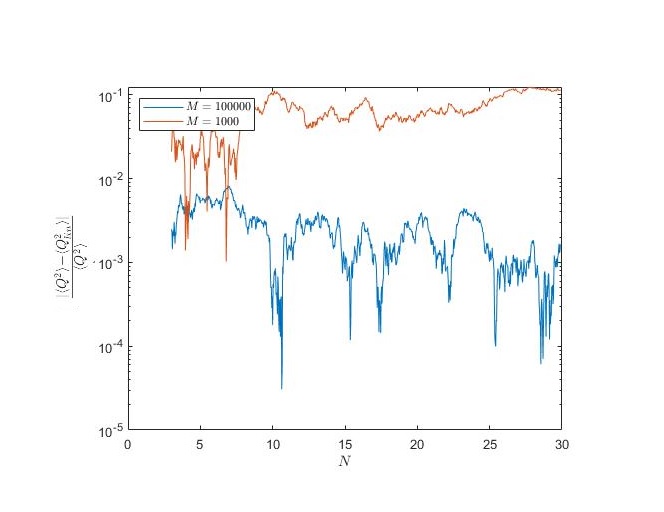}
    \caption{During quadratic slow-roll, the relative difference between the real space correlator at all orders in $\epsilon_1$ calculated with the stochastic formalism and with the linear theory decreases as the number of statistics increases. We expect these two theories to exactly coincide at $M \rightarrow \infty$. \label{Rel_QSR}}
\end{figure}
\end{center}

In Fig. \ref{CORR_QSR} and Fig. \ref{Rel_QSR} we show that there are no appreciable differences between the three different approaches. 

\subsection{Absence of quantum diffusion}
\label{Sec_checks_HT}
In this section we will study the model of Hilltop inflation \cite{Boubekeur:2005zm}, in which inflation is supposed to take place near a maximum of a potential. The potential can then be written as follows
\begin{equation}
V(\phi)=V_0 -\frac{1}{2}m^2\phi^2,
\label{HT_potential}
\end{equation}
where we have chosen $V_0=0.1$ and $m^2=0.001$.

The reason we want to study a potential like \eqref{HT_potential} is because it was claimed in \cite{Vennin:2015hra} that whenever $s\equiv\frac{1}{24 \pi^2}\left|2 V - \frac{V_{\phi\phi} V^2}{V_{\phi}^2} \right| > 1$, stochastic effects will be important. With the initial value of the field at $\phi(0)=0.1$, $s\sim\mathcal{O}(10)$. Note that this is still SR.

\begin{center}
\begin{figure}[h]
    \includegraphics[scale=0.5]{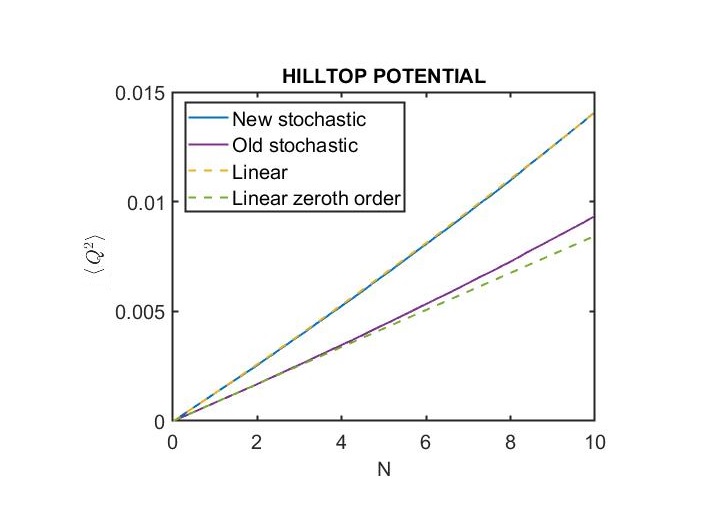}
    \caption{During Hilltop inflation, no important differences are seen between linear perturbation theory at all orders in $\epsilon_1$ (yellow dashed line) and ``new'' stochastic formalism (blue line). Stochastic correlator at zeroth order (purple line) and linear correlator at zeroth order (dashed green line) both slightly differ from the two correlators enumerated at the beginning of the caption, this difference is due to the dependence of numerical noises on the coarse grained scale $\sigma$. Finally, the difference between the purple and the dashed green lines is due to the dynamics of the $IR$ field as explained in the main text. \label{CORR_HT}}
\end{figure}
\end{center}

\begin{center}
\begin{figure}[h]
    \includegraphics[scale=0.5]{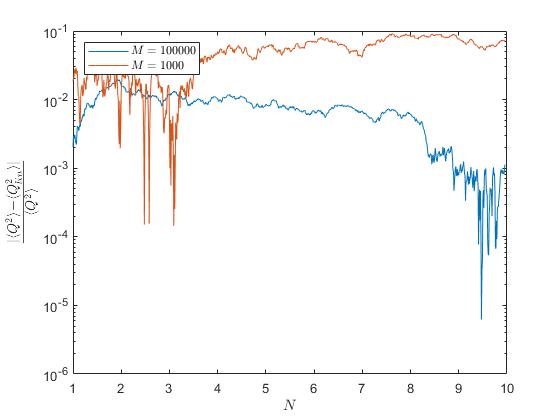}
    \caption{During Hilltop inflation, the relative difference between the real space correlator at all orders in $\epsilon_1$ calculated with the stochastic formalism and with the linear theory decreases as the number of statistics increases. We expect these two theories to exactly coincide at $M \rightarrow \infty$. \label{Rel_HT}}
\end{figure}
\end{center}

In Fig. \ref{CORR_HT} and Fig. \ref{Rel_HT} we show the comparisons between the different approaches. As we can see the correlator from the ``new'' stochastic formalism exactly coincides with linear perturbation theory at all order in SR parameters. Fig. \ref{CORR_HT} present some notable differences between the rest of correlators, we will explore those differences in the following showing that none of them are due to something that could be interpreted as a signal of quantum diffusion.

\subsubsection{Perturbation theory: leading order versus all orders in $\epsilon_i$} 
\label{pert_diff}

The difference between the correlators calculated with the full numerical solution of $Q_{\mathbf{k}}$ (blue and yellow dashed lines) and the correlators at zeroth order in $\epsilon_i$, is mainly due to the coarse-grained scale $\sigma$:

Schematically, the analytical expression for the power spectrum in the Slow-Roll approximation is
\begin{equation}
\mathcal{P}_{Q}(k,N)=\frac{H^2}{4\pi^2} \left(\frac{k}{a H}\right)^{{\cal{O}}(\epsilon_i)}\ .
\label{PS_SR_k}
\end{equation}
The real space correlator is then calculated as
\begin{equation}
\langle \delta \phi^2_{lin}(N,\sigma)\rangle=\int_{\sigma a(N=0) H}^{\sigma a(N)H}\frac{dk}{k} \frac{H^2}{4\pi^2} \left(\frac{k}{a H}\right)^{{\cal{O}}(\epsilon_i)}\ ,
\label{CORR_SR_sigma}
\end{equation}
leading to
\begin{equation}
\frac{|\langle \delta \phi^2_{lin}(N,\sigma)\rangle|-\langle \delta \phi^2_{lin}(N)\rangle}{\langle \delta \phi^2_{lin}(N,\sigma)\rangle} \simeq 1-\sigma^{{\cal{O}}(\epsilon_i)},
\label{Rel_diff_HT}
\end{equation}
where $\langle \delta \phi^2_{lin}(N)\rangle$ is the correlator calculated at leading order in $\epsilon_i$, i.e.
\begin{equation}
\langle \delta \phi^2_{lin}(N)\rangle=\int_{\sigma a(N=0) H}^{\sigma a(N)H}\frac{dk}{k} \frac{H^2}{4\pi^2} \ .
\end{equation}
It is easy to prove that the difference in \eqref{Rel_diff_HT} is much larger than $\epsilon_i$, which is what it is shown in Fig. \ref{CORR_HT}. In fact, $\sigma^{{\cal O}(\epsilon_i)}-1\gg \epsilon_i$ is equivalent to $\Big|\log(\sigma)\Big|\gg {\cal O}(1)$, which is always satisfied. 

One could be lead to think that a result strongly dependent on $\sigma$ is nonphysical. However, both linear theory and the new stochastic approach have the exact same dependence on $\sigma$ which is removed in Fourier space. In other words, the $\sigma$ dependence in real space simply translates to a k-dependence in Fourier space.

\subsubsection{Linear perturbation theory versus stochastic formalism at leading order in $\epsilon_i$}
\label{lin_diff}

The difference between the linear theory at zeroth order and the old stochastic formalism (purple solid and green dashed lines, respectively, observed in Fig. \ref{CORR_HT}) is once again rooted in the $k$ dependence of the power spectrum. We will see that the old stochastic formalism captures, in the correlator,  the next to leading order in $\epsilon_i$ with respect to linear theory. Thus no non-perturbative effects from the stochastic inflation should be searched to explain this difference.

We can define, in our model, the parameter $x=\frac{m^2}{V_0}=0.01 \ll 1$ and compute the real space correlators in linear theory and in the ``old'' stochastic approach, to first order in $x$. Within this approximation, the background equations are

\begin{equation}
\frac{\partial \phi^b}{\partial N} - x \phi^b \simeq 0\,; \qquad \epsilon_1\simeq 0 \,; \qquad \epsilon_2\simeq 2x\,.
\label{HT_xapp}
\end{equation}

Using \eqref{HT_xapp}, the ``old'' stochastic equation \eqref{STO_old_SR} reduces to

\begin{equation}
\frac{\partial \phi^{IR}_{\text{f}}}{\partial N} - x \phi^{IR}_{\text{f}} = \frac{H}{2\pi}\xi(N),
\label{HT_xapp_sto}
\end{equation}
where $\langle\xi(N_1)\xi(N_2)\rangle=\delta (N_1-N_2)$. From \eqref{HT_xapp_sto}, $\text{Var} \left(\phi^{IR}_{\text{f}}\right)$ is then straightforwardly computed:

\begin{equation}
\text{Var} \left(\phi^{IR}_{\text{f}}\right)=\frac{H^2}{4 \pi^2}N\left(1+x N\right) \,.\label{HT_xapp_corr}
\end{equation}

Eq \eqref{HT_xapp_corr} corresponds to the purple solid line of Fig. \ref{CORR_HT}. 

The real space correlator in linear theory, at zeroth order in $x$, would instead give
\begin{equation}
\langle \delta \phi^2_{lin}\rangle\Big|_{x=0}=\int_{\sigma a(N=0) H}^{\sigma a(N)H}\frac{dk}{k}  \frac{H^2}{4 \pi^2}  = \frac{H^2}{4 \pi^2}N\, ,
\end{equation}
missing the additional term $\propto x N^2$ in \eqref{HT_xapp_corr}.

If we now instead use the definition of the real space correlator \eqref{CORR_lin}, and compute it up to first order in $x$, we get:
\begin{equation}
\langle \delta \phi^2_{lin}\rangle\simeq\int_{\sigma a(N=0) H}^{\sigma a(N)H}\frac{dk}{k}  \frac{H^2}{4 \pi^2} \left(\frac{k}{\sigma a H}\right)^{-2 x} = \frac{H^2}{4 \pi^2}N \left(1+x N\right)\,.
\label{HT_xapp_corrlin2}
\end{equation}
which precisely matches the result of the old stochastic framework.

\subsection{Ultra Slow Roll}
\label{Sec_checks_USR}
In an Ultra Slow Roll phase, the inflaton moves in an exactly flat potential ($V=V_0$), this means that its velocity decreases exponentially and so $\epsilon_1$. Due to this exponential decreasing, the procedure explained in section \ref{Sec_num_lin} leads to large numerical errors. Thus, in this subsection we will only use the approximate analytical solutions for $Q_{\mathbf{k}}$ both at zeroth and a first order in $\epsilon_1$. Because of the smallness of $\epsilon_1$ those approximations will be exponentially precise. The way of solving the MS equation up to $\epsilon_1$ precision in a USR regime is explained in Appendix \ref{App_MS_sol}.

\begin{center}
\begin{figure}[h]
    \includegraphics[scale=0.5]{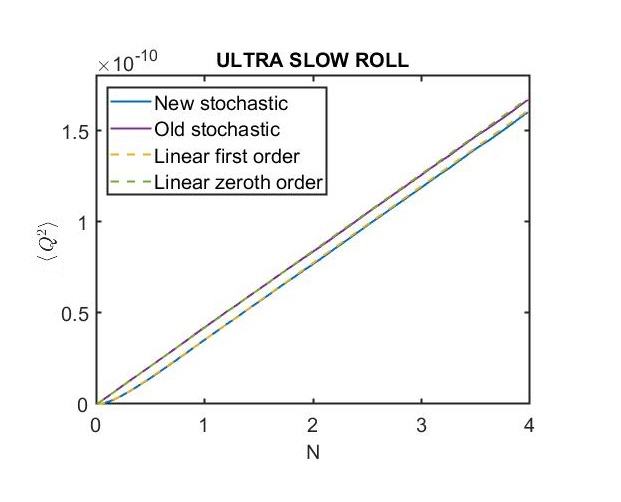}
    \caption{During an Ultra Slow Roll phase, no important differences are seen between linear perturbation theory and stochastic formalism. Differences between zeroth and first order in $\epsilon_1$ correlators are just due to $\epsilon_1 (N=0)$ terms. This dependence on the value of $\epsilon_1$ at $N=0$ is explained (within the framework of linear theory) in appendix \ref{App_MS_sol}. \label{CORR_USR}}
\end{figure}
\end{center}

The real space variances computed both in linear perturbation theory and within the stochastic formalism at zeroth and first order in $\epsilon_1$, are shown in Fig. \ref{CORR_USR}. We see that they give all approximately equal results.

At zeroth order in $\epsilon_1$, one can compute the correlator in the ``old'' stochastic formalism analytically as follows:

The stochastic system of equations to solve is:
\begin{align}
\nonumber
\pi^{IR}_{\text{f}}&=\frac{\partial \phi^{IR}_{\text{f}}}{\partial N} + \frac{1}{2\pi}\sqrt{\frac{V_0}{3}}\xi(N)\\ \nonumber
\frac{\partial \pi^{IR}_{\text{f}}}{\partial N} & = - 3 \pi^{IR}_{\text{f}},
\label{STO_USR}
\end{align}

The second moments of the variables $\phi^{IR}_{\text{f}}$ and $\pi^{IR}_{\text{f}}$ follow a system of stochastic equations:

\begin{eqnarray} \nonumber
&\left(
\begin{array}{cc}
\frac{\partial \langle \pi^{IR}_{\text{f}}(N) \pi^{IR}_{\text{f}}(N)\rangle}{\partial N} &  \frac{\partial \langle \pi^{IR}_{\text{f}}(N) \phi^{IR}_{\text{f}}(N)\rangle}{\partial N}  \\
  \frac{\partial \langle \phi^{IR}_{\text{f}}(N) \pi^{IR}_{\text{f}}(N)\rangle}{\partial N} &  \frac{\partial \langle \phi^{IR}_{\text{f}}(N) \phi^{IR}_{\text{f}}(N)\rangle}{\partial N}  \\
\end{array}
\right) =
\left(
\begin{array}{cc}
-3 &   0  \\
 1 &  0  \\
\end{array}
\right) 
\left(
\begin{array}{cc}
\langle\pi^{IR}_{\text{f}}(N) \pi^{IR}_{\text{f}}(N)\rangle &   \langle\pi^{IR}_{\text{f}}(N) \phi^{IR}_{\text{f}}(N)\rangle  \\
 \langle\phi^{IR}_{\text{f}}(N) \pi^{IR}_{\text{f}}(N)\rangle &  \langle\phi^{IR}_{\text{f}}(N) \phi^{IR}_{\text{f}}(N)\rangle \\
\end{array}
\right) \\ 
& + \left(
\begin{array}{cc}
\langle\pi^{IR}_{\text{f}}(N) \pi^{IR}_{\text{f}}(N)\rangle &   \langle\pi^{IR}_{\text{f}}(N) \phi^{IR}_{\text{f}}(N)\rangle  \\
\langle \phi^{IR}_{\text{f}}(N) \pi^{IR}_{\text{f}}(N)\rangle &  \langle\phi^{IR}_{\text{f}}(N) \phi^{IR}_{\text{f}}(N)\rangle \\
\end{array}
\right)
\left(
\begin{array}{cc}
-3 &   1  \\
 0 &  0  \\
\end{array}
\right) 
+
\left(
\begin{array}{cc}
0 &   0  \\
 0 &  \frac{V_0}{12 \pi^2}\\
\end{array}
\right) \ .
\label{moments_USR}
\end{eqnarray}
Using that $\langle\phi^{IR}_{\text{f}}(N)\rangle=\phi^b(N)$ and $\langle\pi^{IR}_{\text{f}}(N)\rangle=\frac{\partial\phi^b(N)}{\partial N}$ we can easily obtain:

\begin{equation}
{\rm Var}\left(\phi^{IR}_{\text{f}}(N)\right) = \frac{V_0}{12 \pi^2} N.
\label{STO_anal_USR}
\end{equation}

\begin{center}
\begin{figure}[h]
    \includegraphics[scale=0.5]{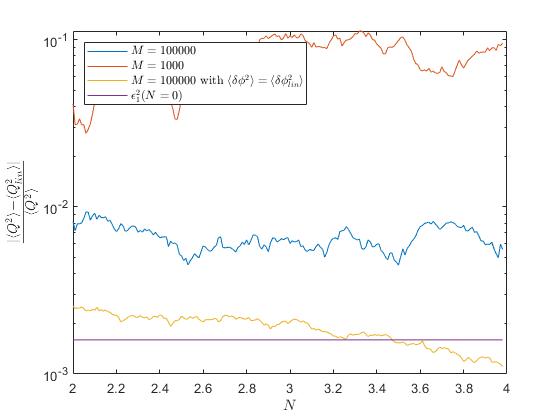}
    \caption{During an Ultra Slow Roll phase, the relative difference between the real space correlator at first order in $\epsilon_1$ calculated with the ``new'' stochastic formalism and the real space correlator at first order in $\epsilon_1$ calculated with linear theory decreases as the number of statistics increases up to the yellow line, where we have used the fact that the ``old'' stochastic formalism and linear theory must coincide at zeroth order to eliminate most of the statistical fluctuactions. We can see that the yellow line is $\mathcal{O}\left(\epsilon_1^2 (N=0)\right)$, which is expected since we are only computing noises up to $\mathcal{O}\left(\epsilon_1(N=0)\right)$. \label{Rel_USR}}
\end{figure}
\end{center}

The correlator \eqref{STO_anal_USR} exactly coincides with \eqref{CORR_lin_zeroth} in a USR regime as already shown in \cite{Cruces:2018cvq}. There is no doubt then that, at zeroth order in $\epsilon_1$ and during a USR phase, both the stochastic formalism and linear perturbation theory give the exact same real space correlator and hence the same power spectrum. This fact also allow us to eliminate statistical errors when comparing the ``new'' stochastic formalism with linear perturbation theory at first order in $\epsilon_1$ as shown in Fig \ref{Rel_USR}.

\subsection{Transition between SR and USR}
\label{Sec_checks_TRANS}

Finally, we will study the more realistic case in which a SR phase is followed by an USR. The transition between these two phases is quite interesting as it is the regime in which we could expect some difference between the $IR$ part of ``old'' and ``new'' stochastic equations. This is because the inflaton field is overshoot \cite{Germani:2017bcs} making $\epsilon_1$ only slightly smaller than one. 

The potential used to simulate the SR-USR-SR transition is a cubic potential containing an inflection point at $\phi=\phi_0=1$, i.e.

\begin{equation}
V(\phi)=V_0\left(1+\beta\left(\phi-\phi_0\right)^3\right),
\label{Tran_potential}
\end{equation}
where the parameters chosen are $V_0=1 \times 10^{-8}$ and $\beta=0.8$.

\begin{center}
\begin{figure}[h]
    \includegraphics[scale=0.5]{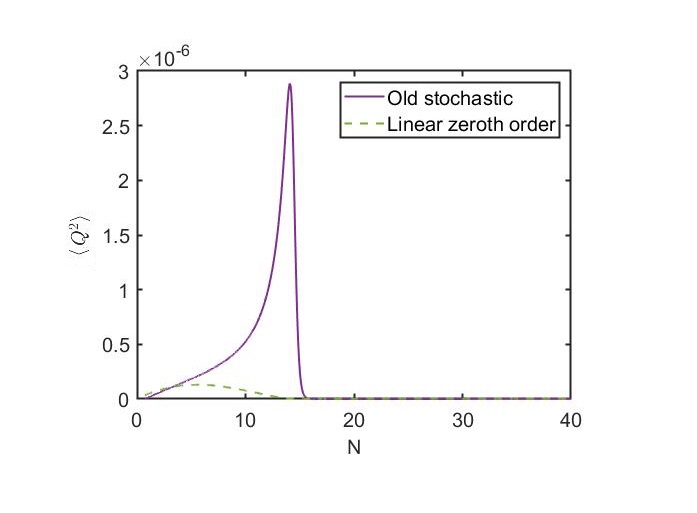}
    \caption{At leading order in $\epsilon_1$ and during the transition between SR and USR, there seems to be huge stochastic effects in the two-point correlator (purple solid line) with respect to the one computed within linear perturbation theory (green dashed line). However, results at zeroth order are not a good approximation at all in this case so they should not be trusted. We show this plot just for completeness  \label{CORR_tran2}}
\end{figure}
\end{center}

\begin{center}
\begin{figure}[h]
    \includegraphics[scale=0.25]{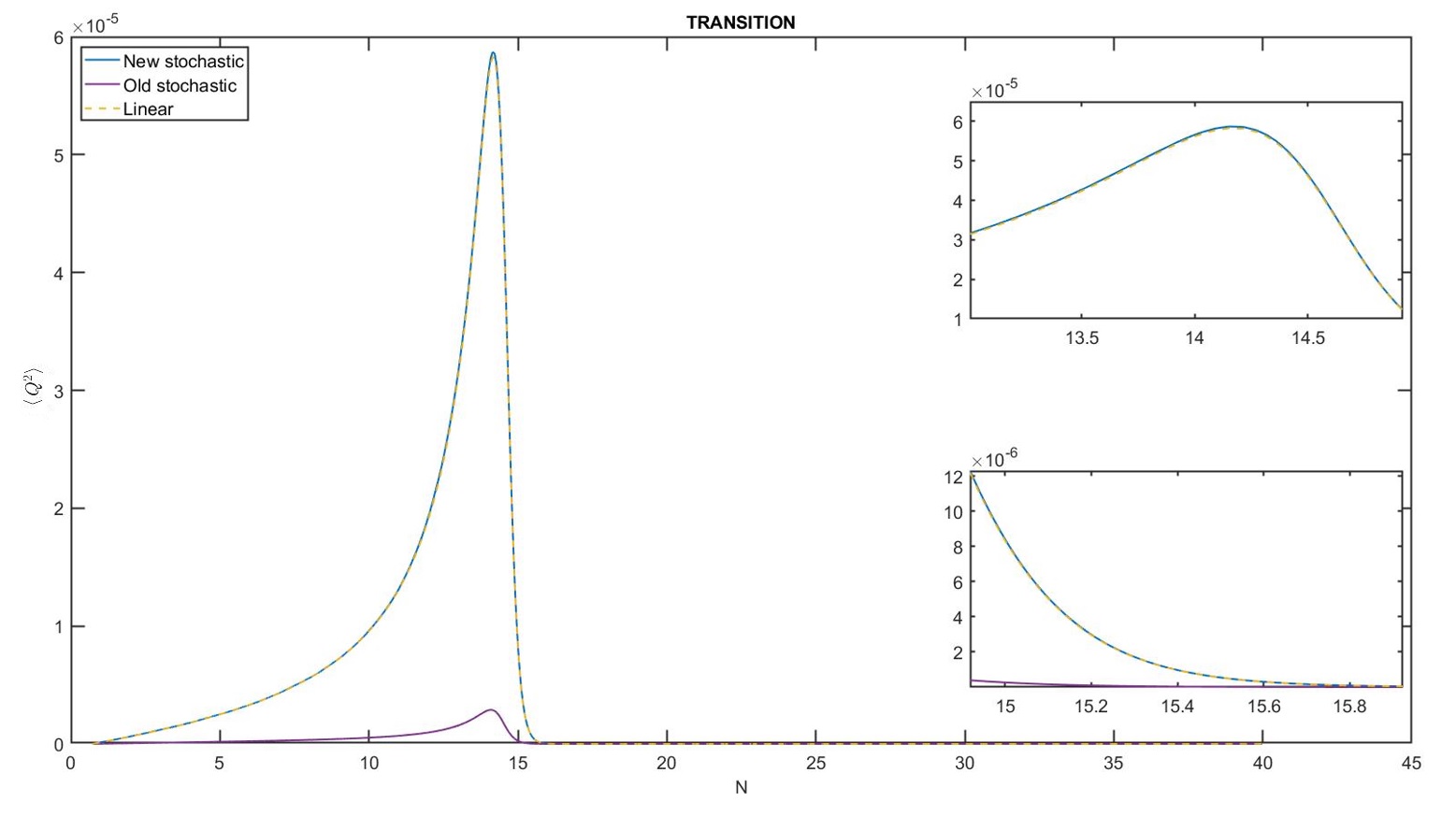}
    \caption{During the transition between SR and USR, the correlator coming from the ``new'' stochastic equations (blue solid line) exactly coincides with the one got from linear perturbation theory (yellow line). It has been also plotted the zeroth order stochastic correlator of Fig. \ref{CORR_tran2} to better visualize the huge difference. \label{CORR_tran}}
\end{figure}
\end{center}

\begin{center}
\begin{figure}[h]
    \includegraphics[scale=0.5]{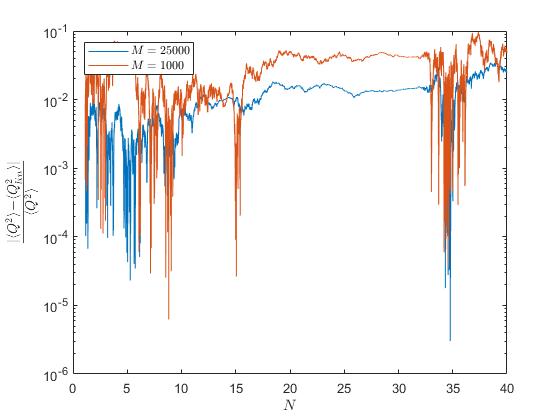}
    \caption{During the transition between SR and USR, the relative difference between the real space correlator at all orders in $\epsilon_1$ calculated with the stochastic formalism and with the linear theory decreases as the number of statistics increases. We expect these two theories to exactly coincide at $M \rightarrow \infty$. \label{Rel_Tran}}
\end{figure}
\end{center}

In this regime, the zeroth order solution for $Q_{\mathbf{k}}$ (the one obtained using the SR approximation) has quite poor precision. We illustrate this in Fig. \ref{CORR_tran2} where this bad approximation would show huge stochastic effects. In Fig. \ref{CORR_tran} we show instead that the fully numerical linear perturbation theory correlator, exactly coincides with the correlator from the ``new'' stochastic formalism while disagreeing with the old one. In Fig. \ref{Rel_Tran}, as we expect, we plot the relative difference between linear theory and stochastic formalism and show that it decreases with the number of realizations.

\section{Conclusions}
\label{Sec_concl}

In this paper we have firstly elucidated that, under the Starobinsky approximation or, equivalently, under the white noise construction, a stochastic framework to inflation might only be developed in the linear perturbation regimes. Thus, the stochastic approach to inflation is incapable to give more information than perturbation theory. 

In fact, in its standard form, it is even less precise than perturbation theory. The reason is that what we called ``old'' stochastic inflation is only consistent at leading order in slow-roll parameters. 

By introducing the momentum constraint of Einstein equations in the stochastic framework, we have developed a ``new'' stochastic formalism valid at all orders in $\epsilon_i$. We have shown that the ``new'' stochastic formalism exactly reproduces the power spectrum calculated via linear perturbation theory in different inflationary scenarios. In particular, we have shown that there are no ``quantum diffusion'' dominated regimes in the realm of slow-roll and ultra-slow-roll inflation and, in the case of a transition between a slow-roll and an ultra-slow-roll regime (a relevant case for primordial black hole formation) the old stochastic formalism would lead to largely un-physical results, while, once again, our ``new'' stochastic formalism would reproduce linear theory very accurately.   

Finally, we would like to stress that our framework is only valid whenever the perturbations are in the linear regime. Nevertheless, any discordance between our new stochastic inflation and linear theory, would point out to non-perturbative effects that, unfortunately, cannot be captured by any stochastic methods. 

\acknowledgments
We would like to thank Misao Sasaki and Vincent Vennin for the many correspondences and help in understanding the gradient expansion method and the ``old'' stochastic formalism. We also thank Aichen Li for re-checking all the computations.  In the first part of the development of this paper CG was supported by the Ramon y Cajal program.  DC is supported by the Spanish MECD fellowship PRE2018-086135. We are also supported by the Unidad de Excelencia Maria de Maeztu Grants No. MDM-2014-0369 and CEX2019-000918-M, and the Spanish national grants FPA2016-76005-C2-2-P, PID2019-105614GB-C22 and PID2019-106515GB-I00. 
\appendix

\section{Appendix A: ADM equations}
\label{AppADM}
Here we present the basic equations for non-linear quantities coming from the ADM formalism \cite{Arnowitt:1962hi}. The Klein-Gordon equation is
\begin{equation}
\frac{1}{\sqrt{-g}}\frac{\partial}{\partial x^{\mu}}\left[\sqrt{-g}g^{\mu\nu}\frac{\partial \phi}{\partial x^{\nu}}\right]-V_{\phi}=0,
\label{ADM_KG}
\end{equation}
where $V_{\phi}=\frac{d V(\phi)}{d \phi}$.

As it is written in the main text, $\alpha$ and $\beta_i$ are Lagrange multipliers, the constraints associated to them are the energy and momentum constraints:
\begin{equation}
R^{(3)}-\tilde{A}_{ij}\tilde{A}^{ij}+\frac{2}{3}K^2=16\pi G E,
\label{ADM_HAM}
\end{equation}
\begin{equation}
D^j\tilde{A}_{i j}-\frac{2}{3}D_iK=8\pi G J_i,
\label{ADM_MOM}
\end{equation}
where $E\equiv T_{\mu\nu}n^{\mu}n^{\nu}$ and $J_i\equiv -T_{\mu\nu}n^{\mu}\gamma^{\nu}_i$. 

The evolution equations for the dynamic variable $\gamma_{ij}$ are 
\begin{equation}
(\partial_t-\beta^k\partial_k)\zeta+\frac{\dot{a}}{a}=-\frac{1}{3}(\alpha K-\partial_k\beta^k),
\label{ADM_EVSPA}
\end{equation}
\begin{equation}
(\partial_t-\beta^k\partial_k)\tilde{\gamma}_{ij}=-2\alpha\tilde{A}_{ij}+\tilde{\gamma}_{ik}\partial_j\beta^k+\tilde{\gamma}_{jk}\partial_i\beta^k-\frac{2}{3}\tilde{\gamma}_{ij}\partial_k\beta^k.
\label{ADM_EVSPA2}
\end{equation}

Finally, the evolution equations for $K_{ij}$ are:

\begin{equation}
(\partial_t-\beta^k\partial_k)K=\alpha\left(\tilde{A}_{ij}\tilde{A}^{ij}+\frac{1}{3}K^2\right)-D_k D^k \alpha+4\pi G \alpha(E+S_k^k),
\label{ADM_EVEXT}
\end{equation}
\begin{align} \nonumber
(\partial_t-\beta^k\partial_k)\tilde{A}_{ij} & =\frac{e^{-2\zeta}}{a^2}\left[\alpha\left(R_{ij}^{(3)}-\frac{\gamma_{ij}}{3}R^{(3)}\right)-\left(D_i D_j\alpha-\frac{\gamma_{ij}}{3}D_k D^k \alpha\right)\right] \\
&+ \alpha(K\tilde{A}_{ij}-2\tilde{A}_{ik}\tilde{A}^k_j)+\tilde{A}_{ik}\partial_j\beta^k+\tilde{A}_{jk}\partial_i\beta^k-\frac{2}{3}\tilde{A}_{ij}\partial_k\beta^k\\ \nonumber
&-\frac{8\pi G \alpha e^{-2\zeta}}{a^2}\left(S_{ij}-\frac{\gamma_{ij}}{3}S_k^k\right),
\label{ADM_EVEXT2}
\end{align}
where $S_{ij}=T_{ij}$ and $S^k_k=\gamma^{kl}S_{lk}$.

\section{Appendix B: Linear perturbation theory}
\label{AppLin_uniH}
In this appendix we will explore the well known linear perturbation theory during inflation \cite{Mukhanov:1990me}, paying special attention to the uniform-Hubble gauge.

\subsection{Linear perturbation theory in a generic gauge}

The scalar sector of the perturbed FLRW metric is
\begin{equation}
ds^2=-(1+2 A)dt^2+2a\partial_iBdx^idt+a^2\left[(1+2 D)\delta_{ij} - 2 E_{ij}^s\right]dx^idx^j,
\label{Lin_metric}
\end{equation}
where
\begin{equation}
E_{ij}^s=\left(\partial_i\partial_j-\frac{1}{3}\delta_{ij}\nabla^2\right)E.
\label{Traceless_metric}
\end{equation}

We can now define the linear curvature perturbation as
\begin{equation}
\psi\equiv D + \frac{1}{3}\nabla^2 E.
\label{Lin_curv_def}
\end{equation}

The usual gauge-invariant Mukhanov-Sasaki (MS) variable is
\begin{equation}
Q\equiv\delta\phi + \frac{\dot{\phi}^b}{H^b}\left(D+\frac{1}{3}\nabla^2 E\right) =\delta\phi + \frac{\dot{\phi}^b}{H^b}\psi
\label{Lin_Q_def}
\end{equation}

With metric \eqref{Lin_metric} perturbed Einstein equations in an universe filled with a single scalar field in an arbitrary gauge are:
\begin{changemargin}{-2cm}{-1cm}
\begin{align}
3H^b(H^b A-\dot{D})+\frac{\nabla^2}{a^2}\left[D+\frac{1}{3}\nabla^2 E + H^b aB\right]&=-\frac{1}{2M_{PL}^2}\left[\dot{\phi}^b(\delta\dot{\phi}-\dot{\phi}^b A)+V_{\phi}\delta\phi\right], \label{Lin_HAM}\\ 
\partial_i\left(H^b A-\dot{D}-\frac{1}{3}\nabla^2 \dot{E}\right)&=\frac{1}{2M_{PL}^2}\dot{\phi}^b\partial_i\delta\phi, \label{Lin_MOM} \\
\nonumber
 \Bigg\{ H^b\dot{A}+2\dot{H}^bA+3\left(H^b\right)^2 A-\ddot{D}-3H^b\dot{D}-H^b\nabla^2\dot{E}+\frac{1}{2}\frac{\nabla^2}{a^2} &  \left[A+D+\frac{1}{3}\nabla^2E+2H^b(aB+a^2\dot{E})+a\frac{d}{dt}(B+a\dot{E})\right]\Bigg\}\delta^j_i\\ 
-\frac{1}{2a^2}\partial_i\partial^j\bigg[A+D+\frac{1}{3}\nabla^2 E+2H^b(aB+a^2\dot{E})+a\frac{d}{dt}&(B+a\dot{E})\bigg]=\frac{1}{2M_{PL}^2}\left[\dot{\phi}^b(\delta\dot{\phi}-\dot{\phi}^b\alpha)-V_{\phi}\delta\phi\right]\delta_i^j.  \label{Lin_extra}
\end{align}
\end{changemargin}

Perturbed KG equation is
\begin{equation}
\delta\ddot{\phi}+3H^b\delta\dot{\phi}+\left(V_{\phi\phi}-\frac{\nabla^2}{a^2}\right)\delta\phi=-2V_{\phi} A+\dot{\phi}^b\left[\dot{ A}-3\dot{D}+\frac{\nabla^2}{a}B\right].
\label{Lin_KG}
\end{equation}

After some manipulations, one can get an equation of motion for the MS variable \eqref{Lin_Q_def}:
\begin{equation}
\ddot{Q}+3H\dot{Q}+\left(-\frac
{\nabla^2}{a^2}+H^2\left(-\frac{3}{2}\epsilon_2+\frac{1}{2}\epsilon_1\epsilon_2-\frac{1}{4}\epsilon_2^2-\frac{1}{2}\epsilon_2\epsilon_3\right)\right)Q=0.
\label{MS_Q_t}
\end{equation}

Finally, we will introduce the overall expansion rate (or trace of extrinsic curvature) as:
\begin{equation}
K=\nabla_{\mu} \mathfrak{n}^{\mu} = 3\left(H^b-H^b A +\dot{D}-\frac{1}{3}\frac{\nabla^2}{a}B\right),
\label{Lin_H_def}
\end{equation}
where $\mathfrak{n}$ is the linearized unit time-like vector. 

\subsection{Linear perturbation theory in $\delta K=0$ gauge}
Since the trace of the extrinsic curvature $K$ is defined as $K \equiv 3H$, it is now trivial to write down the linear version for the uniform Hubble gauge (where $H=H^b$):
\begin{equation}
H^b A_{\delta K=0} -\dot{D}_{\delta K=0} +\frac{1}{3}\frac{\nabla^2}{a}B_{\delta K=0}=0.
\label{Lin_uniformH}
\end{equation}

As also done when studying $\delta K=0$ gauge in gradient expansion, we can use the residual gauge freedom to set $B=0$. This is the so-called time-slice-orthogonal threading \cite{Naruko:2012fe}
\begin{align}
\nonumber
H^b A_{\delta K=0}-\dot{D}_{\delta K=0}=0,\\
B_{\delta K=0}=0.
\label{Lin_uniformH2}
\end{align}

Once specified the gauge, equations \eqref{Lin_HAM}-\eqref{Lin_extra} are:
\begin{align}
\frac{\nabla^2}{a^2}\left[D_{\delta K=0}+\frac{1}{3}\nabla^2 E_{\delta K=0}\right]&=-\frac{1}{2M_{PL}^2}\left[\dot{\phi}^b(\delta\dot{\phi}_{\delta K=0}-\dot{\phi}^b A_{\delta K=0})+V_{\phi}\delta\phi_{\delta K=0}\right], \label{Lin_HAM_uniH}\\ 
\frac{1}{3}\nabla^2 \dot{E}_{\delta K=0}&=-\frac{1}{2M_{PL}^2}\dot{\phi}^b\delta\phi_{\delta K=0}, \label{Lin_MOM_uniH} \\
\dot{H}^bA_{\delta K=0}-\frac{1}{3}\nabla^2 \ddot{E}_{\delta K=0} - H^b\nabla^2 \dot{E}_{\delta K=0}&=-\frac{1}{2M_{PL}^2}\left[\dot{\phi}^b(\delta\dot{\phi}_{\delta K=0}-\dot{\phi}^b A_{\delta K=0})-V_{\phi}\delta\phi_{\delta K=0}\right],  \label{Lin_extra_uniH} \\
A_{\delta K=0}+D_{\delta K=0}+\frac{1}{3}\nabla^2 & E_{\delta K=0}+3H^ba^2\dot{E}_{\delta K=0}+a^2\ddot{E}_{\delta K=0}=0, \label{Lin_extra_uniH2} 
\end{align}
where \eqref{Lin_extra_uniH} and \eqref{Lin_extra_uniH2} are the diagonal and non-diagonal part of \eqref{Lin_extra}, respectively. Now we can use $\dot{H}^b=-\frac{\left(\dot{\phi}^b\right)^2}{2M_{PL}^2}$ in \eqref{Lin_extra_uniH} and after manipulating \eqref{Lin_HAM_uniH}, \eqref{Lin_extra_uniH} and \eqref{Lin_extra_uniH2} we arrive to a result for $A_{\delta K=0}$:

\begin{equation}
\frac{\nabla^2}{a^2}A_{\delta K=0}=\frac{1}{2M_{PL}^2}\left(4\dot{\phi}^b\delta\dot{\phi}_{\delta K=0}-2V_{\phi}\delta\phi_{\delta K=0}-\left(\dot{\phi}^b\right)^2A_{\delta K=0}\right),
\label{Lin_lapse}
\end{equation}
which can be written in Fourier space as \cite{Tanaka:2007gh}:
\begin{equation}
\left.A_\mathbf{k}\right|_{\delta K=0}=\frac{\frac{1}{2 M_{PL}^2}\left(4\dot{\phi}^b\left.\delta\dot{\phi}_{\mathbf{k}}\right|_{\delta K=0}-2 V_{\phi}\left.\delta\phi_{\mathbf{k}}\right|_{\delta K=0}\right)}{\frac{\left(\phi^b\right)^2}{2M_{PL}^2}-\frac{k^2}{a^2}}.
\label{Lin_lapse_fourier}
\end{equation}
 
It is also convenient to use the gauge condition \eqref{Lin_uniformH2} in the perturbed KG equation.
\begin{equation}
\delta\ddot{\phi}_{\delta K=0}+3H^b\delta\dot{\phi}_{\delta K=0}+\left(V_{\phi\phi}-\frac{\nabla^2}{a^2}\right)\delta\phi_{\delta K=0}=-2V_{\phi} A_{\delta K=0}+\dot{\phi}^b\left[\dot{ A}_{\delta K=0}-3 H^b A_{\delta K=0} \right].
\label{Lin_KG_uniH}
\end{equation}

Finally, and for completeness, we will write an evolution equation for $\psi$ using \eqref{Lin_uniformH2} and \eqref{Lin_MOM_uniH}:
\begin{equation}
\dot{\psi}_{\delta K=0}=-\frac{1}{2M_{PL}^2}\dot{\phi}^b\delta\phi_{\delta K=0} + H^b A_{\delta K=0}. 
\label{Lin_evpsi_uniH}
\end{equation}

\subsection{Linear gauge transformation between spatially flat and $\delta K=0$ gauges}
In this appendix we will also calculate the variables $\delta\phi_{\delta K=0}$ and $\psi_{\delta K=0}$ that we use in ``new'' stochastic formalism. In order to do so, we will make a gauge transformation between spatially flat gauge and uniform Hubble gauge. The reason is that we know that $\delta\phi_{\text{f}}=Q$ (because of \eqref{Lin_Q_def}).This means that to perform the gauge transformation $\delta\phi_{\text{f}} \rightarrow \delta\phi_{\delta K=0}$ allow us to write $\delta\phi_{\delta K=0}$ in terms of the gauge invariant quantity $Q$, as we will see.

In order to do so we define an infinitesimal vector as $\lambda=\left(\lambda^0,\lambda^i\right)$, where we decompose $\lambda^i=\lambda^i_{\perp}+\partial^i\eta$, where $\lambda^i_{\perp}$ is a 3-vector with zero divergence and $\eta$ is a scalar function, then a scalar quantity transforms under an infinitesimal gauge transformation as: 
\begin{equation}
\delta\phi\rightarrow\tilde{\delta\phi}=\delta\phi+\phi'\lambda^0.
\label{gauge_transf_phi}
\end{equation}
 
Imposing gauge invariance of \eqref{Lin_metric} we get the transformation rules for the different parameters in the metric
\begin{align} \nonumber
D\rightarrow\tilde{D}=D+\mathcal{H}\lambda^0+\frac{1}{3}\nabla^2 \eta,\\ \nonumber
A \rightarrow \tilde{A}=A+\mathcal{H}\lambda^0+\lambda^{0\prime}\\ \nonumber
E\rightarrow \tilde{E}=E-\eta, \\ \label{gauge_transf_metric}
B\rightarrow \tilde{B}=B+\eta'-\lambda^0, 
\end{align}
where, as before, $(')$ means a derivative with respect to the conformal time ($dt=a d\tau$), and $\mathcal{H}=\frac{a'}{a}=a H^b$.

From the definition of the overall expansion rate $\theta= H^b + \delta K$ and \eqref{Lin_H_def} we deduce that $\delta K=-H^b A+\dot{D}-\frac{1}{3}\frac{\nabla^2}{a} B$ or $\delta \mathcal{H}=-\mathcal{H}A+D'-\frac{1}{3}\nabla^2 B$. Performing a gauge transformation we get:

\begin{equation}
\delta \mathcal{H}\rightarrow \tilde{\delta \mathcal{H}}=\delta \mathcal{H}-\mathcal{H}^2\lambda^0+\mathcal{H}'\lambda^0+\frac{1}{3}\nabla^2\lambda^0,
\label{gauge_transf_H}
\end{equation}

In order to find the gauge transformation parameter $\lambda^0_{\text{f}\rightarrow \delta K=0}$ between flat (f) and $\delta K=0$ gauges we need to set $\tilde{\delta \mathcal{H}}=0$ and $\delta \mathcal{H}=\delta \mathcal{H}_{\text{f}}=-\mathcal{H}A_{\text{f}}-\frac{1}{3}\nabla^2B_{\text{f}}$ (spatially flat gauge, or $\psi=0$, implies, from \eqref{Lin_curv_def}, that both $D$ and $E$ are 0) in \eqref{gauge_transf_H} obtaining:
\begin{equation}
-3\lambda^0_{\text{f}\rightarrow \delta K=0}\mathcal{H}^2\epsilon_1+\nabla^2\lambda^0_{\text{f}\rightarrow \delta K=0}=S,
\label{gauge_transf_eq}
\end{equation}
where $S=-3\delta\mathcal{H}_{\text{f}}$ we have used the definition of $\epsilon_1$ in conformal time i.e. $\epsilon_1=1-\frac{\mathcal{H}'}{\mathcal{H}^2}$.

The only thing left to do is to specify $S$, in order to do so we use energy and momentum constraints \eqref{Lin_HAM} and \eqref{Lin_MOM} in spatially flat gauge (i.e setting $D=0$ and $E=0$).
\begin{align}
\mathcal{H}\left(3\mathcal{H} A_{\text{f}}+\nabla^2 B_{\text{f}}\right)&=-\frac{1}{2M_{PL}^2}\left[\phi^{b\prime}(\delta\phi'_{\text{f}}-\phi^{b\prime} A_{\text{f}})+a^2V_{\phi}\delta\phi_{\text{f}}\right], \label{Lin_HAM_psi}\\ 
\mathcal{H} A_{\text{f}}&=\frac{1}{2M_{PL}^2}\phi^{b\prime}\delta\phi_{\text{f}}, \label{Lin_MOM_psi} 
\end{align}

We now can see that the left-hand side of \eqref{Lin_HAM_psi} is nothing more than $\mathcal{H}S$. Using \eqref{Lin_MOM_psi} to substitute $A_{\text{f}}$ in \eqref{Lin_HAM_psi} we get a solution for S:

\begin{equation}
S=-\frac{Q}{2M_{PL}^2\mathcal{H}}\left(a^2 V_{\phi}+\phi^{b\prime}\frac{Q'}{Q}-\phi^{b\prime}\mathcal{H}\epsilon_1\right),
\label{gauge_transf_eq2}
\end{equation}
where we have already used $\delta\phi_{\text{f}}=Q$.

The next step is to solve \eqref{gauge_transf_eq} for $\left.\lambda^0_{\mathbf{k}}\right|_{\text{f}\rightarrow \delta K=0}$ in Fourier space:

\begin{equation}
\left.\lambda^0_{\mathbf{k}}\right|_{\text{f}\rightarrow \delta K=0}=-\frac{S_{\mathbf{k}}}{3\mathcal{H}^2\epsilon_1+k^2}=\frac{\frac{Q_{\mathbf{k}}}{2M_{PL}^2\mathcal{H}}\left(a^2 V_{\phi}+\phi^{b\prime}\frac{Q_{\mathbf{k}}'}{Q_{\mathbf{k}}}-\phi^{b\prime}\mathcal{H}\epsilon_1\right)}{\mathcal{H}^2\left(\left(\frac{k}{\mathcal{H}}\right)^2+3\epsilon_1\right)}.
\label{gauge_transf_par}
\end{equation}
where $k=|\mathbf{k}|$.

Finally, using \eqref{gauge_transf_phi}, the field perturbation in uniform Hubble gauge is
\begin{equation}
\left.\delta\phi_{\mathbf{k}}\right|_{\delta K=0}=Q_{\mathbf{k}}+\phi^{b\prime}\left.\lambda^0_{\mathbf{k}}\right|_{\text{f}\rightarrow \delta K=0}=Q_{\mathbf{k}}\left[1-\frac{3\epsilon_1+\frac{1}{2}\epsilon_1\epsilon_2-\epsilon_1\frac{Q'_{\mathbf{k}}}{\mathcal{H}Q_{\mathbf{k}}}}{\left(\frac{k}{\mathcal{H}}\right)^2+3\epsilon_1}\right].
\label{gauge_transf_field}
\end{equation}

This result together with the definition of the MS variable gives the value of $\left.\psi_{\mathbf{k}}\right|_{\delta K=0}$

\begin{equation}
\left.\psi_{\mathbf{k}}\right|_{\delta K=0}=\frac{\mathcal{H}}{\phi^{b\prime}}\left(Q_{\mathbf{k}}-\left.\delta\phi_{\mathbf{k}}\right|_{\delta K=0}\right)=\frac{\mathcal{H}Q_{\mathbf{k}}}{\phi^{b\prime}}\left[\frac{3\epsilon_1+\frac{1}{2}\epsilon_1\epsilon_2-\epsilon_1\frac{Q'_{\mathbf{k}}}{\mathcal{H}Q_{\mathbf{k}}}}{\left(\frac{k}{\mathcal{H}}\right)^2+3\epsilon_1}\right].
\label{gauge_transf_psi}
\end{equation}

In order to conclude this appendix we are going to study some properties of \eqref{gauge_transf_field} and \eqref{gauge_transf_psi} in the long-wavelength limit i.e. when $\frac{k}{\mathcal{H}}=\frac{k}{a H}=\sigma\ll 1$. We want to study this limit because as one can clearly see, it is the same limit we have used both in the gradient expansion as in the stochastic formalism.

In this limit we can use the expansion parameter $\frac{\sigma^2}{3\epsilon_1}\ll 1$.\footnote{The condition $\sigma\ll 1$ is automatically satisfied as soon as our coarse-grained scale is big enough, however, $\epsilon_1$ is also very small during inflation so this expansion could cease to be valid at some point. In order to use our stochastic formalism one has to be sure that $\frac{\sigma^2}{3\epsilon_1}\ll 1$, this does not represent a problem since it is easy to check that one can always choose a value for $\sigma$ such that:

\begin{equation}
\exp\left(-\frac{1}{4\epsilon_1}\right)\ll \sigma \ll \sqrt{3\epsilon_1},
\label{lower_limit_sigma}
\end{equation}
where the lower limit was first obtained in \cite{Starobinsky:1994bd} and later refined in \cite{Vennin:2020kng}. \label{foot_sigma}}
The result is:
\begin{align}
\nonumber
\left.\delta\phi_{\mathbf{k}}\right|_{\delta K=0}&= \frac{Q_{\mathbf{k}}}{6}\left[\left(2\frac{Q'_{\mathbf{k}}}{\mathcal{H}Q_{\mathbf{k}}}-\epsilon_2 \right)+\left(6+\epsilon_2 -2\frac{Q'_{\mathbf{k}}}{\mathcal{H}Q_{\mathbf{k}}}\right)\left(\frac{\sigma^2}{3\epsilon_1}\right)\right],\\ 
\left.\psi_{\mathbf{k}}\right|_{\delta K=0}&=\frac{Q_{\mathbf{k}}\mathcal{H}}{6 \phi^{b\prime}}\left[\left(6+\epsilon_2 -2\frac{Q'_{\mathbf{k}}}{\mathcal{H}Q_{\mathbf{k}}}\right)+\left(2\frac{Q'_{\mathbf{k}}}{\mathcal{H}Q_{\mathbf{k}}}-6-\epsilon_2 \right)\left(\frac{\sigma^2}{3\epsilon_1}\right)\right]. \label{psiphi_series} 
\end{align}

Equation \eqref{psiphi_series} is enough for the numerical implementation of the stochastic formalism we present in the main text. Note that the expansion done in \eqref{psiphi_series} requires $\epsilon_1 \neq 0$, indeed, if we impose $\epsilon_1=0$ in \eqref{gauge_transf_field} and \eqref{gauge_transf_psi}, we get that uniform Hubble and spatially flat gauges are equivalent and hence we recover the noises from the ``old'' stochastic formalism.

As an example we can write an analytical expression for $\left.\delta\phi_{\mathbf{k}}\right|_{\delta K=0}$ and $\left.\psi_{\mathbf{k}}\right|_{\delta K=0}$ in SR at first order in $\epsilon_1$. 

If we write the MS equation \eqref{MS_Q_t} using the conformal time $\tau$ defined as  $\left(\tau=\int\frac{dt}{a}\right)$ as time variable, we get:
\begin{equation}
Q_{\mathbf{k}}''+2\mathcal{H}Q_{\mathbf{k}}'+\left(k^2+\mathcal{H}^2(2-\epsilon_1)+\frac{z''}{z}\right)Q_{\mathbf{k}}=0,
\label{MS_Q_tau}
\end{equation}
where (') denotes a derivative with respect to $\tau$ and we have defined $z=a \frac{\phi^{b\prime}}{\mathcal{H}^b}=a\sqrt{2\epsilon_1}M_{PL}$ such that $\frac{z''}{z}$ can be written in terms of SR parameters:

\begin{equation}
\frac{z''}{z}=a^2 H^2 \left(2-\epsilon_1+\frac{3}{2}\epsilon_2+\frac{1}{4}\epsilon_2^2-\frac{1}{2}\epsilon_1\epsilon_2+\frac{1}{2}\epsilon_2\epsilon_3\right).
\label{zppz}
\end{equation}

The solution of \eqref{MS_Q_tau} provided that $\nu^2=\frac{1}{4}+\tau^2\frac{z''}{z}$ is constant up to the level of precision we are looking for (in our case $\nu$ must be constant up to $\mathcal{O}(\epsilon_1)$) is:
\begin{equation}
Q_{\mathbf{k}}=\left.\delta\phi_{\mathbf{k}}\right|_{\text{f}}=\frac{e^{\frac{i}{2}\pi\left(\nu+\frac{1}{2}\right)}}{a}\frac{\sqrt{\pi}}{2}\sqrt{- \tau}H_{\nu}^{(1)}(-k\tau),
\label{Q_hankel}
\end{equation}
where we have used the Bunch-Davies vacuum  \cite{Bunch:1978yq} as a initial condition. $H_{\nu}^{(1)}$ is the Hankel function of first class.

Using the expansion of the Henkel function when $(-k\tau)\simeq \sigma \ll 1$ and $\nu>1$ (always the case in SR) we have
\begin{equation}
H_{\nu}^{(1)}(-k\tau)=\frac{i}{\pi}2^{\nu}(-k\tau)^{-\nu}\left(-\Gamma[\nu]+\frac{1}{4}\Gamma[\nu-1](-k\tau)^2\right),
\label{hankel_exp}
\end{equation}
we obtain the following expression for $Q_{\mathbf{k}}$ up to $(-k\tau)^2$:
\begin{equation}
Q_{\mathbf{k}}=\left.\delta\phi_{\mathbf{k}}\right|_{\text{f}}= - i \frac{e^{\frac{i}{2}\pi\left(\nu+\frac{1}{2}\right)}2^{\nu-1}}{a \sqrt{\pi} }\sqrt{- \tau}(-k\tau)^{-\nu}\left(\Gamma[\nu]-\frac{1}{4}\Gamma[\nu-1](-k\tau)^2\right),
\label{Q_hankel_exp}
\end{equation}

and hence for $\frac{Q'_{\mathbf{k}}}{\mathcal{H}Q_{\mathbf{k}}}$ up to $(-k\tau)^2$ is:
\begin{equation}
\frac{Q'_{\mathbf{k}}}{\mathcal{H}Q_{\mathbf{k}}}\simeq \frac{1-2\nu-2\mathcal{H}\tau}{2\mathcal{H}\tau} + \frac{1}{2\mathcal{H}\tau(\nu-1)}(-k\tau)^2,
\label{QdotQ_exp}
\end{equation}

Finally one can integrate by parts $\tau=\int\frac{dt}{a}$ up to order $\epsilon_1$ as done in appendix \ref{App_MS_sol}. This gives a result in SR of 
\begin{equation}
\tau^{SR} \simeq -\frac{1}{\mathcal{H}}\left(1+\epsilon_1\right),
\label{tau_exp_SR}
\end{equation}
which, together with the definition of $\nu$, give us the expression of $\nu$ in SR up to $\mathcal{O}(\epsilon_i)$ :
\begin{equation}
\nu^{SR}=\frac{3}{2}+\epsilon_1+\frac{\epsilon_2}{2}.
\label{nu_SR}
\end{equation}

After inserting \eqref{QdotQ_exp}, \eqref{tau_exp_SR} and \eqref{nu_SR} into \eqref{psiphi_series} one gets:

\begin{equation}
  \left\{
        \begin{array}{ll}
            \left.\delta\phi^{SR}_{\mathbf{k}}\right|_{\delta K=0} \simeq \mathcal{O}\left(\frac{\sigma^2}{3\epsilon_1}\right)  \\
            \left.\psi^{SR}_{\mathbf{k}}\right|_{\delta K=0} \simeq \frac{\mathcal{H} Q^{(\epsilon)}_{\mathbf{k}}}{\phi^{b\prime}} + \mathcal{O}\left(\frac{\sigma^2}{3\epsilon_1}\right)
        \end{array}
    \right.
\end{equation}

where $Q^{(\epsilon)}_{\mathbf{k}}$ stands for $Q_{\mathbf{k}}$ but expanded at first order in $\epsilon_1$. In a SR regime, comoving gauge ($\delta \phi=0$) and uniform Hubble gauge are equivalent up to $\mathcal{O}(\epsilon_i)$ and $\mathcal{O}(\sigma^2)$, provided that $\epsilon_1 \ll 1$ but $\epsilon_1 \neq 0$.

Note that in order to study analytical solutions in other regimes like USR one must know the solution of equation \eqref{MS_Q_tau} up to precision $\mathcal{O}(\epsilon_1)$, however, there is not analytical solution for \eqref{MS_Q_tau} when $\nu$ is not a constant. This is precisely the case in USR up to precision $\mathcal{O}(\epsilon_1)$, this is because in USR we have $\frac{d \epsilon_1}{d N} \sim \mathcal{O}(\epsilon_1)$. In appendix \ref{App_MS_sol} we will present an alternative approximation in order to get an analytical  solution for the MS equation for regimes beyond SR and up to $\mathcal{O}(\epsilon_1)$. Using results from appendix \ref{App_MS_sol} together with $\epsilon_2^{USR}\simeq -6$, we can write \eqref{psiphi_series} in USR:

\begin{equation}
  \left\{
        \begin{array}{ll}
            \left.\delta\phi^{USR}_{\mathbf{k}}\right|_{\delta K=0} \simeq  Q_{\mathbf{k}} + \mathcal{O}\left(\frac{\sigma^2}{3\epsilon_1}\right)  \\
            \left.\psi^{USR}_{\mathbf{k}}\right|_{\delta K=0} \simeq \mathcal{O}\left(\frac{\sigma^2}{3\epsilon_1}\right)
        \end{array}
    \right.
\end{equation}

\section{Appendix C: Derivation of the ``old'' stochastic formalism.}
\label{AppOld}
In this appendix we will derive equations \eqref{STO_old_SR} and \eqref{STO_old_USR} by using spatially flat gauge with the further assumption of $\partial_i\left(_{(0)}\beta_{\text{f}}\right)^i=0$. In this case, the Hamiltonian constraint \eqref{ADM_HAM_flat_final} simplifies considerably:

\begin{equation}
\left(\frac{H^b}{_{(0)}\alpha^{IR}_{\text{f}}}\right)^2 = \frac{V\left(_{(0)}\phi^{IR}_{\text{f}}\right)}{3 M_{PL}^2 - \frac{1}{2}\left(\frac{\partial \,_{(0)}\phi^{IR}_{\text{f}}}{\partial N}\right)^2-\dot{\phi^b}\left.\xi_1(N)\right|_{\text{f}}} = \frac{V\left(_{(0)}\phi^{IR}_{\text{f}}\right)}{3 M_{PL}^2 - \frac{\left(\,_{(0)}\pi^{IR}_{\text{f}}\right)^2}{2}},
\label{STO_HAM_flat}
\end{equation}
where in the last equality we have defined the auxiliary variable $\,_{(0)}\pi^{IR}_{\text{f}}$ as:
\begin{equation}
\,_{(0)}\pi^{IR}_{\text{f}} \equiv \frac{\partial \,_{(0)}\phi^{IR}_{\text{f}}}{\partial N} + \left.\xi_1(N)\right|_{\text{f}}
\label{STO_redef_flat}
\end{equation}

Once we have seen how the stochastic equation for the Hamiltonian constraint is derived we can follow the same procedure to write the whole set of ADM equations of appendix \ref{AppADM}. After a straightforward calculation, one can see that the only equations of interest at leading order in gradient expansion are:

\begin{itemize}
\item The Klein-Gordon equation for the field \eqref{ADM_KG}
\begin{equation}
\frac{\partial \,_{(0)}\pi^{IR}_{\text{f}}}{\partial N}  = - \left(3 + \frac{\frac{\partial }{\partial N}\left(\frac{H^b}{\,_{(0)}\alpha^{IR}_{\text{f}}}\right)}{\frac{H^b}{\,_{(0)}\alpha^{IR}_{\text{f}}}}\right)\,_{(0)}\pi^{IR}_{\text{f}} - \frac{V_{\phi}\left(_{(0)}\phi^{IR}_{\text{f}}\right)}{\left(\frac{H^b}{\,_{(0)}\alpha^{IR}_{\text{f}}}\right)^2} - \left.\xi_2(N)\right|_{\text{f}} + \frac{\partial \phi^b}{\partial N}\left.\xi_3(N)\right|_{\text{f}} \,,
\label{STO_KG_flat}
\end{equation}

where, in the same way as we did with $\xi_1$ we define the white noises $\xi_2$ and $\xi_3$ (Note that they are now written using $N$ as time variable).

\begin{align}
\nonumber
\left.\xi_2(N)\right|_{\text{f}}=-\sigma a H^b (1-\epsilon_1)\int \frac{d^3 k}{(2\pi)^{3/2}}\delta(k-\sigma a H^b)\left.\frac{\delta\varphi_{\mathbf{k}}}{\partial N}\right|_{\text{f}},\\ \nonumber
\left.\xi_3(N)\right|_{\text{f}}=-\sigma a H^b (1-\epsilon_1)\int \frac{d^3 k}{(2\pi)^{3/2}}\delta(k-\sigma a H^b)\left.\mathcal{A}_{\mathbf{k}}\right|_{\text{f}},\\
\label{Noises_old}
\end{align}

\item Evolution equation for the trace of the extrinsic curvature $K$ \eqref{ADM_EVEXT}
\begin{equation}
\frac{\frac{\partial }{\partial N}\left(\frac{H^b}{\,_{(0)}\alpha^{IR}_{\text{f}}}\right)}{\frac{H^b}{\,_{(0)}\alpha^{IR}_{\text{f}}}} = - \frac{\left(\,_{(0)}\pi^{IR}_{\text{f}}\right)^2}{2 M_{PL}^2}
\label{STO_EVEX_flat}
\end{equation}
\end{itemize}

Note that we have not taken into account the momentum constraint because we are at leading order in $\epsilon_1$. It is now easy to realize that \eqref{STO_HAM_flat}, \eqref{STO_KG_flat} and \eqref{STO_EVEX_flat} can be written in a compact way:

\begin{changemargin}{-2cm}{-1cm}
\begin{align}
\nonumber
\,_{(0)}\pi^{IR}_{\text{f}}&=\frac{\partial \,_{(0)}\phi^{IR}_{\text{f}}}{\partial N} + \left.\xi_1(N)\right|_{\text{f}}\\ \nonumber
\frac{\partial \,_{(0)} \pi^{IR}_{\text{f}}}{\partial N} & = - \left(3 - \frac{\left(\,_{(0)}\pi^{IR}_{\text{f}}\right)^2}{2 M_{PL}^2}\right)\,_{(0)}\pi^{IR}_{\text{f}} - \frac{V_{\phi}\left(\,_{(0)}\phi^{IR}_{\text{f}}\right)}{V\left(\,_{(0)}\phi^{IR}_{\text{f}}\right)}\left(3M_{PL}^2-\frac{\left(\,_{(0)}\pi^{IR}_{\text{f}}\right)^2}{2}\right) - \left.\xi_2(N)\right|_{\text{f}} + \frac{\partial \phi^b}{\partial N}\left.\xi_3(N)\right|_{\text{f}}\\
\label{STO_old}
\end{align}
\end{changemargin}

In order to solve \eqref{STO_old} it is necessary to calculate the variance of the noises \eqref{Field_noi_old} and \eqref{Noises_old}.

\begin{changemargin}{-1.5cm}{-1cm} 
\begin{align} \nonumber
\langle\left.\xi_1(N_1)\right|_{\text{f}}\left.\xi_1(N_2)\right|_{\text{f}}\rangle &=\frac{\left(\sigma a H^b\right)^3}{2\pi^2}(1-\epsilon_1)\left|Q_{\mathbf{k}}\right|^2_{k=\sigma a H^b}\delta (N_1-N_2),\\ \nonumber
\langle\left.\xi_2(N_1)\right|_{\text{f}}\left.\xi_2(N_2)\right|_{\text{f}}\rangle &=\frac{\left(\sigma a H^b\right)^3}{2\pi^2}(1-\epsilon_1)\left|\frac{\partial Q_{\mathbf{k}}}{\partial N}\right|^2_{k=\sigma a H^b}\delta (N_1-N_2), \\ \nonumber
\langle\left.\xi_3(N_1)\right|_{\text{f}}\left.\xi_3(N_2)\right|_{\text{f}}\rangle &=\frac{\left(\sigma a H^b\right)^3}{2\pi^2}(1-\epsilon_1)\left|\left. A_{\mathbf{k}}\right|_{\psi=0}\right|^2_{k=\sigma a H^b}\delta (N_1-N_2) =\epsilon_1 \langle\left.\xi_1(N_1)\right|_{\text{f}}\left.\xi_1(N_2)\right|_{\text{f}}\rangle,\\ 
\langle\left.\xi_1(N_1)\right|_{\text{f}}\left.\xi_2(N_2)\right|_{\text{f}}\rangle &=\frac{\left(\sigma a H^b\right)^3}{2\pi^2}(1-\epsilon_1)\left(Q_{\mathbf{k}}^{\star}\frac{\partial Q_{\mathbf{k}}}{\partial N}\right)_{k=\sigma a H^b}\delta (N_1-N_2),
\label{Noises_old_variance}
\end{align}
\end{changemargin}
where the last equality in the third line comes from using momentum constraint \eqref{Lin_MOM_psi} in spatially flat gauge. Note that we have also written the correlators in terms of the Mukhanov-Sasaki variable $Q_{\textbf{k}}=\delta\phi_{\textbf{k}} + \frac{\partial \phi^b}{\partial N} \psi_{\textbf{k}}$ because in spatially flat gauge we have $Q_{\mathbf{k}}=\left.\delta\phi_{\mathbf{k}}\right|_{\text{f}}$.

Another important aspect to remark from \eqref{Noises_old_variance} is that they are completely correlated (or completely anti-correlated, depending on the values of the correlation matrix whose entries are \eqref{Noises_old_variance}) noises. This is because they all come from linear perturbation theory. This means that the sign before the noises in the stochastic formalism is important, which is a crucial aspect to take into account when doing numerics with the stochastic formalism as in section \ref{Sec_checks}.

Eq \eqref{STO_old} is not consistent because it should not include any term of leading order in $\epsilon_1$. This is why we should rewrite \eqref{STO_old} as:

\begin{align} \nonumber
\,_{(0)}\pi^{IR}_{\text{f}}&=\frac{\partial \,_{(0)}\phi^{IR}_{\text{f}}}{\partial N} + \left.\xi_1(N)\right|_{\text{f}} \,, & \\ \nonumber
\frac{\partial \,_{(0)} \pi^{IR}_{\text{f}}}{\partial N} & = - 3\,_{(0)}\pi^{IR}_{\text{f}} - \left.\xi_2(N)\right|_{\text{f}} \,, & \text{in USR}\\ \nonumber
& \\
3\frac{\partial \phi^{IR}_{\text{f}}}{\partial N}&=-3M_{PL}^2\frac{V_{\phi}\left(\phi^{IR}_{\text{f}}\right)}{V\left(\phi^{IR}_{\text{f}}\right)} - 3\left.\xi_1(N)\right|_{\text{f}}- \left.\xi_2(N)\right|_{\text{f}}\,, & \text{in SR}
\label{STO_old_fullnoi}
\end{align}

Finally, $\xi_1$ and $\xi_2$ must also be computed at leading order in $\epsilon_1$, which results into $\langle\xi_1(N_1)\xi_1(N_2)\rangle = \left(\frac{H^b}{2\pi}\right)^2 \delta(N_1-N_2)$ and $\langle\xi_2(N_1)\xi_2(N_2)\rangle = 0$. This allow us to perform the noise redefinition $\left.\xi_1(N)\right|_{\text{f}}\rightarrow \frac{H^b}{2\pi }\xi(N)$, where $\langle\xi(N_1)\xi(N_2)\rangle=\delta (N_1-N_2)$. Inserting the definition of $\xi(N)$ into \eqref{STO_old_fullnoi} we arrive to:

\begin{align} \nonumber
\,_{(0)}\pi^{IR}_{\text{f}}&=\frac{\partial \,_{(0)}\phi^{IR}_{\text{f}}}{\partial N} + \left(\frac{H^b}{2\pi }\right)\left.\xi(N)\right|_{\text{f}} \,, & \\ \nonumber
\frac{\partial \,_{(0)} \pi^{IR}_{\text{f}}}{\partial N} & = - 3\,_{(0)}\pi^{IR}_{\text{f}} \,, & \text{in USR}\\ \nonumber
& \\ 
\frac{\partial \phi^{IR}_{\text{f}}}{\partial N}&=-M_{PL}^2\frac{V_{\phi}\left(\phi^{IR}_{\text{f}}\right)}{V\left(\phi^{IR}_{\text{f}}\right)} - \left(\frac{H^b}{2\pi }\right)\left.\xi_1(N)\right|_{\text{f}}\,, & \text{in SR}
\label{STO_old_app}
\end{align}
which are Eq \eqref{STO_old_USR} and \eqref{STO_old_SR}, respectively. 

\section{Appendix D: Derivation of the ``new'' stochastic formalism}
\label{AppNew}
In this appendix we will derive each one of the equations of the ``new'' stochastic formalism. Since we are always using the Starobinski approximation we will directly write the linear perturbation theory equation for the $UV$ side. We will follow the same order as in the main text:

\begin{itemize}
\item The evolution equation for the spatial metric \eqref{ADM_EVSPA} in uniform Hubble gauge is:
\begin{equation}
\frac{\partial \,_{(n)}\zeta}{\partial N} - \left(\,_{(n)}\alpha-1\right)=0,
\end{equation}
where, as in the main text, a subindex $\,_{(n)}$ means that we are at all orders in $\sigma$ so the above equation is exact. When we split it between $IR$ and $UV$ we get:
\begin{equation}
\frac{\partial \zeta}{\partial N} - \left(\alpha-1\right) = -\frac{\partial D}{\partial N} + A,
\label{EVSPA_app}
\end{equation}
where, as already noted, we have adopted the notation of linear perturbation theory ($D$ and $A$ instead of $\zeta^{UV}$ and $\alpha^{UV}$) because we are already assuming Starobinski approximation.

Finally, when including the Fourier splitting \eqref{Splitting} into \eqref{EVSPA_app} we get:

\begin{equation}
\frac{\partial \zeta}{\partial N} - \left(\alpha-1\right) = -\left.\xi_4(N)\right|_{\delta K=0},
\end{equation}
where $\left.\xi_4(N)\right|_{\delta K=0}$ is defined as:
\begin{equation}
\left.\xi_4(N)\right|_{\delta K=0} \equiv -\sigma a H^b (1-\epsilon_1)\int \frac{d^3 k}{(2\pi)^{3/2}}\delta(k-\sigma a H^b)\left.\mathcal{D}_{\mathbf{k}}\right|_{\delta K=0},
\label{EVSPA_app_sto}
\end{equation}

\item The scalar field equation of motion \eqref{ADM_KG} in uniform Hubble gauge once the splitting between $IR$ and $UV$ has been done is:
 
\begin{align} \nonumber
\frac{1}{\alpha}\left[\frac{\partial^2 \phi}{\partial N^2} + \left(3-\epsilon_1 -\frac{1}{\alpha}\frac{\partial \alpha}{\partial N}-\frac{\partial \zeta}{\partial N}\right)\frac{\partial \phi}{\partial N}\right]+& \alpha \frac{V_{\phi}}{H^2} = \\
 \frac{\partial ^2\delta\phi}{\partial N^2}+  (3-\epsilon_1)\frac{\partial \delta \phi}{\partial N} & +\left(V_{\phi\phi}-\frac{\nabla^2}{a^2}\right)\delta\phi + 2 V_{\phi} A+\frac{\partial \phi^b}{\partial N}\left[\frac{\partial A}{\partial N}-3 A \right]
 \label{KG_app}
\end{align}

If we now include the Fourier splitting \eqref{Splitting} into \eqref{KG_app} we get:

\begin{align} \nonumber
\frac{1}{\alpha}\left[\frac{\partial^2 \phi}{\partial N^2} + \left(3-\epsilon_1 -  \frac{1}{\alpha}\frac{\partial \alpha}{\partial N}+3\frac{\partial \zeta}{\partial N}\right)\frac{\partial \phi}{\partial N}\right] & + \alpha \frac{V_{\phi}}{H^2} = \\ - \left(3-\epsilon_1\right)\left.\xi_1(N)\right|_{\delta K=0} &  - \frac{\partial \left.\xi_1(N)\right|_{\delta K=0}}{\partial N} - \left.\xi_2(N)\right|_{\delta K=0} + \frac{\partial \phi^b}{\partial N}\left.\xi_3(N)\right|_{\delta K=0},
\end{align}
where $\left.\xi_2(N)\right|_{\delta K=0}$, $\left.\xi_2(N)\right|_{\delta K=0}$ and $\left.\xi_3(N)\right|_{\delta K=0}$ are defined as:

\begin{align} \nonumber
\left.\xi_1(N)\right|_{\delta K=0} \equiv -\sigma a H^b (1-\epsilon_1)\int \frac{d^3 k}{(2\pi)^{3/2}}\delta(k-\sigma a H^b)\left.\delta\varphi_{\mathbf{k}}\right|_{\delta K=0}, \\  \nonumber
\left.\xi_2(N)\right|_{\delta K=0} \equiv -\sigma a H^b (1-\epsilon_1)\int \frac{d^3 k}{(2\pi)^{3/2}}\delta(k-\sigma a H^b)\left.\frac{\delta\varphi_{\mathbf{k}}}{\partial N}\right|_{\delta K=0}, \\
\left.\xi_4(N)\right|_{\delta K=0} \equiv -\sigma a H^b (1-\epsilon_1)\int \frac{d^3 k}{(2\pi)^{3/2}}\delta(k-\sigma a H^b)\left.\mathcal{A}_{\mathbf{k}}\right|_{\delta K=0},
\label{EVSPA_app_sto}
\end{align}

Finally, we can use \eqref{EVSPA_app_sto} in order to eliminate $\frac{\partial \zeta}{\partial N}$ and a redefinition of the velocity of the field in order to eliminate $\frac{\partial \left.\xi_1(N)\right|_{\delta K=0}}{\partial N}$ in \eqref{EVSPA_app_sto} getting:

\begin{changemargin}{-1.5cm}{-1cm} 
\begin{equation}
\frac{\partial \pi}{\partial N} + (3\alpha - \epsilon_1)\frac{\partial \tilde{\phi}}{\partial N} + \alpha \frac{V_{\phi}(\phi)}{H^2}= - \left(3-\epsilon_1\right)\left.\xi_1(N)\right|_{\delta K=0} - \left.\xi_2(N)\right|_{\delta K=0} + \frac{\partial \phi^b}{\partial N}\left(\left.\xi_3(N)\right|_{\delta K=0}+3\left.\xi_4(N)\right|_{\delta K=0}\right),
\label{STO_KG_H_app}
\end{equation}
\end{changemargin}
\begin{equation}
\pi\equiv\frac{1}{\alpha}\frac{\partial \phi}{\partial N} + \left.\xi_1(N)\right|_{\delta K=0}=\frac{\partial \tilde{\phi}}{\partial N} + \left.\xi_1(N)\right|_{\delta K=0}.
\label{STO_redef_H_app}
\end{equation}

\item The Hamiltonian constraint \eqref{ADM_HAM} already separated between $IR$ and $UV$ is:
\begin{changemargin}{-1.5cm}{-1cm} 
\begin{equation}
H^2-\frac{1}{3M_{PL}^2}\left(V(\phi)+H^2\frac{1}{2}\left(\frac{\partial \tilde{\phi}}{\partial N}\right)^2\right)= - \frac{2}{3}\frac{\nabla^2}{a^2}\left[D+\frac{1}{3}\nabla^2 E\right]+\frac{1}{3M_{PL}^2}\left[H^2\frac{\partial \phi^b}{\partial N}\left(\frac{\partial \delta\phi}{\partial N}-\frac{\partial \phi^b}{\partial N} A\right)+V_{\phi}\delta\phi\right]\,.
\end{equation}
\end{changemargin}

Using the splitting in Fourier space once again and using the fact that due to the gauge chosen, the Hamiltonian constraint must coincide with the Hamiltonian constraint of the background system we can write:
\begin{equation}
H^2=\frac{V(\phi^b)}{3M_{PL}^2-\frac{1}{2}\left(\frac{\partial \phi^b}{\partial N}\right)^2}=\frac{V(\phi)}{3M_{PL}^2-\frac{1}{2}\left(\frac{\partial \tilde{\phi}}{\partial N}\right)^2 - \frac{\partial \phi^b}{\partial N}\left.\xi_1(N)\right|_{\delta K=0}} \,.
\label{STO_HAM_H_app}
\end{equation}

\item Now we do the same with the evolution equation for the trace of the extrinsic curvature \eqref{ADM_EVEXT}:

\begin{equation}
-3 H\frac{\partial H}{\partial N}-3H^2\alpha - \frac{\alpha}{M_{PL}^2}\left(H^2\frac{\partial \tilde{\phi}}{\partial N}-V(\phi)\right)=\frac{1}{M_{PL}^2}\left(\frac{3}{2}H^2\frac{\partial \phi^b}{\partial N} A +2H^2\frac{\partial \phi^b}{\partial N}\frac{\partial \delta \phi}{\partial N}-V_{\phi}\partial \phi\right)\,.
\end{equation}

If we write the $UV$ part in Fourier space we get the noise:
\begin{equation}
-3 H\frac{\partial H}{\partial N}-3H^2\alpha - \frac{\alpha}{M_{PL}^2}\left(H^2\frac{\partial \tilde{\phi}}{\partial N}-V(\phi)\right)=\frac{H^2}{M_{PL}^2}\frac{\partial \phi^b}{\partial N}\left.\xi_1(N)\right|_{\delta K=0}\,.
\end{equation}

Finally we can substitute $\frac{\partial H}{\partial N}$ by its background value and eliminate $3H^2\alpha$ using eq \eqref{STO_HAM_H_app}. The result is:
\begin{equation}
\left(\frac{\partial \phi^b}{\partial N}\right)^2=\alpha\left(\frac{\partial \tilde{\phi}}{\partial N}\right)^2 + \frac{2}{3}\left(2+\alpha\right)\frac{\partial \phi^b}{\partial N}\left.\xi_1(N)\right|_{\delta K=0} \,.
\end{equation}

\item The last equation to derive is the momentum constraint. However, since we have derived it in the main text we will only write here the value for the noise $\left.\xi_5(N)\right|_{\delta K=0}$:

\begin{equation}
\left.\xi_5(N)\right|_{\delta K=0} \equiv -\sigma a H^b (1-\epsilon_1)\int \frac{d^3 k}{(2\pi)^{3/2}}\delta(k-\sigma a H^b)k^2\left.\mathcal{E}_{\mathbf{k}}\right|_{\delta K=0},
\end{equation}
\end{itemize}

\section{Appendix E: Solution of the MS equation for USR and CR regimes}
\label{App_MS_sol}

This appendix is devoted to the study of solutions for the MS equation for the cases in which $\nu^2=\frac{1}{4}+\frac{z''}{z}\tau^2$ cannot be assumed to be a constant (see discussion below \eqref{MS_Q_tau}). First of all, we will rewrite \eqref{MS_Q_tau} in terms of $u_{\mathbf{k}}=a Q_{\mathbf{k}}$, where $a$ is the scale factor and $Q_{\mathbf{k}}$ is the MS variable defined in \eqref{Lin_Q_def}.

\begin{equation}
u_{\mathbf{k}}''(\tau)+ \left(k^2-\frac{z''}{z}\right)u_{\mathbf{k}}(\tau)=0,
\label{MS_u_tau}
\end{equation}
where we have defined $z=a \frac{\phi^{b\prime}}{H^b}=a\sqrt{2\epsilon_1}M_{PL}$ such that $\frac{z''}{z}$ can be written in terms of SR parameters:

\begin{equation}
\frac{z''}{z}=a^2 H^2 \left(2-\epsilon_1+\frac{3}{2}\epsilon_2+\frac{1}{4}\epsilon_2^2-\frac{1}{2}\epsilon_1\epsilon_2+\frac{1}{2}\epsilon_2\epsilon_3\right).
\label{zppz}
\end{equation}

In order to have an analytical solution in terms of Henkel functions as in \eqref{Q_hankel} we need $\nu^2=\frac{1}{4}+\frac{z''}{z}\tau^2$ to be a constant. Let us study when this is the case: 

First of all it is very convenient to write $\tau$ in terms of $ a H$ or viceversa to see if the term $\frac{z''}{z}\tau^2$ is a constant. We will do this up to $\mathcal{O}(\epsilon_1)$

From the definition of $\epsilon_1= -\frac{\dot{H}^b}{\left(H^b\right)^2}$ and $\epsilon_2 = \frac{\ddot{H}^b}{\dot{H}^b H^b}-2\frac{\dot{H}^b}{\left(H^b\right)^2}$ together with the background equation of motion of the field we can write:

\begin{equation}
\epsilon_2=-6\left(1+\frac{V_{\phi}}{3H\dot{\phi}}\right)+2\epsilon_1,
\label{epsilon_2}
\end{equation}
and since $\frac{\dot{\epsilon}_1}{\epsilon_1}=H\epsilon_2$ we can write $\epsilon_1$ as
\begin{equation}
\epsilon_1=\epsilon_1^0a^{-6}\exp\left[-6\int\left(\frac{V_{\phi}}{3\dot{\phi}}-\frac{H\epsilon_1}{3}\right)dt\right],
\label{epsilon_1}
\end{equation}
where $\epsilon_1^0$ is the initial value of $\epsilon_1$ and we have used $N=\log a$.

The next step is to use the definition of $\tau$ and integrate by parts:
\begin{equation}
\tau=-\frac{1}{aH}+\int\frac{da}{a^2H}\epsilon_1.
\label{tau_exp}
\end{equation}

We now have to integrate by parts again the last term in \eqref{tau_exp} taking into account the result \eqref{epsilon_1}, after a straightforward computation we get:
\begin{equation}
\int\frac{da}{a^2H}\epsilon_1=-\frac{\epsilon_1}{7Ha}-\frac{6}{7}\int\frac{da}{Ha^2}\left(\frac{V_{\phi}}{3H\dot{\phi}}\right)\epsilon_1+\frac{3}{7}\int\frac{da}{Ha^2}\epsilon_1^2.
\label{tau_exp_step}
\end{equation}

The last term in \eqref{tau_exp_step} is second order in $\epsilon_1$ so we will neglect it. If we keep integrating by parts we will find terms proportional to $\frac{d^n}{dt^n}\left(\frac{V_{\phi}}{3H\dot{\phi}}\right)\epsilon_1$. However, for the regimes of interest (SR, USR or CR) we have $\left(\frac{V_{\phi}}{3H\dot{\phi}}\right)=\frac{\kappa}{3} + \mathcal{O}(\epsilon_1)$, so we can neglect all these terms and write an formula for $\tau$ valid up to first order in $\epsilon_1$.

\begin{equation}
\tau\simeq -\frac{1}{Ha}\Bigg\{1+\left[\sum_{i=0}^{\infty}(-1)^i\frac{6^i}{7^{i+1}}\left(\frac{V_{\phi}}{3H\dot{\phi}}\right)^i\right]\epsilon_1\Bigg\} = -\frac{1}{Ha}\left(1+\frac{1}{7+6\left(\frac{V_{\phi}}{3H\dot{\phi}}\right)}\epsilon_1\right)= -\frac{1}{Ha}\left(1+\frac{1}{7+2\kappa}\epsilon_1\right).
\label{tau_exp_full}
\end{equation}

Once we have the general expansion of $\tau$ it is easy to get $\nu$ from \eqref{zppz}, the result is 
\begin{equation}
\nu=\frac{3}{2}\sqrt{1-\frac{4}{9}\frac{V_{\phi\phi}}{H^2}}-\frac{3\left(15+12\kappa+2\kappa^2\right)}{|3+2\kappa|(7+2\kappa)}\epsilon_1,
\label{nu_exp_full}
\end{equation}
where we have used the definition of $\nu^2=\frac{1}{4}+\frac{z''}{z}\tau^2$ together with the following result:
\begin{equation}
\frac{V_{\phi\phi}}{H^2}=\left(6\epsilon_1-\frac{3}{2}\epsilon_2-2\epsilon_1^2+\frac{5}{2}\epsilon_1\epsilon_2-\frac{1}{4}\epsilon_2^2-\frac{1}{2}\epsilon_2\epsilon_3\right)=-3\kappa-\kappa^2+\mathcal{O}(\epsilon_1)
\label{Vpp}
\end{equation}

Let us finally apply \eqref{nu_exp_full} to SR and USR. 
\begin{itemize}
\item SR: At leading order in SR $\kappa\simeq -3$ and $\frac{V_{\phi\phi}}{H^2}\simeq 6\epsilon_1-\frac{3}{2}\epsilon_2$
\begin{equation}
\nu_{SR}=\frac{3}{2}+\epsilon_1+\frac{1}{2}\epsilon_2 = \frac{3}{2}+2\epsilon_1^0.
\label{nu_exp_SR}
\end{equation}
\item USR: The potential is exactly flat i.e. $\kappa=V_{\phi\phi}=0$
\begin{equation}
\nu_{USR}=\frac{3}{2}-\frac{15}{7}\epsilon_1=\frac{3}{2}-\frac{15}{7}\epsilon_1^0  \tau^6 H^6.
\label{nu_exp_USR}
\end{equation}
\end{itemize} 
In the last equality of each regime we have used the results \eqref{epsilon_2} and \eqref{epsilon_1} together with $\tau=-\frac{1}{a H}(1+\mathcal{O}(\epsilon_1))$ (where $H=constant$). These results obviously coincide with known results (see for example \cite{Pattison:2019hef}).

It is then obvious that the solution in terms of Henkel functions is valid at zeroth and first order in $\epsilon_1$ for SR but it is only valid at zeroth order in $\epsilon_1$ for USR and CR, this is because at order $\epsilon_1$, $\nu_{USR}$ is not a constant. The aim of this appendix is to give a solution for \eqref{MS_u_tau} up to $\epsilon_1$ which is valid for USR. The procedure to follow is very simple:

\begin{enumerate}
\item First we write the MS equation \eqref{MS_u_tau} at first order in $\epsilon_1$ using \eqref{tau_exp_full}
\begin{equation}
u_{\mathbf{k}}''(\tau)+\left(k^2-\frac{1}{\tau^2}\left(2-\frac{45}{7}\epsilon_1\right)\right)u_{\mathbf{k}}(\tau)=0.
\label{MS_u_USR}
\end{equation}
\item We then write explicitly the time dependence of $\epsilon_1$.
\begin{equation}
u_{\mathbf{k}}''(\tau)+\left(k^2-\frac{2}{\tau^2} + \frac{45}{7}\epsilon_1^0 H^6 \tau^4 \right)u_{\mathbf{k}}(\tau) = 0.
\label{MS_u_USR2}
\end{equation}
\item Eq. \eqref{MS_u_USR2} does not have analytical solution, however we know that the solution up to order $\epsilon_1$ must be of the form $u_{\mathbf{k}}(\tau) = u_{\mathbf{k}}^{(0)}(\tau) + \epsilon_1^0 u_{\mathbf{k}}^{(1)}(\tau)$. The equation that will follow each of the parts of the solution is:
\begin{align}
\nonumber
\left(u_{\mathbf{k}}^{(0)}\right)''(\tau)+\left(k^2-\frac{2}{\tau^2}\right)u_{\mathbf{k}}^{(0)}(\tau)=0,  \\
\left(u_{\mathbf{k}}^{(1)}\right)''(\tau)+\left(k^2-\frac{2}{\tau^2}\right)u_{\mathbf{k}}^{(1)}(\tau) + \frac{45}{7} H^6 \tau^4 u_{\mathbf{k}}^{(0)}(\tau) = 0.
\label{MS_u_USR3}
\end{align}

\item The final solution $u_{\mathbf{k}}(\tau)$ for USR is with the Bunch-Davies vacuum as initial condition is:
\begin{align}
\nonumber
u_{\mathbf{k}}(\tau) = \frac{e^{i k \tau}}{\sqrt{2 k}}\Bigg[1+\frac{i}{k \tau} + & \\
\frac{45}{7}\epsilon_0 H_0^6 \tau^6 & \left(\frac{i}{5 k \tau}-\frac{7}{10 (k \tau)^2}-\frac{7 i}{6 (k \tau)^3} + \frac{7}{6 (k \tau)^4} + \frac{7}{4 (k \tau)^6}+ \frac{7 i}{4 (k \tau)^7}\right)\Bigg]
\label{MS_u_USR_sol}
\end{align}
\end{enumerate}

From \eqref{MS_u_USR_sol} we can write the solution for $Q_{\mathbf{k}}=\frac{u_{\mathbf{k}}}{a}$ as:

\begin{align}
\nonumber
Q_{\mathbf{k}}(\tau) = - \frac{e^{i k \tau}H\tau}{\sqrt{2 k}}&\left(1-\frac{1}{7}\epsilon_0 H^6 \tau^6\right)\Bigg[1+\frac{i}{k \tau} +  \\
\frac{45}{7}\epsilon_0 H^6 \tau^6 & \left(\frac{i}{5 k \tau}-\frac{7}{10 (k \tau)^2}-\frac{7 i}{6 (k \tau)^3} + \frac{7}{6 (k \tau)^4} + \frac{7}{4 (k \tau)^6}+ \frac{7 i}{4 (k \tau)^7}\right)\Bigg]
\label{MS_Q_USR_sol}
\end{align}

When evaluating $k=\sigma a(N_{\star}) H$ and applying the limit $\sigma \rightarrow 0$ we are left with the following expression:

\begin{equation}
\nonumber
Q_{\mathbf{k}}(N_{\star}) = - \frac{e^{-i \sigma \left(1+\frac{1}{7}\epsilon_1\right)}H}{\sqrt{2}\left(\sigma a H\right)^{3/2}}\left(1-\frac{1}{7}\epsilon_1 \right)  \Bigg[i + \frac{45}{4}\frac{\epsilon_1}{\sigma^6} \Bigg]
\label{MS_Q_USR_sol_limN}
\end{equation}

Eq \eqref{MS_Q_USR_sol_limN} is written in terms of the time used for the stochastic simulation $N_{\star}$, if we want to relate it with the background we must shift the time variable according to $N=N_{\star}+\log (\sigma)$ (see discussion below \eqref{PS_SR_k}). The final expression is:

\begin{equation}
\nonumber
Q_{\mathbf{k}}(N) = - \frac{e^{-i \sigma}H}{\sqrt{2}(a H)^{3/2}}\Bigg[i + \frac{45}{4}\epsilon_1 \Bigg]
\label{MS_Q_USR_sol_final}
\end{equation}


\newpage
\begin{thebibliography}{99}
%\cite{Chapline:1975ojl}
\bibitem{Chapline:1975ojl}
G.~F.~Chapline,
%``Cosmological effects of primordial black holes,''
Nature \textbf{253} (1975) no.5489, 251-252
doi:10.1038/253251a0
%165 citations counted in INSPIRE as of 21 Jul 2021

\bibitem{ilia}
%\cite{Germani:2018jgr}
C.~Germani and I.~Musco,
%``Abundance of Primordial Black Holes Depends on the Shape of the Inflationary Power Spectrum,''
Phys. Rev. Lett. \textbf{122} (2019) no.14, 141302
doi:10.1103/PhysRevLett.122.141302
[arXiv:1805.04087 [astro-ph.CO]].
%127 citations counted in INSPIRE as of 21 Jul 2021

%\cite{Starobinsky:1994bd}
\bibitem{Starobinsky:1994bd}
A.~A.~Starobinsky and J.~Yokoyama,
%``Equilibrium state of a selfinteracting scalar field in the De Sitter background,''
Phys. Rev. D \textbf{50} (1994), 6357-6368
doi:10.1103/PhysRevD.50.6357
[arXiv:astro-ph/9407016 [astro-ph]].
%472 citations counted in INSPIRE as of 05 Jun 2020 

%\cite{Pattison:2021oen}
\bibitem{Pattison:2021oen}
C.~Pattison, V.~Vennin, D.~Wands and H.~Assadullahi,
%``Ultra-slow-roll inflation with quantum diffusion,''
[arXiv:2101.05741 [astro-ph.CO]].
%7 citations counted in INSPIRE as of 19 Jul 2021

%\cite{Firouzjahi:2018vet}
\bibitem{Firouzjahi:2018vet}
H.~Firouzjahi, A.~Nassiri-Rad and M.~Noorbala,
%``Stochastic Ultra Slow Roll Inflation,''
JCAP \textbf{01} (2019), 040
doi:10.1088/1475-7516/2019/01/040
[arXiv:1811.02175 [hep-th]].
%23 citations counted in INSPIRE as of 19 Jul 2021

%\cite{Prokopec:2019srf}
\bibitem{Prokopec:2019srf}
T.~Prokopec and G.~Rigopoulos,
%``$\Delta\mathcal{N}$ and the stochastic conveyor belt of Ultra Slow-Roll,''
[arXiv:1910.08487 [gr-qc]].
%6 citations counted in INSPIRE as of 19 Jul 2021

%\cite{Ballesteros:2020sre}
\bibitem{Ballesteros:2020sre}
G.~Ballesteros, J.~Rey, M.~Taoso and A.~Urbano,
%``Stochastic inflationary dynamics beyond slow-roll and consequences for primordial black hole formation,''
JCAP \textbf{08} (2020), 043
doi:10.1088/1475-7516/2020/08/043
[arXiv:2006.14597 [astro-ph.CO]].
%9 citations counted in INSPIRE as of 19 Jul 2021

%\cite{Casini:1998wr}
\bibitem{Casini:1998wr}
H.~Casini, R.~Montemayor and P.~Sisterna,
%``Stochastic approach to inflation. 2. Classicality, coarse graining and noises,''
Phys. Rev. D \textbf{59} (1999), 063512
doi:10.1103/PhysRevD.59.063512
[arXiv:gr-qc/9811083 [gr-qc]].
%26 citations counted in INSPIRE as of 19 Jul 2021

%\cite{Pattison:2017mbe}
\bibitem{Pattison:2017mbe}
C.~Pattison, V.~Vennin, H.~Assadullahi and D.~Wands,
%``Quantum diffusion during inflation and primordial black holes,''
JCAP \textbf{10} (2017), 046
doi:10.1088/1475-7516/2017/10/046
[arXiv:1707.00537 [hep-th]].
%75 citations counted in INSPIRE as of 19 Jul 2021

%\cite{Assadullahi:2016gkk}
\bibitem{Assadullahi:2016gkk}
H.~Assadullahi, H.~Firouzjahi, M.~Noorbala, V.~Vennin and D.~Wands,
%``Multiple Fields in Stochastic Inflation,''
JCAP \textbf{06} (2016), 043
doi:10.1088/1475-7516/2016/06/043
[arXiv:1604.04502 [hep-th]].
%42 citations counted in INSPIRE as of 19 Jul 2021

%\cite{Clesse:2010iz}
\bibitem{Clesse:2010iz}
S.~Clesse,
%``Hybrid inflation along waterfall trajectories,''
Phys. Rev. D \textbf{83} (2011), 063518
doi:10.1103/PhysRevD.83.063518
[arXiv:1006.4522 [gr-qc]].
%65 citations counted in INSPIRE as of 19 Jul 2021

%\cite{Arnowitt:1962hi}
\bibitem{Arnowitt:1962hi}
R.~L.~Arnowitt, S.~Deser and C.~W.~Misner,
%``The Dynamics of general relativity,''
Gen.\ Rel.\ Grav.\  {\bf 40} (2008) 1997
doi:10.1007/s10714-008-0661-1
[gr-qc/0405109].
%%CITATION = doi:10.1007/s10714-008-0661-1;%%
%1433 citations counted in INSPIRE as of 24 Mar 2020
 
%\cite{constraints}
\bibitem{constraints}
B.~Carr, K.~Kohri, Y.~Sendouda and J.~Yokoyama,
%``Constraints on Primordial Black Holes,''
[arXiv:2002.12778 [astro-ph.CO]].
%227 citations counted in INSPIRE as of 20 Jul 2021

%\cite{germani-sheth}
\bibitem{hu}
%\cite{Motohashi:2017kbs}
H.~Motohashi and W.~Hu,
%``Primordial Black Holes and Slow-Roll Violation,''
Phys. Rev. D \textbf{96} (2017) no.6, 063503
doi:10.1103/PhysRevD.96.063503
[arXiv:1706.06784 [astro-ph.CO]].
%141 citations counted in INSPIRE as of 21 Jul 2021

%\cite{germani-sheth}
\bibitem{germani-sheth}
C.~Germani and R.~K.~Sheth,
%``Nonlinear statistics of primordial black holes from Gaussian curvature perturbations,''
Phys. Rev. D \textbf{101} (2020) no.6, 063520
doi:10.1103/PhysRevD.101.063520
[arXiv:1912.07072 [astro-ph.CO]].
%47 citations counted in INSPIRE as of 20 Jul 2021

%\cite{Germani:2017bcs}
\bibitem{Germani:2017bcs}
C.~Germani and T.~Prokopec,
%``On primordial black holes from an inflection point,''
Phys. Dark Univ. \textbf{18} (2017), 6-10
doi:10.1016/j.dark.2017.09.001
[arXiv:1706.04226 [astro-ph.CO]].
%146 citations counted in INSPIRE as of 20 Jul 2021
  
%\cite{Kinney:2005vj}
\bibitem{Kinney:2005vj}
W.~H.~Kinney,
%``Horizon crossing and inflation with large eta,''
Phys. Rev. D \textbf{72} (2005), 023515
doi:10.1103/PhysRevD.72.023515
[arXiv:gr-qc/0503017 [gr-qc]].
%221 citations counted in INSPIRE as of 20 Jul 2021

%\cite{Martin:2012pe}
\bibitem{Martin:2012pe}
J.~Martin, H.~Motohashi and T.~Suyama,
%``Ultra Slow-Roll Inflation and the non-Gaussianity Consistency Relation,''
Phys. Rev. D \textbf{87} (2013) no.2, 023514
doi:10.1103/PhysRevD.87.023514
[arXiv:1211.0083 [astro-ph.CO]].
%174 citations counted in INSPIRE as of 20 Jul 2021

%\cite{Martin:2012pe}
\bibitem{atal}
V.~Atal and C.~Germani,
%``The role of non-gaussianities in Primordial Black Hole formation,''
Phys. Dark Univ. \textbf{24} (2019), 100275
doi:10.1016/j.dark.2019.100275
[arXiv:1811.07857 [astro-ph.CO]].
%62 citations counted in INSPIRE as of 20 Jul 2021
  
%\cite{Salopek:1990jq}
\bibitem{Salopek:1990jq}
D.~S.~Salopek and J.~R.~Bond,
%``Nonlinear evolution of long wavelength metric fluctuations in inflationary models,''
Phys.\ Rev.\ D {\bf 42} (1990) 3936.
doi:10.1103/PhysRevD.42.3936
%%CITATION = doi:10.1103/PhysRevD.42.3936;%%
%854 citations counted in INSPIRE as of 24 Mar 2020
  
%\cite{Lyth:2004gb}
\bibitem{Lyth:2004gb}
D.~H.~Lyth, K.~A.~Malik and M.~Sasaki,
%``A General proof of the conservation of the curvature perturbation,''
JCAP {\bf 0505} (2005) 004
doi:10.1088/1475-7516/2005/05/004
[astro-ph/0411220].
%%CITATION = doi:10.1088/1475-7516/2005/05/004;%%
%604 citations counted in INSPIRE as of 24 Mar 2020
 
%\cite{Vennin:2015hra}
\bibitem{Vennin:2015hra}
V.~Vennin and A.~A.~Starobinsky,
%``Correlation Functions in Stochastic Inflation,''
Eur. Phys. J. C \textbf{75} (2015), 413
doi:10.1140/epjc/s10052-015-3643-y
[arXiv:1506.04732 [hep-th]].
%72 citations counted in INSPIRE as of 05 Jun 2020   
  
%\cite{Kunze:2006tu}
\bibitem{Kunze:2006tu}
K.~E.~Kunze,
%``Perturbations in stochastic inflation,''
JCAP \textbf{07} (2006), 014
doi:10.1088/1475-7516/2006/07/014
[arXiv:astro-ph/0603575 [astro-ph]].
%15 citations counted in INSPIRE as of 05 Jun 2020  
 
%\cite{Grain:2017dqa}
\bibitem{Grain:2017dqa}
J.~Grain and V.~Vennin,
%``Stochastic inflation in phase space: Is slow roll a stochastic attractor?,''
JCAP \textbf{05} (2017), 045
doi:10.1088/1475-7516/2017/05/045
[arXiv:1703.00447 [gr-qc]].
%52 citations counted in INSPIRE as of 19 Jul 2021

%\cite{Clesse:2015wea}
\bibitem{Clesse:2015wea}
S.~Clesse and J.~Garc\'\i{}a-Bellido,
%``Massive Primordial Black Holes from Hybrid Inflation as Dark Matter and the seeds of Galaxies,''
Phys. Rev. D \textbf{92} (2015) no.2, 023524
doi:10.1103/PhysRevD.92.023524
[arXiv:1501.07565 [astro-ph.CO]].
%253 citations counted in INSPIRE as of 19 Jul 2021 
  
%\cite{Maldacena:2002vr}
\bibitem{Maldacena:2002vr}
J.~M.~Maldacena,
%``Non-Gaussian features of primordial fluctuations in single field inflationary models,''
JHEP \textbf{05} (2003), 013
doi:10.1088/1126-6708/2003/05/013
[arXiv:astro-ph/0210603 [astro-ph]].
%2261 citations counted in INSPIRE as of 19 Jul 2021   
  
%\cite{Sugiyama:2012tj}
\bibitem{Sugiyama:2012tj}
N.~S.~Sugiyama, E.~Komatsu and T.~Futamase,
%``$\delta$N formalism,''
Phys.\ Rev.\ D {\bf 87} (2013) no.2,  023530
doi:10.1103/PhysRevD.87.023530
[arXiv:1208.1073 [gr-qc]].
%%CITATION = doi:10.1103/PhysRevD.87.023530;%%
%42 citations counted in INSPIRE as of 24 Mar 2020
  
%\cite{Tanaka:2021dww}
\bibitem{Tanaka:2021dww}
T.~Tanaka and Y.~Urakawa,
%``Anisotropic separate universe and Weinberg's adiabatic mode,''
JCAP \textbf{07} (2021), 051
doi:10.1088/1475-7516/2021/07/051
[arXiv:2101.05707 [astro-ph.CO]].
%0 citations counted in INSPIRE as of 28 Oct 2021
 
%\cite{Garriga:2015tea}
\bibitem{Garriga:2015tea}
J.~Garriga, Y.~Urakawa and F.~Vernizzi,
%``$\delta N$ formalism from superpotential and holography,''
JCAP \textbf{02} (2016), 036
doi:10.1088/1475-7516/2016/02/036
[arXiv:1509.07339 [hep-th]].
%22 citations counted in INSPIRE as of 28 Oct 2021   
  
%\cite{Starobinsky:1986fx}
\bibitem{Starobinsky:1986fx}
A.~A.~Starobinsky,
%``Stochastic De Sitter (inflationary) Stage In The Early Universe,''
Lect.\ Notes Phys.\  {\bf 246} (1986) 107.
doi:10.1007/3-540-16452-9\_6
%%CITATION = doi:10.1007/3-540-16452-9_6;%%
%218 citations counted in INSPIRE as of 30 Mar 2020  
  
%\cite{Figueroa:2020jkf}
\bibitem{Figueroa:2020jkf}
D.~G.~Figueroa, S.~Raatikainen, S.~Rasanen and E.~Tomberg,
%``Non-Gaussian tail of the curvature perturbation in stochastic ultra-slow-roll inflation: implications for primordial black hole production,''
[arXiv:2012.06551 [astro-ph.CO]].
%5 citations counted in INSPIRE as of 19 Jul 2021  
  
%\cite{Kiefer:2008ku}
\bibitem{Kiefer:2008ku}
C.~Kiefer and D.~Polarski,
%``Why do cosmological perturbations look classical to us?,''
Adv. Sci. Lett. \textbf{2} (2009), 164-173
doi:10.1166/asl.2009.1023
[arXiv:0810.0087 [astro-ph]].
%135 citations counted in INSPIRE as of 01 Feb 202

%\cite{Grishchuk:1990bj}
\bibitem{Grishchuk:1990bj}
L.~P.~Grishchuk and Y.~V.~Sidorov,
%``Squeezed quantum states of relic gravitons and primordial density fluctuations,''
Phys. Rev. D \textbf{42} (1990), 3413-3421
doi:10.1103/PhysRevD.42.3413
%312 citations counted in INSPIRE as of 01 Feb 2021  

%\cite{Ramos:2013nsa}
\bibitem{Ramos:2013nsa}
R.~O.~Ramos and L.~A.~da Silva,
%``Power spectrum for inflation models with quantum and thermal noises,''
JCAP \textbf{03} (2013), 032
doi:10.1088/1475-7516/2013/03/032
[arXiv:1302.3544 [astro-ph.CO]].
%82 citations counted in INSPIRE as of 19 Jul 2021
  
%\cite{Pattison:2019hef}
\bibitem{Pattison:2019hef}
C.~Pattison, V.~Vennin, H.~Assadullahi and D.~Wands,
%``Stochastic inflation beyond slow roll,''
JCAP {\bf 1907} (2019) 031
doi:10.1088/1475-7516/2019/07/031
[arXiv:1905.06300 [astro-ph.CO]].
%%CITATION = doi:10.1088/1475-7516/2019/07/031;%%
%12 citations counted in INSPIRE as of 30 Mar 2020    
  
%\cite{Tanaka:2007gh}
\bibitem{Tanaka:2007gh}
Y.~Tanaka and M.~Sasaki,
%``Gradient expansion approach to nonlinear superhorizon perturbations. II. A Single scalar field,''
Prog.\ Theor.\ Phys.\  {\bf 118} (2007) 455
doi:10.1143/PTP.118.455
[arXiv:0706.0678 [gr-qc]].
%%CITATION = doi:10.1143/PTP.118.455;%%
%25 citations counted in INSPIRE as of 25 Mar 2020  
  
%\cite{Hamazaki:2008mh}
\bibitem{Hamazaki:2008mh}
T.~Hamazaki,
%``Long wavelength limit of evolution of nonlinear cosmological perturbations,''
Phys. Rev. D \textbf{78} (2008), 103513
doi:10.1103/PhysRevD.78.103513
[arXiv:0811.2366 [astro-ph]].
%9 citations counted in INSPIRE as of 19 Jul 2021  
 
%\cite{Takamizu:2013gy}
\bibitem{Takamizu:2013gy}
Y.~i.~Takamizu and T.~Kobayashi,
%``Nonlinear superhorizon curvature perturbation in generic single-field inflation,''
PTEP \textbf{2013} (2013) no.6, 063E03
doi:10.1093/ptep/ptt033
[arXiv:1301.2370 [gr-qc]].
%9 citations counted in INSPIRE as of 03 Jan 2021

%\cite{Takamizu:2010xy}
\bibitem{Takamizu:2010xy}
Y.~i.~Takamizu, S.~Mukohyama, M.~Sasaki and Y.~Tanaka,
%``Non-Gaussianity of superhorizon curvature perturbations beyond $\delta$ N formalism,''
JCAP \textbf{06} (2010), 019
doi:10.1088/1475-7516/2010/06/019
[arXiv:1004.1870 [astro-ph.CO]].
%45 citations counted in INSPIRE as of 03 Jan 2021  
 
%\cite{Langlois:2005ii}
\bibitem{Langlois:2005ii}
D.~Langlois and F.~Vernizzi,
%``Evolution of non-linear cosmological perturbations,''
Phys. Rev. Lett. \textbf{95} (2005), 091303
doi:10.1103/PhysRevLett.95.091303
[arXiv:astro-ph/0503416 [astro-ph]].
%130 citations counted in INSPIRE as of 19 Jul 2021

%\cite{Rigopoulos:2004gr}
\bibitem{Rigopoulos:2004gr}
G.~I.~Rigopoulos and E.~P.~S.~Shellard,
%``Non-linear inflationary perturbations,''
JCAP \textbf{10} (2005), 006
doi:10.1088/1475-7516/2005/10/006
[arXiv:astro-ph/0405185 [astro-ph]].
%77 citations counted in INSPIRE as of 19 Jul 2021

%\cite{Wang:2013ic}
\bibitem{Wang:2013ic}
J.~Wang,
%``Construction of the conserved \ensuremath{\zeta} via the effective action for perfect fluids,''
Annals Phys. \textbf{362} (2015), 223-238
doi:10.1016/j.aop.2015.07.013
[arXiv:1301.7089 [gr-qc]].
%2 citations counted in INSPIRE as of 26 Jul 2021 
 
%\cite{Robler:2010}
\bibitem{Robler:2010}
A.~Rößler.
%``Runge-Kutta methods for the strong approximation of solutions of stochastic differential equations,''
SIAM J. Numer. Anal., \textbf{48} (3) : 922–952, 2010.
doi.org/10.1137/09076636X

%\cite{Kloeden:1992}
\bibitem{Kloeden:1992}
P.~E.~Kloeden,  E.~Platen. 
%``Numerical Solution of Stochastic Differential Equations. ''
Springer-Verlag, Berlin, 1992.
doi:10.1007/978-3-662-12616-5

%\cite{Burrage:2006}
\bibitem{Burrage:2006}
K.~Burrage, P.~Burrage, D.~J.~Higham, P.~E.~Kloeden and E.~Platen, 
%``Comment on ”numerical methods for stochastic differential equations”.''
Phys. Rev. E \textbf{74}, 068701 (2006). 1
doi:10.1103/PhysRevE.74.068701

%\cite{Boubekeur:2005zm}
\bibitem{Boubekeur:2005zm}
L.~Boubekeur and D.~H.~Lyth,
%``Hilltop inflation,''
JCAP \textbf{07} (2005), 010
doi:10.1088/1475-7516/2005/07/010
[arXiv:hep-ph/0502047 [hep-ph]].
%338 citations counted in INSPIRE as of 18 Mar 2021 
  
%\cite{Cruces:2018cvq}
\bibitem{Cruces:2018cvq}
D.~Cruces, C.~Germani and T.~Prokopec,
%``Failure of the stochastic approach to inflation beyond slow-roll,''
JCAP {\bf 1903} (2019) 048
doi:10.1088/1475-7516/2019/03/048
[arXiv:1807.09057 [gr-qc]].
%%CITATION = doi:10.1088/1475-7516/2019/03/048;%%
%20 citations counted in INSPIRE as of 25 Mar 2020
  
%\cite{Mukhanov:1990me}
\bibitem{Mukhanov:1990me}
V.~F.~Mukhanov, H.~A.~Feldman and R.~H.~Brandenberger,
%``Theory of cosmological perturbations. Part 1. Classical perturbations. Part 2. Quantum theory of perturbations. Part 3. Extensions,''
Phys.\ Rept.\  {\bf 215} (1992) 203.
doi:10.1016/0370-1573(92)90044-Z
%%CITATION = doi:10.1016/0370-1573(92)90044-Z;%%
%2889 citations counted in INSPIRE as of 27 Mar 2020
  
%\cite{Naruko:2012fe}
\bibitem{Naruko:2012fe}
A.~Naruko, Y.~i.~Takamizu and M.~Sasaki,
%``Beyond \delta N formalism,''
PTEP {\bf 2013} (2013) 043E01
doi:10.1093/ptep/ptt008
[arXiv:1210.6525 [astro-ph.CO]].
%%CITATION = doi:10.1093/ptep/ptt008;%%
%17 citations counted in INSPIRE as of 27 Mar 2020

%\cite{Vennin:2020kng}
\bibitem{Vennin:2020kng}
V.~Vennin,
%``Stochastic inflation and primordial black holes,''
[arXiv:2009.08715 [astro-ph.CO]].
%5 citations counted in INSPIRE as of 22 Jul 2021
  
%\cite{Bunch:1978yq}
\bibitem{Bunch:1978yq}
T.~S.~Bunch and P.~C.~W.~Davies,
%``Quantum Field Theory in de Sitter Space: Renormalization by Point Splitting,''
Proc.\ Roy.\ Soc.\ Lond.\ A {\bf 360} (1978) 117.
doi:10.1098/rspa.1978.0060
%%CITATION = doi:10.1098/rspa.1978.0060;%%
%907 citations counted in INSPIRE as of 30 Mar 2020
\end{thebibliography}
\end{document}